\documentclass[useAMS,usenatbib]{mn2e}
\usepackage{amsmath, amssymb}
\usepackage{natbib}
\usepackage{amssymb}
\usepackage{graphicx}
\usepackage{multirow}
\usepackage{times}
\usepackage[normalem]{ulem}
\usepackage{color}
\usepackage[T1]{fontenc}
\def\gsim{ \lower .75ex \hbox{$\sim$} \llap{\raise .27ex \hbox{$>$}} }
\def\lsim{ \lower .75ex\hbox{$\sim$} \llap{\raise .27ex \hbox{$<$}} }
\def\rH{r_{\rm H}}

\title[Simulations of Magnetized Advection Dominated Accretion]
  {GRMHD Simulations of Magnetized Advection
  Dominated Accretion on a Non-Spinning Black Hole: Role of Outflows} 
\author[R. Narayan, A. S\k{a}dowski, R. F. Penna, A. K. Kulkarni] {Ramesh
  Narayan$^1$\thanks{E-mail: rnarayan@cfa.harvard.edu (RN);
    asadowski@cfa.harvard.edu (AS); rpenna@cfa.harvard.edu (RFP)},
  Aleksander S\k{a}dowski$^1$, Robert F. Penna$^1$, Akshay
  K. Kulkarni\\ \\
$^1${Harvard-Smithsonian Center for Astrophysics, 60 Garden Street, Cambridge, MA 02138, USA}\\
}

\begin{document}
\date{\today}
\maketitle
\label{firstpage}

\begin{abstract}

We present results from two long-duration GRMHD simulations of
advection-dominated accretion around a non-spinning black hole. The
first simulation was designed to avoid significant accumulation of
magnetic flux around the black hole. This simulation was run for a
time of $200,000GM/c^3$ and achieved inflow equilibrium out to a
radius $\sim90GM/c^2$. Even at this relatively large radius, the mass
outflow rate $\dot{M}_{\rm out}$ is found to be only 60\% of the net
mass inflow rate $\dot{M}_{\rm BH}$ into the black hole. The second
simulation was designed to achieve substantial magnetic flux
accumulation around the black hole in a magnetically arrested
disc. This simulation was run for a shorter time of $100,000GM/c^3$.
Nevertheless, because the mean radial velocity was several times
larger than in the first simulation, it reached inflow equilibrium out
to a radius $\sim170GM/c^2$.  Here, $\dot{M}_{\rm out}$ becomes equal
to $\dot{M}_{\rm BH}$ at $r\sim 160GM/c^2$. Since the mass outflow
rates in the two simulations do not show robust convergence with time,
it is likely that the true outflow rates are lower than our
estimates. The effect of black hole spin on mass outflow remains to be
explored.  Neither simulation shows strong evidence for convection,
though a complete analysis including the effect of magnetic fields is
left for the future.

\end{abstract}


\section{Introduction}
\label{sec:intro}

Black hole (BH) accretion occurs via at least two distinct modes: (1)
a standard thin accretion disc \citep*{SS73,NT73,FKR02}, and (2) an
advection-dominated accretion flow (ADAF,
\citealt{NY94,NY95,Abramowicz+95,Ichimaru77}; see \citealt*{NMQ98};
\citealt*{FKR02,NM08}; \citealt*{Kato+08} for reviews). Thin discs are
present around stellar-mass and supermassive BHs that accrete at a
substantial fraction $\sim$ few--100\% of the Eddington rate, while
ADAFs are typically found at lower accretion rates
$\dot{M}$.\footnote{Actually, two distinct ADAF modes are possible,
  one in which optically thin two-temperature gas accretes with a
  highly sub-Eddington $\dot{M}$, and a second in which very optically
  thick radiation-trapped gas accretes at rates well above the
  Eddington rate. We are concerned in this paper with the former kind
  of ADAF, which our GRMHD code is capable of simulating. The latter
  variety of ADAF is referred to as a ``slim disc''
  \citep{Abramowicz+88} and requires a radiation MHD code to simulate
  (see \citealt{Ohsuga+09,Ohsuga+11}; and references therein).}

The accreting gas in an ADAF is radiatively inefficient; hence an ADAF
is also referred to as a radiatively inefficient accretion flow
(RIAF). The low radiative efficiency, on top of the already low
accretion rate, makes ADAFs highly underluminous and difficult to
observe.  On the other hand, the vast majority of both stellar-mass
and supermassive BHs in the universe are in the ADAF state, a notable
example being Sgr A$^*$, the supermassive BH at the center of our own
Galaxy \citep*{NYM95,YQN03}.

A simple one-dimensional model of gas dynamics in an ADAF \citep{NY94}
reveals two interesting complications. First, the Bernoulli parameter
of the gas tends to be positive. This means that the gas is not
gravitationally bound to the BH, or at best is only weakly bound.
Therefore, an ADAF is likely to have powerful jets and mass outflows,
as recognized in the very first papers \citep{NY94,NY95b}.  The
connection between ADAFs and relativistic jets has become increasingly
clear over the years (e.g., \citealt{NM08} and references
therein). However, it is presently unknown whether or not ADAFs have
quasi- or non-relativistic winds, and if so how much mass they lose
via these winds.

Some authors (e.g., \citealt{BB99,Begelman12}) have suggested that
winds in ADAFs are so powerful that the mass accretion rate
$\dot{M}_{\rm BH}$ on the BH is as much as $\sim5$ orders of magnitude
less than the mass supply rate $\dot{M}_{\rm supply}$ at the outer
edge of the accretion flow, say at the Bondi radius. In effect, these
authors took the Bernoulli argument for strong outflows proposed in
the original ADAF papers \citep{NY94,NY95b}, and prostulated that
ADAFs would have not just strong outflows, but overwhelmingly strong
outflows.  Other authors (\citealt*{Ogilvie99,Abramowicz+00}),
however, argued that the Bernoulli parameter is not a good diagnostic
for mass loss, especially in the case of viscous non-steady flows.

\citet*{YQN03} attempted to constrain the mass loss in Sgr A$^*$ using
radio data on Faraday rotation
\citep{Aitken+00,QG00a,Agol00,Bower+03,Marrone+07}. They concluded
that, for this source, the decrease of $\dot{M}$ between the Bondi
radius and the BH is on the order of one to two orders of magnitude.
More recently, a few studies (e.g., \citealt{Allen+06};
\citealt*{McNamara+11}) have shown that many radio-loud active
galactic nuclei require a power source comparable to or even greater
than what Bondi accretion can supply.  Even if the power source of the
jet is BH spin energy, one still requires a significant mass accretion
rate on to the BH to tap this spin power
\citep*{NF11,Tchekhovskoy+11}. Therefore, in the above radio sources,
there cannot be significant mass loss between the Bondi radius and the
BH horizon.

The second potential complication in the dynamics of ADAFs is that the
entropy gradient is large and highly unstable according to the
Schwarzschild criterion \citep{NY94}. One might thus suspect that
ADAFs will be very convective. On the other hand, the angular momentum
profile has a stable gradient. It is thus not clear whether the flow
is ultimately stable or unstable to convection.  Analytical models of
convection-dominated accretion flows (CDAFs; \citealt*{NIA00};
\citealt{QG00b}) have been developed, but their relevance to real
ADAFs is unclear (see \citealt{Narayan+02,BH02}; for conflicting
views).

Both mass-loss and convection involve multi-dimensional flows, which
are best studied via numerical simulations. In addition, since the
``viscosity'' that drives accretion originates in the
magnetorotational instability (MRI, \citealt{BH91,BH98}), magnetic
fields play a critical role. This makes analytical studies even less
tractable. Fortunately, multidimensional numerical MHD simulations are
now feasible. Indeed, the limit of a non-radiative ADAF is relatively
easy to simulate, since there is no radiation physics
involved. Moreover, ADAFs are geometrically thick and are less
demanding in terms of spatial resolution. We briefly review here the
large literature on ADAF simulations.

Early numerical simulations of ADAFs employed pseudo-Newtonian codes
with purely hydrodynamic viscosity. Pioneering work by \citet*{SPB99}
indicated that such flows are convective and that a significant
fraction of the inflowing mass near the equatorial plane flows out
along the poles in a strong outflow. Similar results, viz.,
convection, equatorial inflow and bipolar outflow, were obtained also
by \citet{IA99,IA00}. In the latter paper, the authors found that
bipolar outflows required high values of the viscosity parameter
$\alpha$, while low-viscosity models exhibited weaker outflows but had
strong convection. Very recently, \citet*[][see also
  \citealt*{YBW12}]{YWB12} have carried out 2D hydrodynamic
simulations of ADAFs which cover a very large range of radius and show
fairly strong outflows. Most of the outflowing gas is bound to the BH
in the sense that it has a negative Bernoulli parameter, yet it
reaches the outer boundary of the simulation without turning
around. \citet*{LOS12} have carried out hydrodynamic simulations of
ADAFs including the effects of bremsstrahlung cooling and electron
thermal conduction.

Although interesting, hydrodynamic $\alpha$-viscosity simulations are
ultimately not realistic since accretion flows have magnetic fields
and MRI-driven turbulence.  It is thus necessary to include magnetic
fields consistently.  Pseudo-Newtonian magneto-hydrodynamic (MHD)
simulations have been performed by a number of authors.
\citet*{Machida+00} and \citet*{Machida+01} observed temporary
outflows of mass in their MHD simulations and showed that substantial
accretion energy can be released in the vicinity of the BH via
magnetic reconnection. They also claimed that the initial
configuration of the magnetic field may play an important role in
determining the mass outflow rate.  Using axisymmetric (2D) models,
\citet{SP01} showed that significant outflows originate at radii
beyond $r\sim10$ (we express lengths in BH mass units: $GM/c^2$).
Similarly, \citet{HB02} observed outflows for all radii outside the
innermost stable circular orbit (ISCO), though they used a definition
of inflow and outflow based on cyclindrical coordinates (all other
authors use spherical coordinates) which makes their outflow estimates
somewhat ambiguous.

Convective motions were evident in MHD simulations performed by
\citet{Machida+01}, indicating, according to the authors, that
convection is a rather general phenomenon in radiatively inefficient
accretion flows. On the other hand, \citet{SP01} concluded that the
turbulence seen in their MHD simulations was driven by the MRI, not
convection. Similarly, \citet{HB02} noted that, although their models
were unstable according to the classical Hoiland criteria, the flows
appeared not to be convective.  On the other hand, a simulation by
\citet*{INA03}, which was initialized with purely toroidal magnetic
field, showed significant convection, and appeared to be similar to a
CDAF. The same authors found that, if they initialized the simulation
with a poloidal magnetic field, the disc structure was completely
different from the toroidal case.  The poloidal case led to a
configuration in which the magnetic field strongly resisted the
accreting gas, leading to what the authors later called a
``magnetically arrested disc'' (MAD, \citealt*{NIA03}).  In a series
of numerical MHD simulations, \citet*{Pen+03} and \citet*{Pang+11}
found little evidence for either outflows or convection.  Even though
the entropy gradient was unstable the gas was apparently prevented
from becoming convective by the magnetic field.  They coined the term
``frustrated convection'' to describe this behavior.

Beginning with the work of \citet*{DeVilliers+03}, accretion flows
have been studied using general relativistic magneto-hydrodynamic
(GRMHD) codes. \citet{DeVilliers+03} observed two kinds of outflows:
bipolar unbound jets and bound coronal flow.  The coronal flow
supplied gas and magnetic field to the coronal envelope, but
apparently did not have sufficient energy to escape to infinity.  The
jets on the other hand were relativistic and escaped easily, though
carrying very little mass. Jets have been studied in detail by a
number of authors \citep{McKinney_Gammie_04,DeVilliers+05,McKinney06}.
\citet{Beckwith+08,Beckwith+09} and \citet{MB09} noted that the power
emerging in the jets depended strongly on the assumed magnetic field
configuration. While dipolar fields produced strong jets, a
quadrupolar field led to only weak, turbulent outflows.

\citet{Tchekhovskoy+11} simulated a MAD system around a rapidly
spinning BH, and obtained very powerful jets with energy efficiency
$\eta>100\%$, i.e., jet power greater than 100\% of $\dot{M}_{\rm
  BH}c^2$, where $\dot{M}_{\rm BH}$ is the mass accretion rate on to
the BH. Their work showed beyond doubt that at least some part of the
jet power had to be extracted from the spin energy of the BH.  The
jet-spin connection for MAD systems has been explored in greater
detail by \citet*{MTB12}. These authors coined the term ``magnetically
choked accretion flow'' (MCAF) to describe the MAD configuration.

Returning to the present paper, the goal here is to use GRMHD
simulations of ADAFs around BHs to investigate the importance of mass
outflows, and if possible convection. Our simulations are run for a
longer duration than most previous work.  The questions we address
require us to analyze the properties of the accretion flow over as
wide a range of radius as possible.  The only way to obtain converged
results over such large volumes is by running simulations for a very
long time. We introduce a new measure of convergence, or more
accurately a test of internal consistency.  As per this criterion, our
simulations give converged time-steady flows over a range of up to 100
in radius. This turns out to be still not as large as we would
like. Nevertheless, it permits us to reach some interesting
conclusions.

Within the realm of ADAFs, we expect answers to depend on several
factors. One important factor has already been mentioned, viz., the
magnetic field topology in the accreting gas.  The role of field
topology for mass outflows (as distinct from relativistic jets) has
been largely unexplored. The recent work of \citet{MTB12} is one of
the first studies in this area.

In this paper we consider two distinct magnetic topologies and
describe one long-duration simulation for each topology. In one
simulation, we carefully arrange the initial seed magnetic field
(which is later amplified via the MRI) such that the accreting gas
does not become magnetically arrested despite the long duration of the
simulation. We call this the ADAF/SANE simulation (where SANE stands
for ``standard and normal evolution'').  In the second simulation, we
set up the magnetic field topology such that the accretion flow very
quickly becomes magnetically arrested and then remains in this state
for the duration of the run. We call this the ADAF/MAD simulation
(where, as stated earlier, MAD stands for ``magnetically arrested
disc'').

A second obvious parameter that will affect the properties of an ADAF
is the spin of the central BH. Numerical studies of jets, for
instance, clearly show that jet power correlates strongly with BH spin
\citep{McKinney06,Tchekhovskoy+11,Tchekhovskoy+12,TM12}. Observationally
too, there is evidence for such a correlation \citep{NM12}. In this
paper we focus on the case of a non-spinning BH: $a_* \equiv a/M
=0$. We view such a system as the purest form of an ADAF, where the
only available energy source is gravitational potential energy
released via accretion. By simulating an ADAF around a non-spinning BH
using a GRMHD code, we can more easily relate our results to
analytical studies as well as previous non-relativistic
simulations. In the future we plan to run long-duration GRMHD
simulations of ADAFs around spinning BHs. Those simulations will have
two sources of energy, accretion and BH spin. By comparing them with
the simulations described here we should be able to evaluate the role
of BH spin.

The plan of the paper is as follows.  In \S2, we briefly describe the
simulation methods we employ, which are similar to those we have used
in previous work.  In \S3, we discuss in detail our results from the
ADAF/SANE and ADAF/MAD simulations, focusing in particular on mass
outflows. In \S4, we bring together the results of the previous
sections and try to assess the nature of the accretion flow in the two
simulations. In \S5, we conclude with a discussion.

\section{Details of the Simulations}
\label{sec:sim}

\subsection{Computation Method}

The simulations described here were done with the 3D GRMHD code HARM
\citep*{Gammie+03,McKinney06,MB09}, which solves the ideal MHD
equations of motion of magnetized gas in the fixed general
relativistic metric of a stationary BH. The equation of state of the
gas is taken to be $u=p/(\Gamma-1)$, where $u$ and $p$ are the
internal energy and pressure, and $\Gamma$ is the adiabatic index. The
code conserves energy to machine precision, hence any energy lost at
the grid scale, e.g., through turbulent dissipation or numerical
reconnection, is returned as entropy of the gas.  There is no
radiative cooling. The code works in dimensionless units where
$GM=c=1$.  Thus, all lengths and times in this paper are given in
units of $GM/c^2$ and $GM/c^3$, respectively.

\begin{figure*}
\begin{center}
\includegraphics[width=0.449\textwidth]{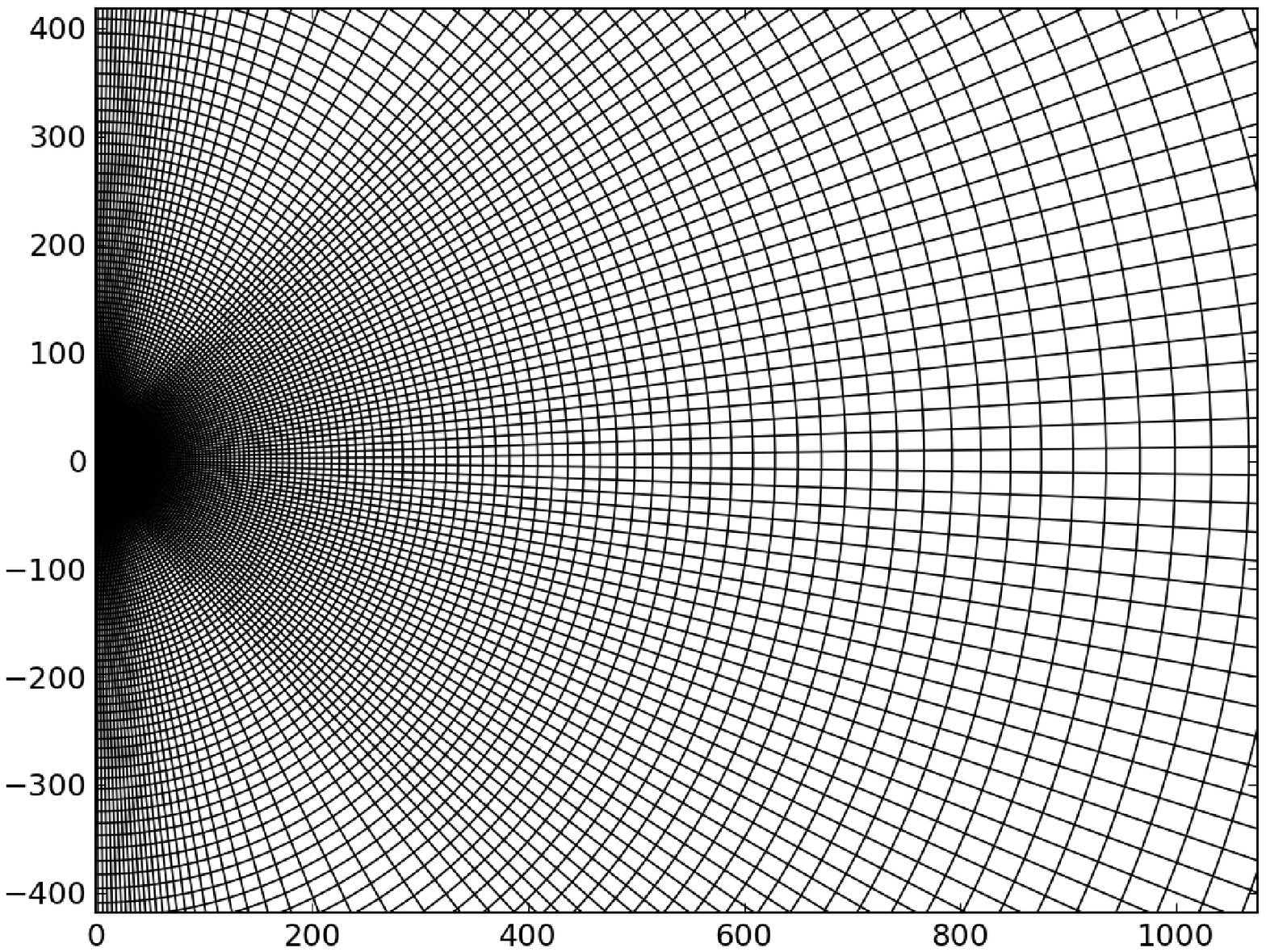}
\includegraphics[width=0.449\textwidth]{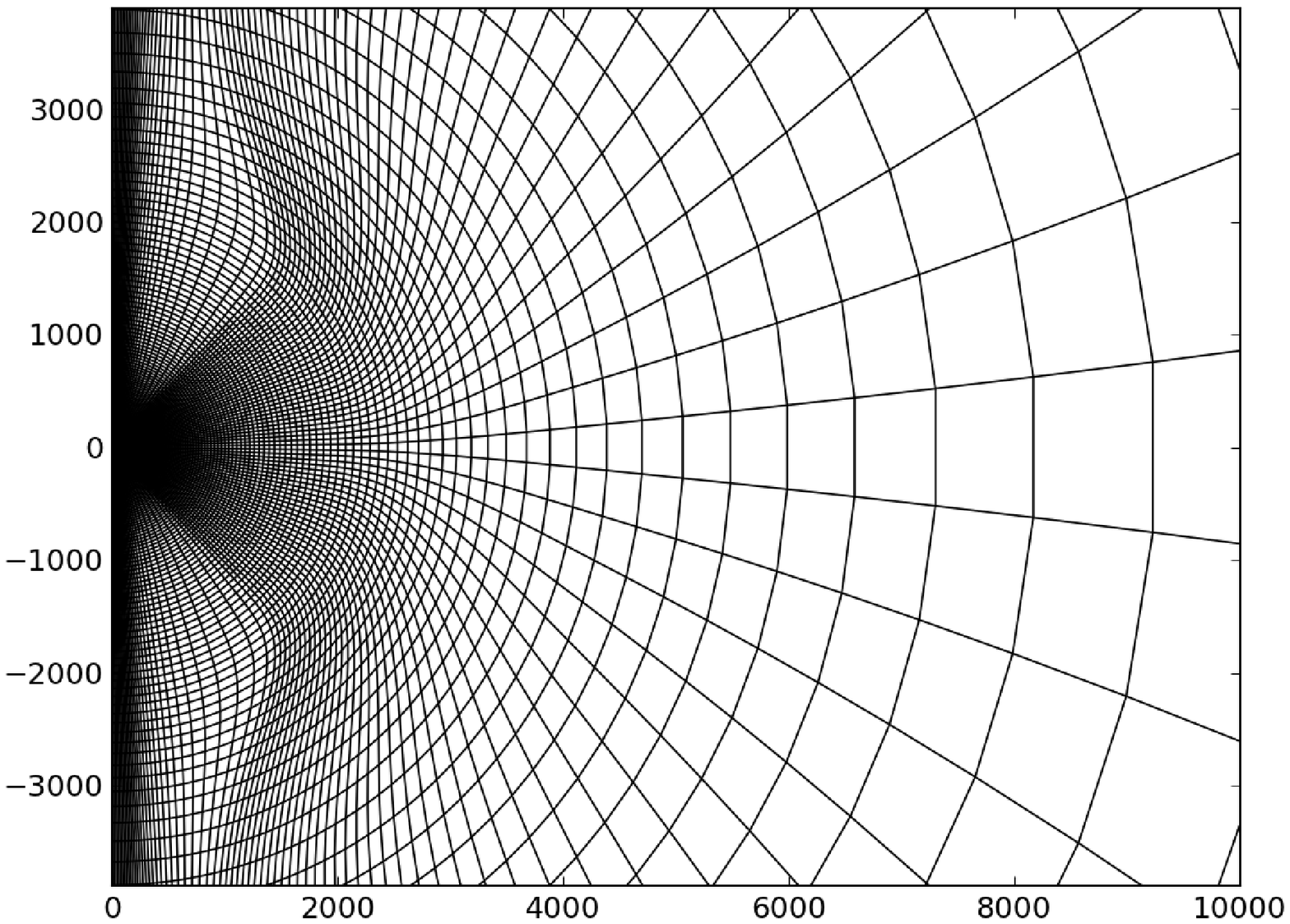}
\end{center}
\caption{Poloidal plane of the grid used in the simulations, shown at
  two zoom levels. }
\label{fig:grid}
\end{figure*}

A key feature of our simulations is the extremely long run time:
$200,000$ time units for the ADAF/SANE simulation, and $100,000$ time
units for the ADAF/MAD simulation.  To avoid spurious signals reaching
the region of interest from the boundary of the simulation, our grid
extends out to a very large radius $\sim10^5$. At the same time, we
require good resolution in the inner regions in order to study the
structure of the flow. To satisfy both requirements, we use a grid
with 256 cells in the radial direction, where the cells are
distributed uniformly in $\log r$ at smaller radii and spaced
hyper-logarithmically near the outermost radii.

In the $\theta$ direction, we employ 128 cells, distributed
non-uniformly so as to provide adequate resolution both in the
geometrically thick equatorial region, where the bulk of the gas
accretes, and the polar region, where a relativistic jet might flow
out. In order to follow such a jet as it collimates at large distance,
we use the grid developed by \citet{Tchekhovskoy+11} in which the
$\theta$ resolution near the pole increases with increasing radius
(see Fig.~\ref{fig:grid})\footnote{As it happens there is no
  significant jet in the simulations described here. However, we plan
  to use the same grid setup and initial conditions in future work
  with spinning BHs, where we do expect to see strong jets.}.

Finally, we use a uniform grid of 64 cells in the azimuthal direction,
covering the full range of $\phi$ from $0$ to $2\pi$.

\subsection{Initial Conditions}

\begin{figure*}
\begin{center}
\includegraphics[width=0.449\textwidth]{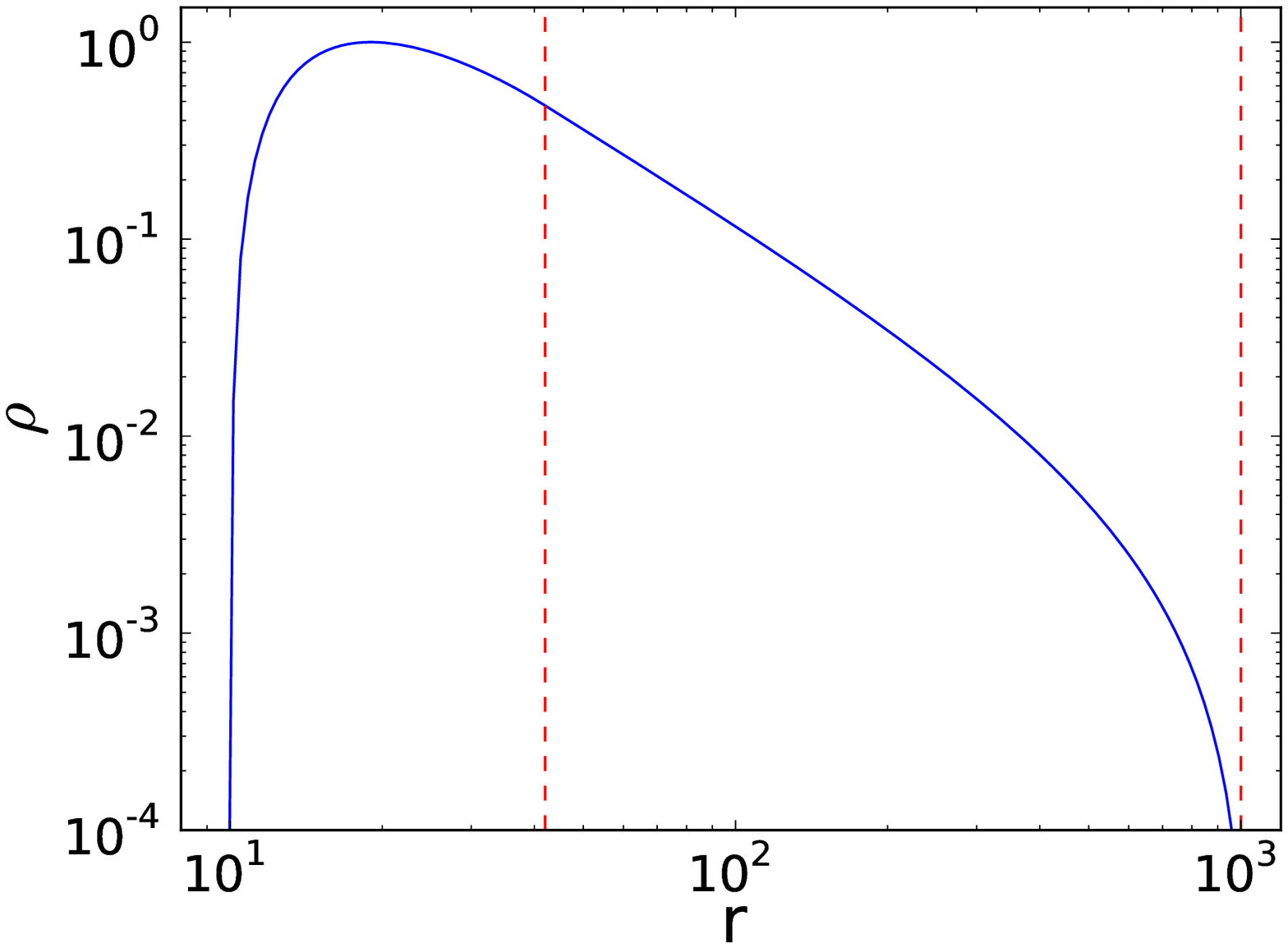}
\includegraphics[width=0.449\textwidth]{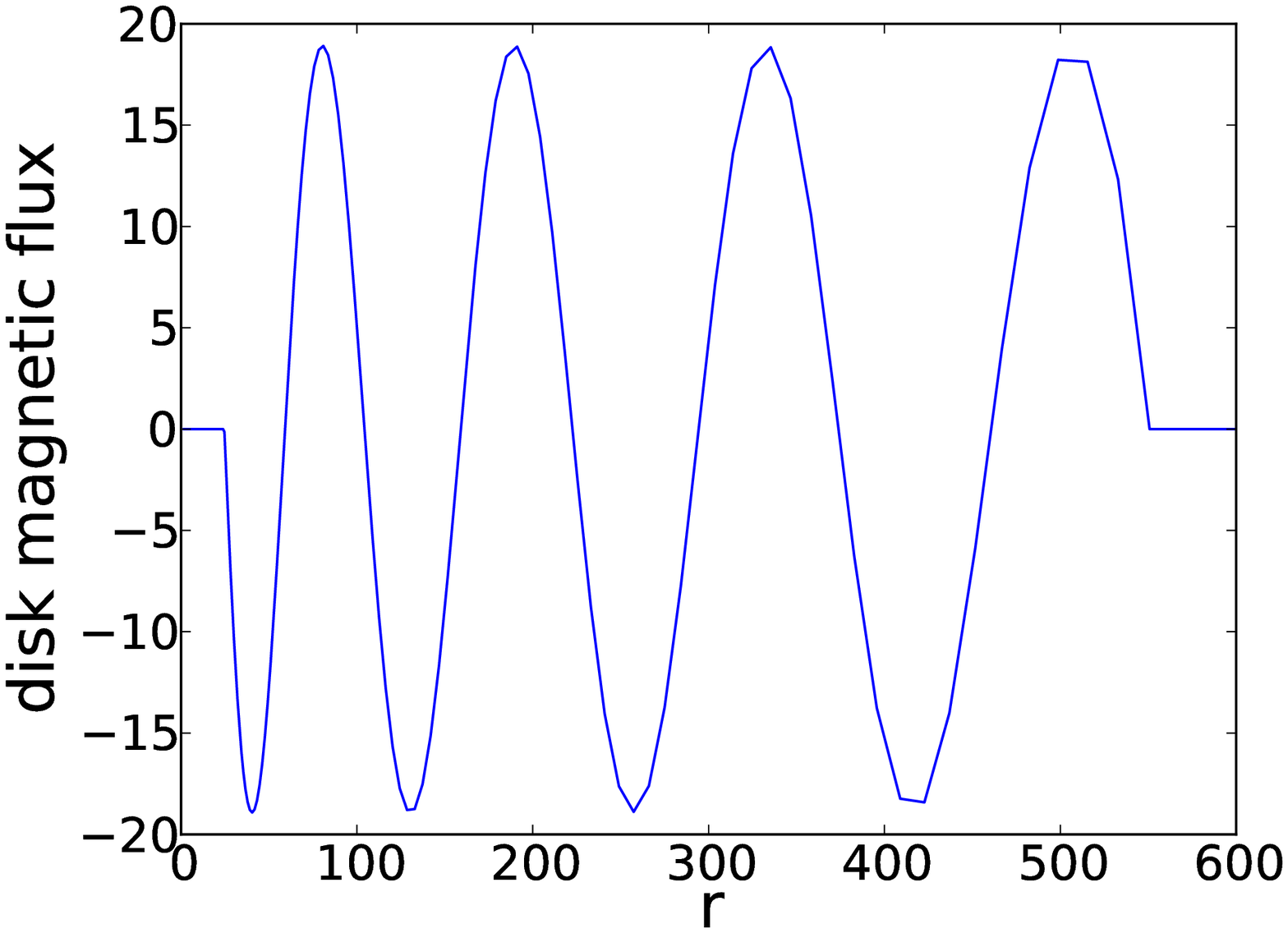}
\includegraphics[width=0.449\textwidth]{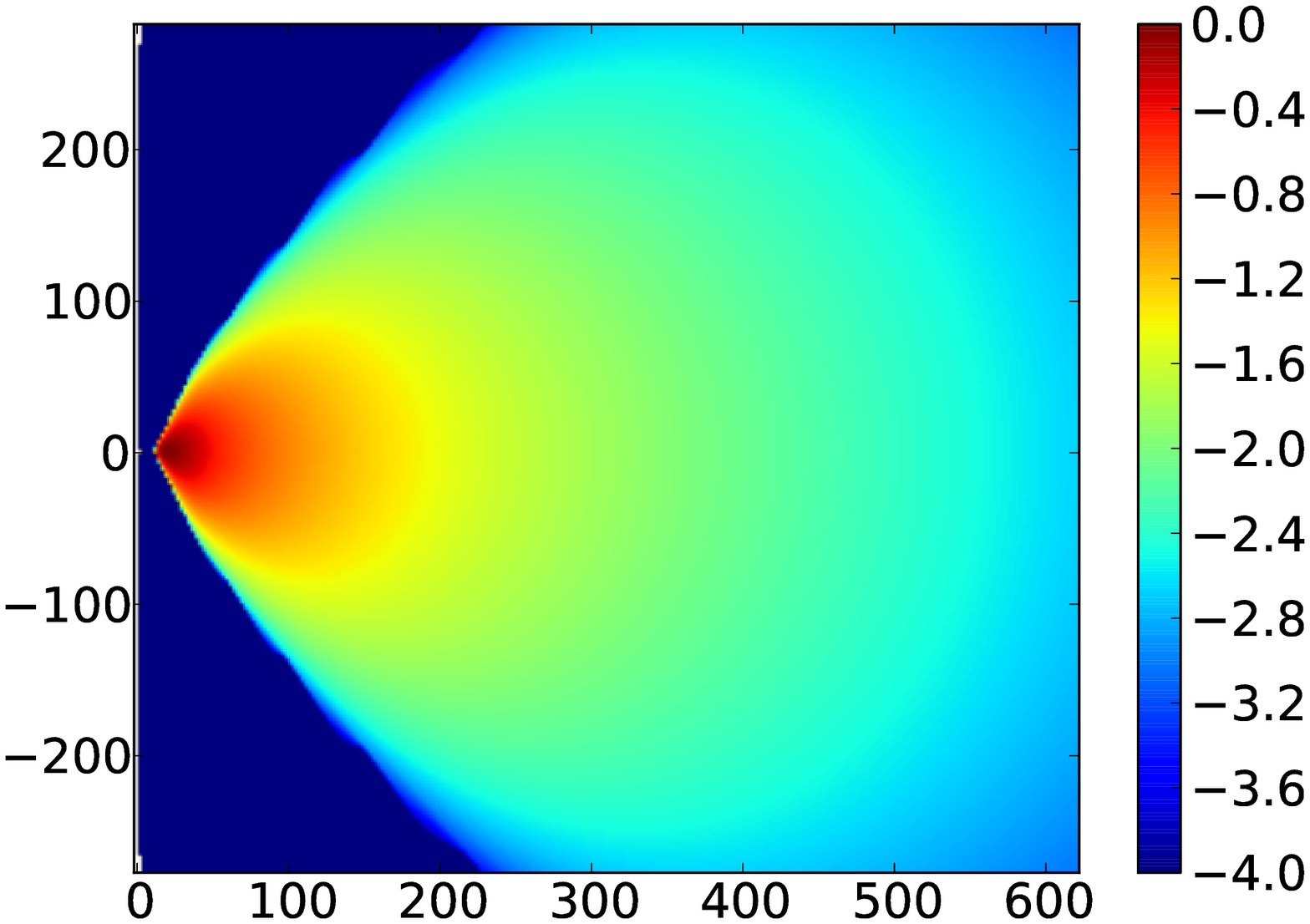}
\includegraphics[width=0.449\textwidth]{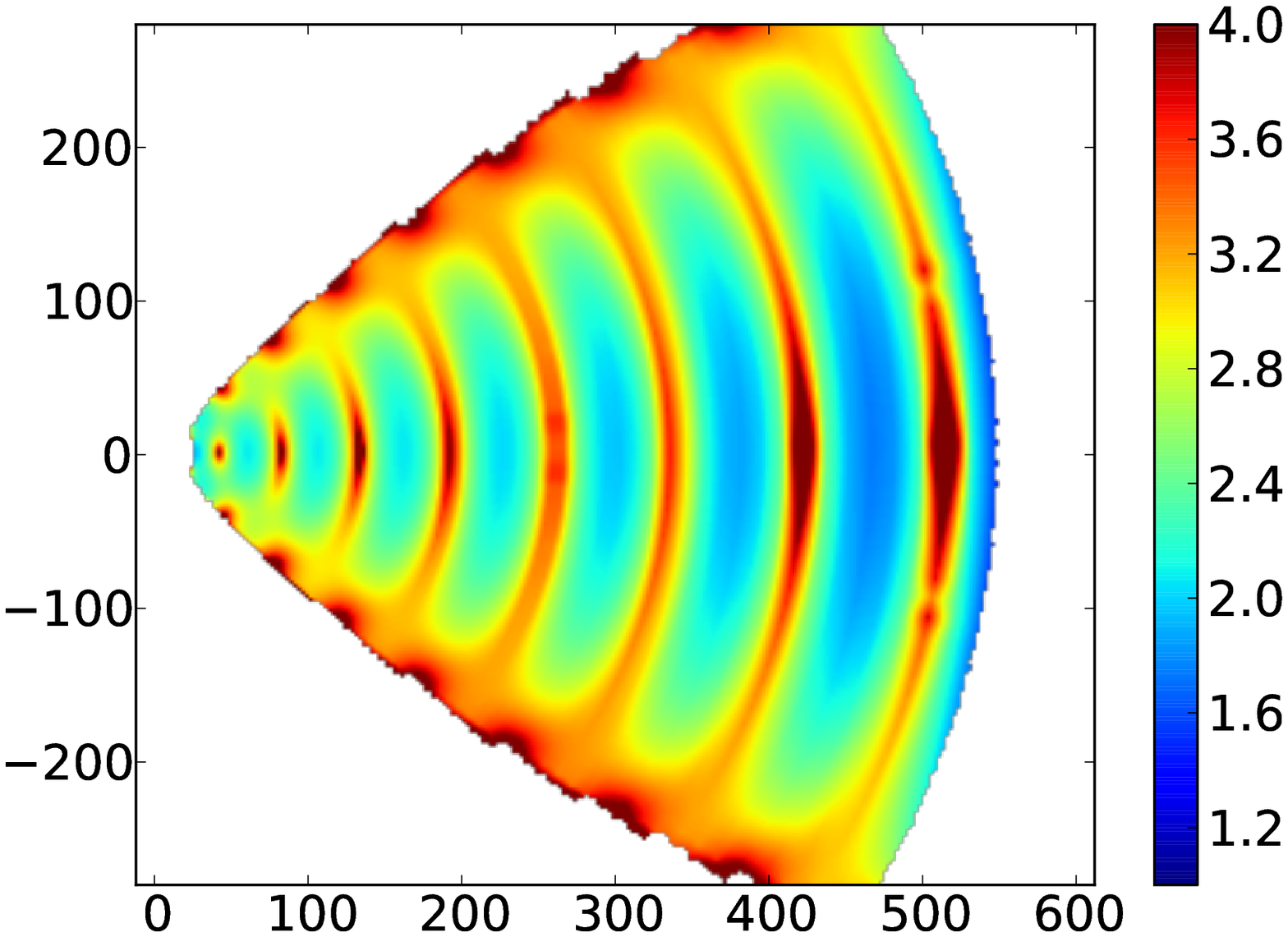}
\end{center}
\caption{Initial configuration of the ADAF/SANE simulation. The top
  two panels show the mid-plane density and the magnetic flux
  threading the equatorial plane as a function of radius. Note the
  extended size of the initial torus, which is required for the
  extremely long duration of this simulation.  Note also the multiple
  oscillations in the magnetic flux, which prevents the accreting gas
  from reaching the MAD state.  The lower two panels show the
  logarithm of the density $\rho$ and the gas-to-magnetic pressure
  ratio $\beta$ of the initial torus in the poloidal plane.}
\label{fig:SANE_init}
\end{figure*}

\begin{figure*}
\begin{center}
\includegraphics[width=0.449\textwidth]{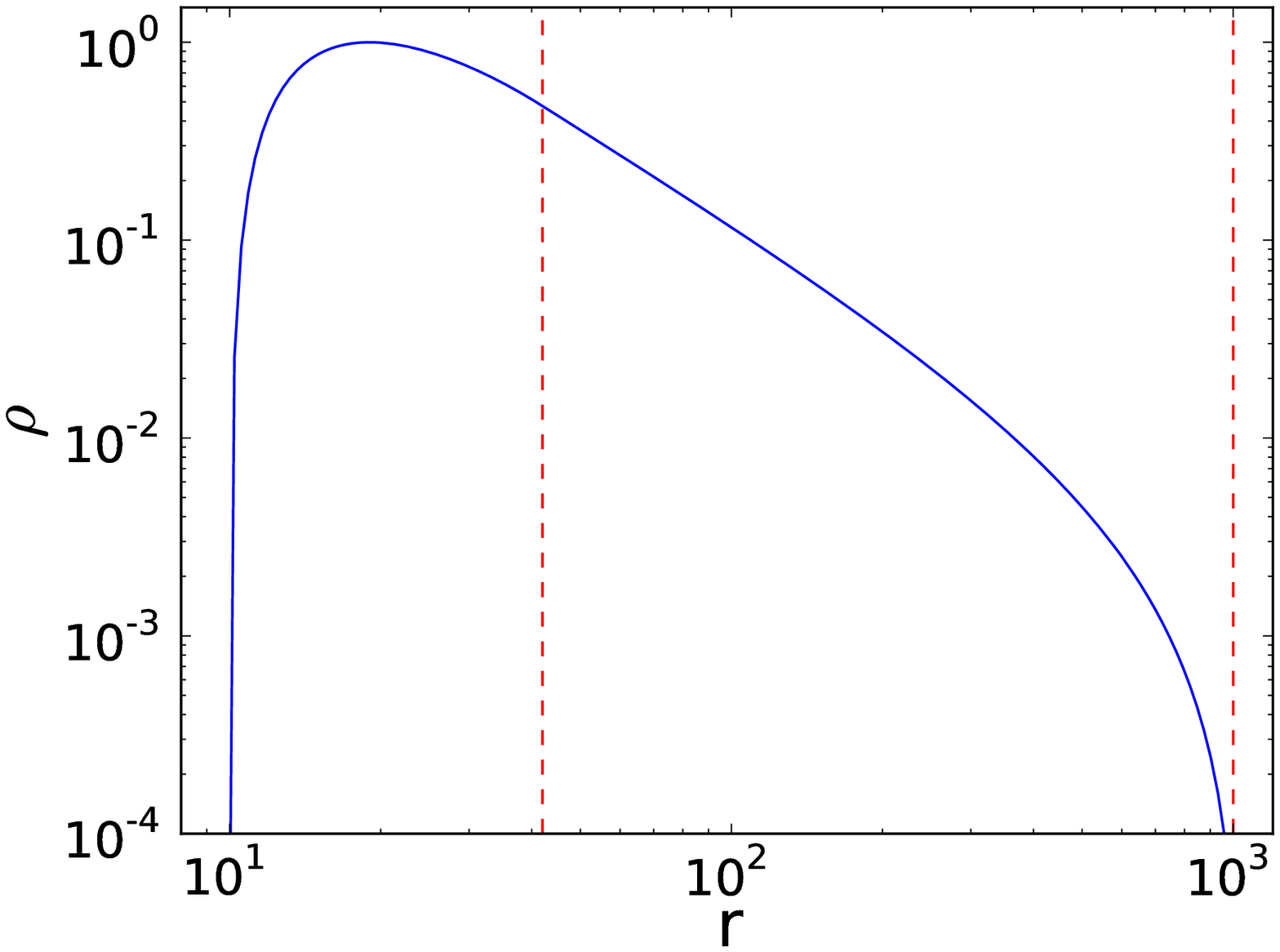}
\includegraphics[width=0.449\textwidth]{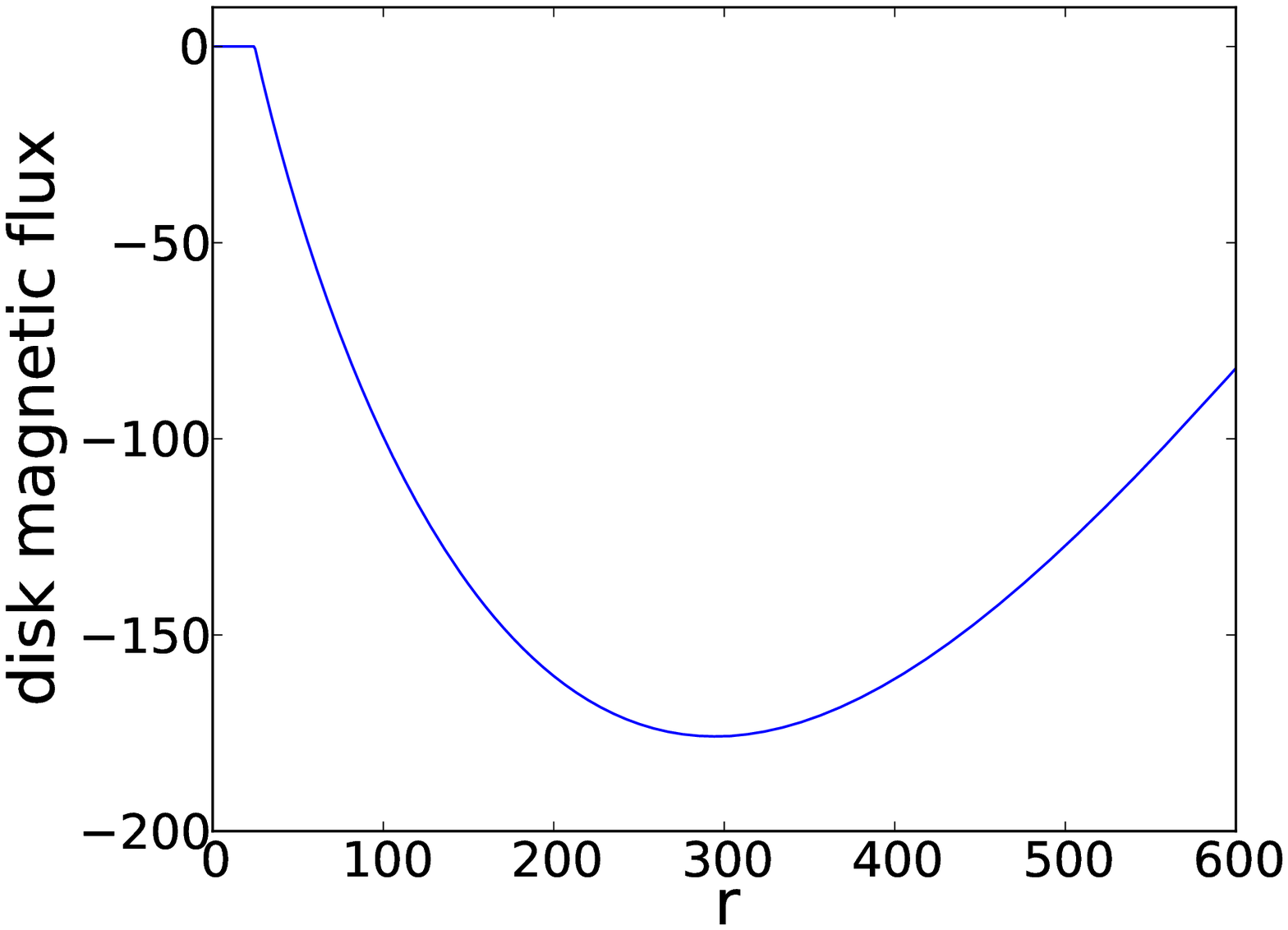}
\includegraphics[width=0.449\textwidth]{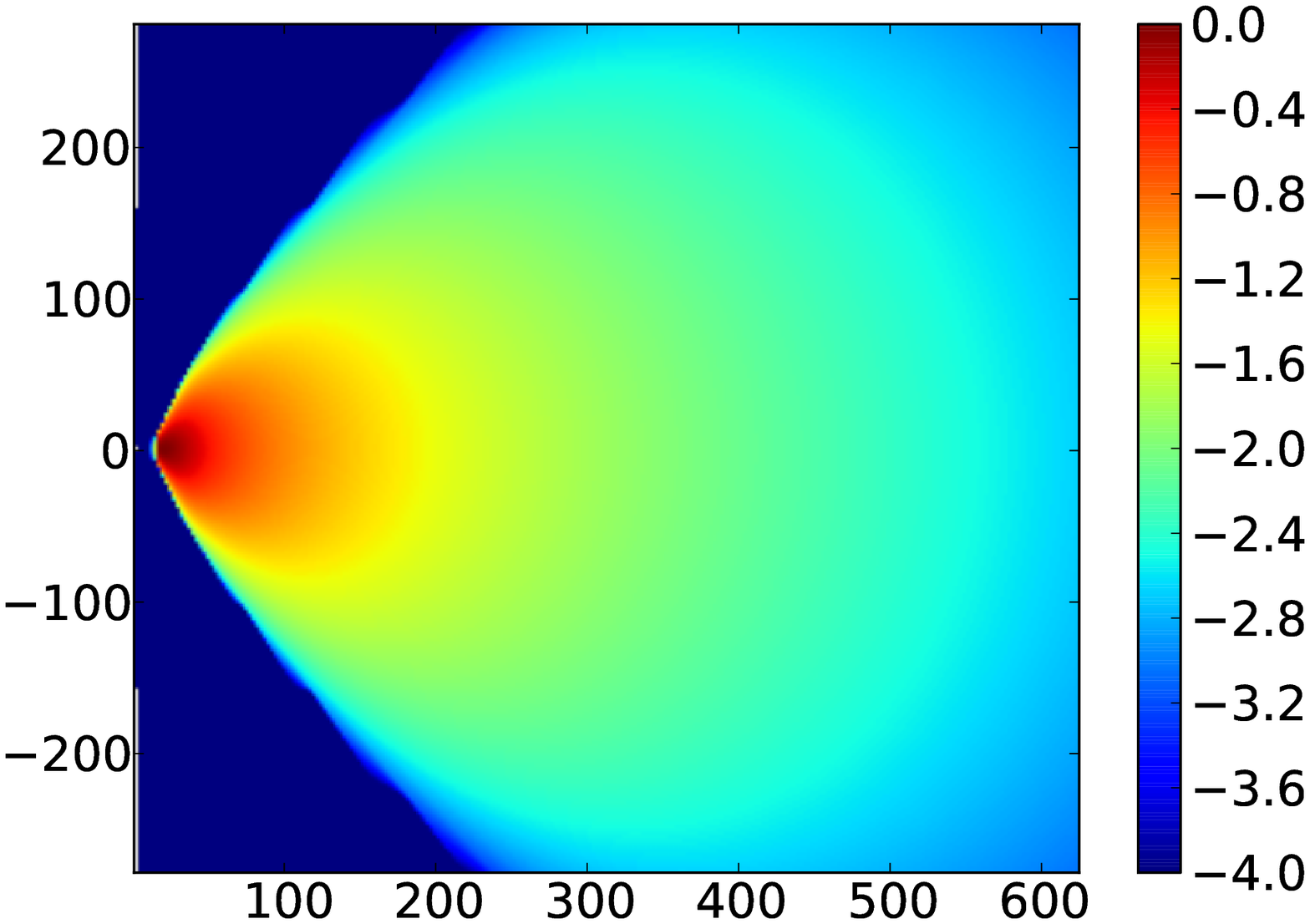}
\includegraphics[width=0.449\textwidth]{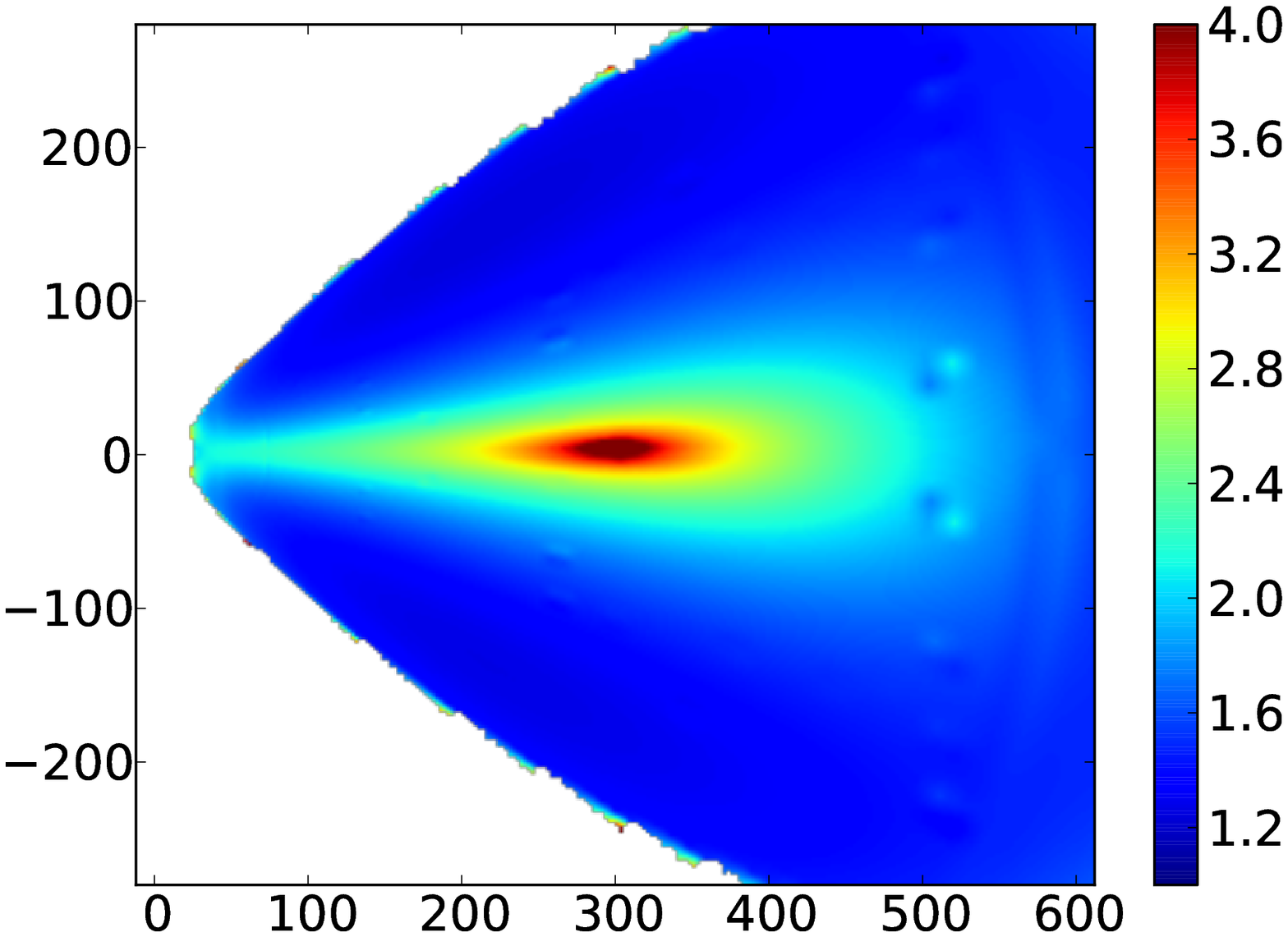}
\end{center}
\caption{Similar to Fig.~\ref{fig:SANE_init} but for the ADAF/MAD
  simulation. The main difference is that the torus here has a single
  loop of field centered at radius $r=300$. As a result, accretion
  causes magnetic flux of one sign to accumulate around the BH,
  leading to the MAD state.}
\label{fig:MAD_init}
\end{figure*}

The fluid initially rotates around the BH in a torus in hydrostatic
equilibrium: a ``Polish doughnut'' \citep*{kozlowski78}.  The
ADAF/SANE and ADAF/MAD simulations begin with the same torus.  It has
inner edge at $r_{\rm in}=10$ and extends to $r\sim 1000$
(Figs.~\ref{fig:SANE_init}, \ref{fig:MAD_init}).  The angular momentum
of the torus is constant inside $r_{\rm break}=42$.  Outside $r_{\rm
  break}$, the angular momentum is $71\%$ of the Keplerian value and
is constant on von Zeipel cylinders.  The entropy is constant
everywhere, $p/\rho^\Gamma=0.00766$, and the Bernoulli is small and
negative, $-Be \sim 10^{-2}-10^{-3}$ (in units of $c^2$).  The torus
is described in detail in \citet*{penna2012}.

The initial magnetic field is purely poloidal.  The magnetic field in
the case of the ADAF/SANE simulation is broken into eight poloidal
loops of alternating polarity (Fig.~\ref{fig:SANE_init}).  Each loop
carries the same amount of magnetic flux, so the BH is unable to
acquire a large net flux over the course of the simulation.  The
normalization of the magnetic field is adjusted such that the
gas-to-magnetic pressure ratio, $\beta$, in the equatorial plane has a
minimum value $\sim100$ for each of the eight loops. Instead of using
multiple poloidal loops, another way of setting up an ADAF/SANE
simulation is to use a toroidal initial field (e.g., Model A in
\citealt{INA03} and Model A0.0BtN10 in \citealt{MTB12}).

The initial magnetic field of the ADAF/MAD simulation forms a single
poloidal loop centered at $r=300$ (Fig.~\ref{fig:MAD_init}). The gas
accreted by the BH in this simulation has the same orientation of the
poloidal magnetic field throughout the run, so the net flux around the
BH increases rapidly and remains at a high value. The accretion flow
is thus maintained in the MAD state.  The minimum value of $\beta$ in
the initial torus is $\sim 50$.

The magnetic field construction is described in detail in
\citet{penna2012}.\footnote{In the notation of \citet{penna2012}, the
  ADAF/SANE magnetic field has $r_{\rm start}=25M$, $r_{\rm
    end}=550M$, and $\lambda_B=2.5$.  The ADAF/MAD magnetic field has
  $r_{\rm start}=25M$, $r_{\rm end}=810M$, and $\lambda_B=25$.}

\subsection{Preliminary Discussion of the Simulations}
\label{sec:preliminary}

\begin{figure*}
\begin{center}
\includegraphics[width=0.449\textwidth]{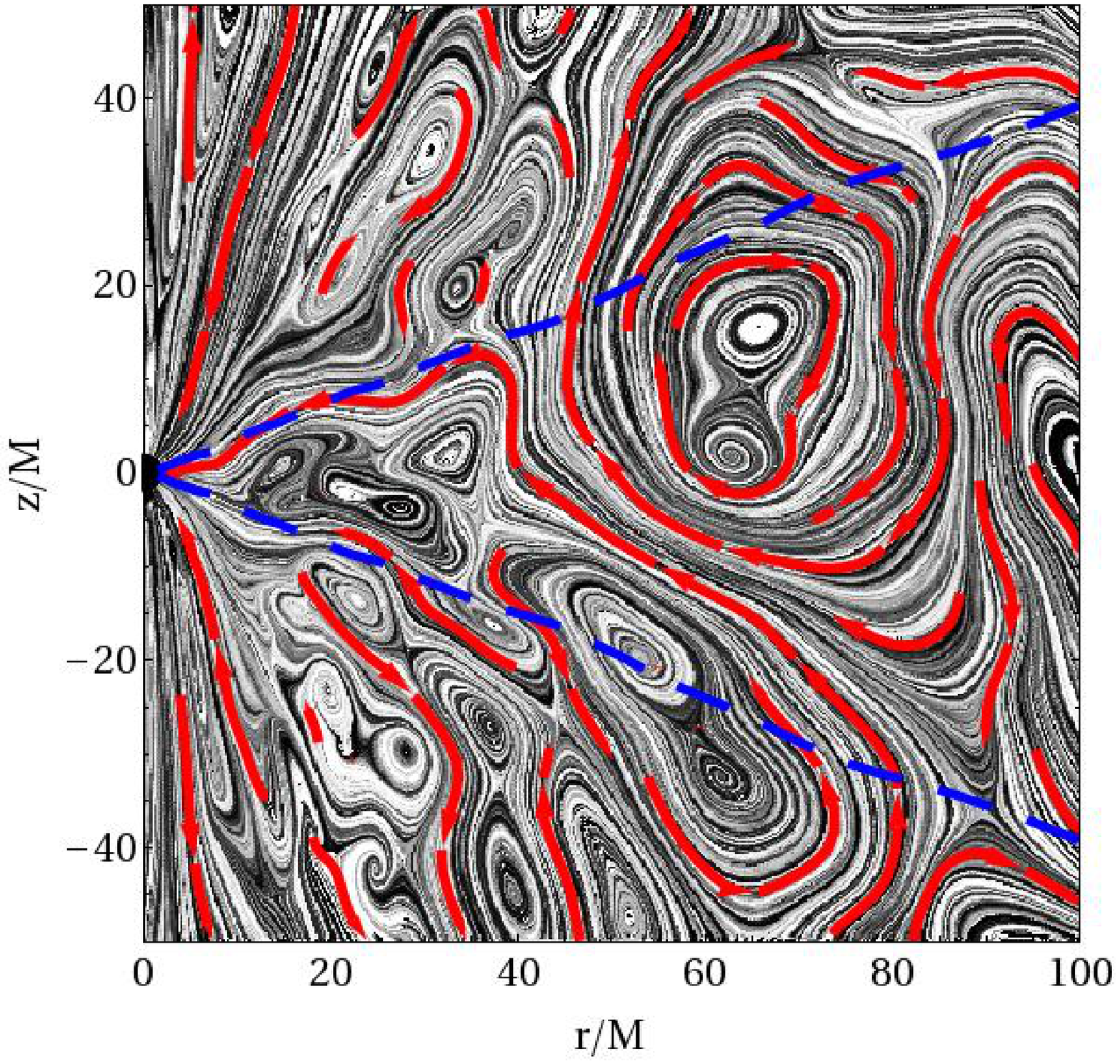}
\includegraphics[width=0.449\textwidth]{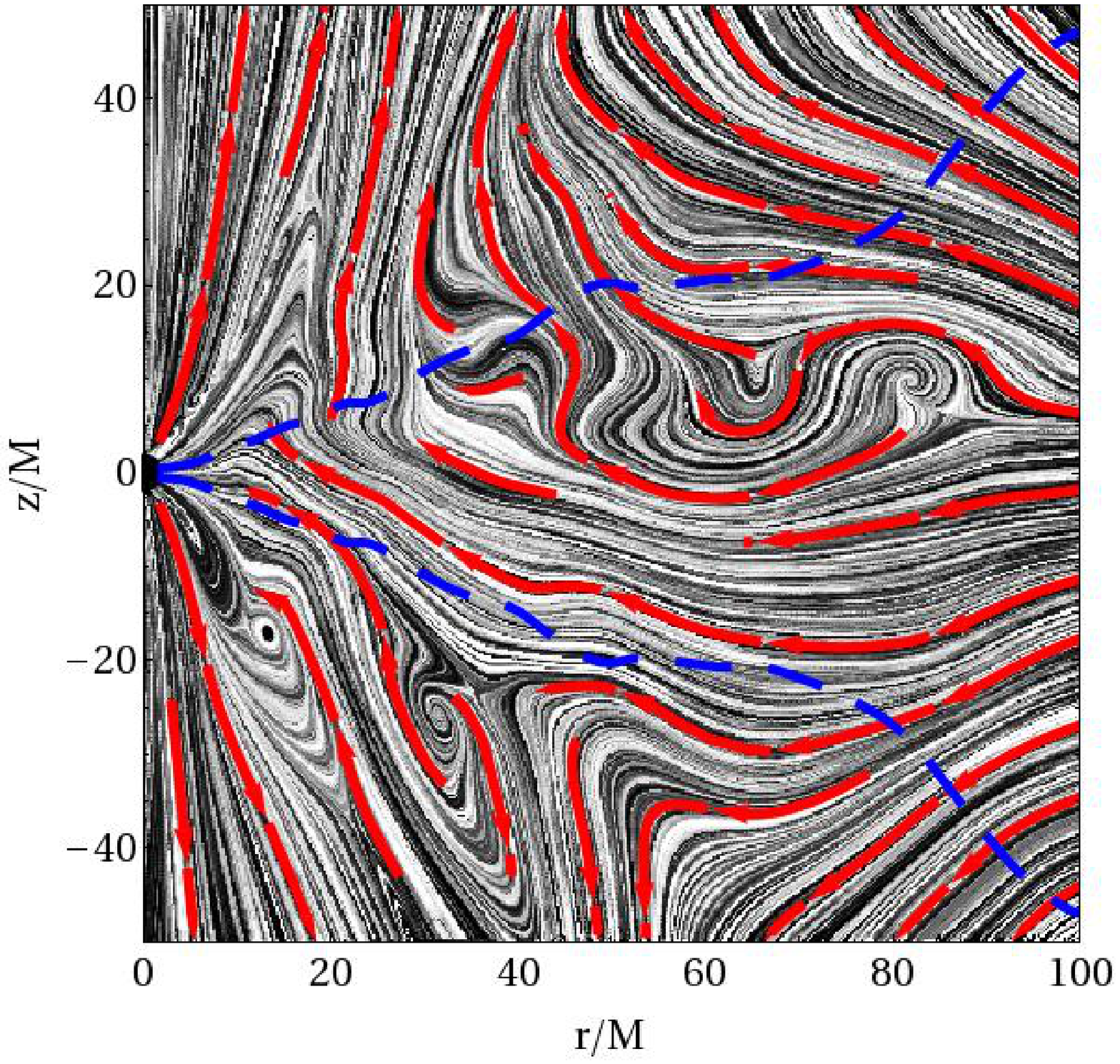}
\end{center}
\caption{Left: Snapshot of the ADAF/SANE simulation at
  $t=200,000$. Black and white streaks as well as red arrows represent
  flow streamlines. Note the turbulent eddies. The blue dashed lines
  indicate the density scale height. Right: Snapshot of the ADAF/MAD
  simulation at $t=100,000M$. There is much less turbulence.}
\label{fig:snapshot}
\end{figure*}

The two panels in Fig.~\ref{fig:snapshot} show snapshots from the end
of the ADAF/SANE and ADAF/MAD simulations. In each panel, the black
and white streaks and red arrows show velocity streamlines in the
poloidal plane at azimuthal angle $\phi=0$, and the dashed lines
correspond to one density scale height. The main difference between
the two simulations is that the SANE run exhibits more turbulence
compared to the MAD run.

Following \citet{Penna+10}, we define the mass accretion rate
$\dot{M}$, the accreted specific energy $e$, and the accreted specific
angular momentum $j$, at radius $r$ and time $t$, as follows:
\begin{eqnarray}
\dot{M}(r,t) &=& -\int_\theta\int_\phi \rho u^r \,dA_{\theta\phi}, 
\label{eq:mdot} \\
e(r,t) &=& \frac{\dot{E}(r,t)}{\dot{M}(r,t)} = \frac{1}{\dot{M}(r,t)}
\int_\theta\int_\phi T_t^r \,dA_{\theta\phi}, 
\label{eq:e} \\
j(r,t) &=& \frac{\dot{J}(r,t)}{\dot{M}(r,t)} = -\frac{1}{\dot{M}(r,t)}
\int_\theta\int_\phi T_\phi^r \,dA_{\theta\phi}, \label{eq:j}
\end{eqnarray}
where $dA_{\theta\phi}=\sqrt{-g}d\theta d\phi$ is an area element in
the $\theta$-$\phi$ plane, $\rho$ is rest mass density, $u^\mu$ is the
four-velocity, and $T_t^r$ and $T_\phi^r$ are components of the
stress-energy tensor describing the radial flux of energy and angular
momentum, respectively:
\begin{eqnarray}
T_t^r &=& (\rho+\Gamma u +b^2)u^ru_t - b^r b_t, \label{eq:Trt} \\ 
T_\phi^r &=& (\rho+\Gamma u +b^2)u^ru_\phi - b^r b_\phi. \label{eq:Trphi}
\end{eqnarray}
The quantity $u$ is the internal energy of the gas, $\Gamma$ is its
adiabatic index which is set to 5/3 in both simulations, and $b^\mu$
is a four-vector which describes the fluid frame magnetic field (see
\citealt{Gammie+03} for details).  In equations
(\ref{eq:mdot})--(\ref{eq:j}), the integrals are over the entire
sphere ($\theta=0$ to $\pi$, $\phi=0$ to $2\pi$), and the signs are
chosen such that $\dot{M}$, $\dot{E}$, $\dot{J}$ are positive when the
corresponding fluxes are pointed inward.  More useful than $e$ is the
quantity $(1-e)$, which is the ``binding energy'' of the accreting gas
relative to infinity.

In addition, we define $\phi_{\rm BH}$ to be the normalized and
averaged magnetic flux threading each hemisphere of the BH horizon
(see \citealt{Tchekhovskoy+11}),
\begin{equation}
\phi_{\rm BH}(t) = \frac{1}{2\sqrt{\dot{M}}} \int_\theta\int_\phi
|B^r(\rH,t)|\,dA_{\theta\phi},
\end{equation}
where $B^r$ is the radial component of the magnetic field and $\rH$ is
the radius of the horizon. The integral is again over the whole
sphere, and the factor of $1/2$ is to convert the result to one
hemisphere.  An accretion flow transitions to the MAD state once
$\phi_{\rm BH}$ crosses a critical value $\sim50$
\citep{Tchekhovskoy+11,Tchekhovskoy+12}. Thus, by monitoring this
quantity, we can evaluate whether a particular simulation is in the
SANE or MAD state.

\begin{figure}
\begin{center}
\includegraphics[angle=270,width=\columnwidth]{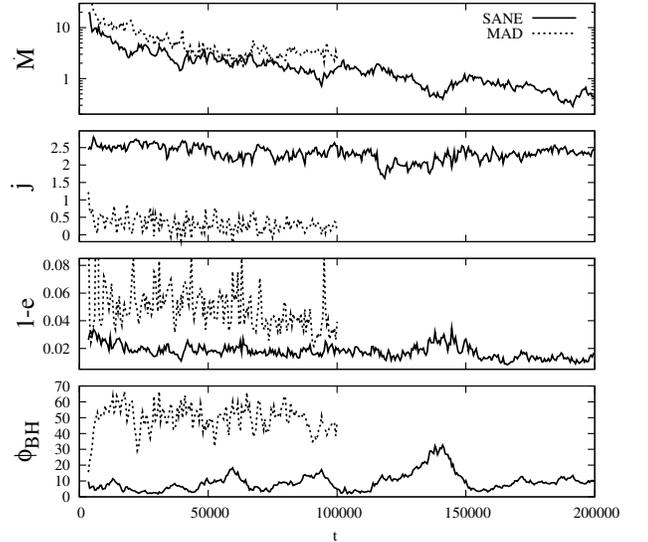}
\end{center}
\caption{Variations of $\dot{M}$, $j$ and $(1-e)$ at $r=10$, and
  $\phi_{\rm BH}$ at $r=\rH$, as a function of time.  Solid lines
  correspond to the ADAF/SANE simulation and dotted lines to the
  ADAF/MAD simulation. Note the very different behaviors of the two
  simulations. The decrease of $\dot{M}$ with increasing time is
  explained in Fig.~\ref{fig:Mdot_Sigma} and the text.}
\label{fig:time_evolve}
\end{figure}

Figure \ref{fig:time_evolve} shows the time evolution of $\dot{M}$,
$j$, $(1-e)$ and $\phi_{\rm BH}$ as a function of time for the
ADAF/SANE and ADAF/MAD simulations. The first three quantities are
measured at $r=10$,\footnote{The reason for choosing $r=10$ rather
  than $r=\rH$ is to avoid small deviations that sometimes arise near
  the horizon because of the activation of floors in the HARM
  code. Since $r=10$ is well inside the inflow equilibrium zone at all
  times of interest, it is a safe choice.} while the fourth is (by
definition) evaluated at the horizon $r=\rH$.  We see that the
magnetization parameter $\phi_{\rm BH}$ behaves very differently in
the two simulations. In the ADAF/SANE simulation, $\phi_{\rm BH}$
remains small, except for one spike at time $t\sim140,000$. In
contrast, in the ADAF/MAD simulation, the magnetization quickly rises
to a value $\sim50$ and remains at this high value for the rest of the
run. As explained in \citet{Tchekhovskoy+11}, the plateau in
$\phi_{\rm BH}$ corresponds to the MAD state where the BH has accepted
as much magnetic flux as it can hold for the given mass accretion
rate. Any additional flux brought in by the accreting gas remains
outside the horizon, where it ``arrests'' the accretion flow.

Corresponding to the dramatic difference in $\phi_{\rm BH}$ in the two
simulations, there are related differences in both the binding energy
flux $(1-e)$ and the specific angular momentum flux $j$. The quantity
$(1-e)$ is about two to three times larger in the MAD simulation,
which indicates that the MAD system has more energy flowing out to
infinity compared to the SANE simulation. Coincident with the spike in
$\phi_{\rm BH}$ in the ADAF/SANE simulation at $t\sim140,000$, there
is a corresponding spike in $(1-e)$.  During this period, the SANE
simulation seems to have made a brief detour close to the MAD limit.

The specific angular momentum flux $j$ is about an order of magnitude less
in the MAD simulation compared to the SANE simulation. Once the gas
has attained the MAD state, it transfers very little angular momentum
to the BH. Instead, angular momentum is transported out, largely
through the magnetic field. This implies that an ADAF/MAD accretion
flow will cause little spin-up of the BH. Indeed, as
\citet{Tchekhovskoy+12} and \citet{MTB12} have shown, if the BH has
virtually any non-zero value of $a_*$, an ADAF/MAD flow will cause
{\it spin-down} rather than spin-up.

\begin{figure}
\begin{center}
\includegraphics[width=0.49\columnwidth]{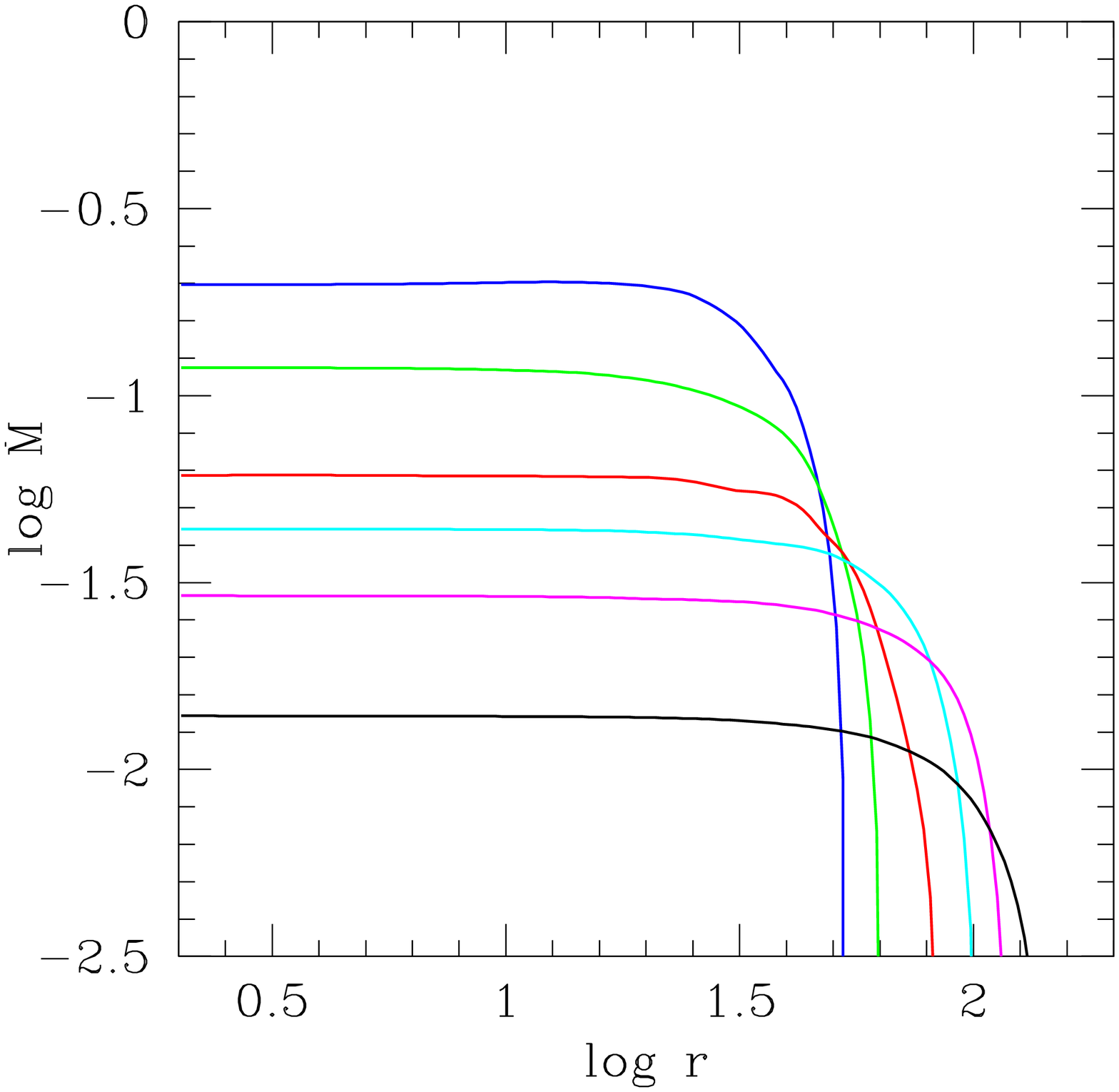}
\includegraphics[width=0.49\columnwidth]{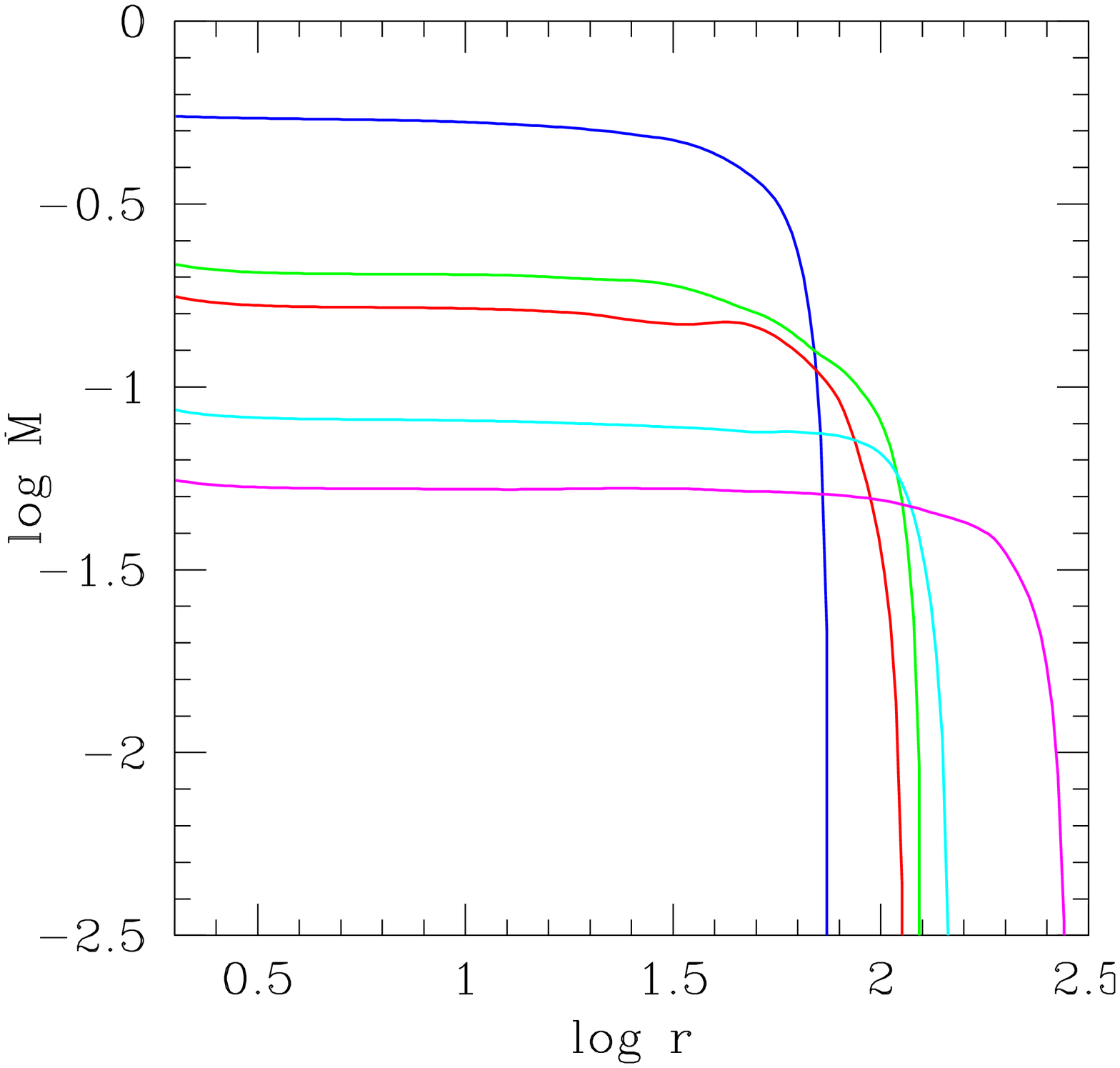}
\includegraphics[width=0.49\columnwidth]{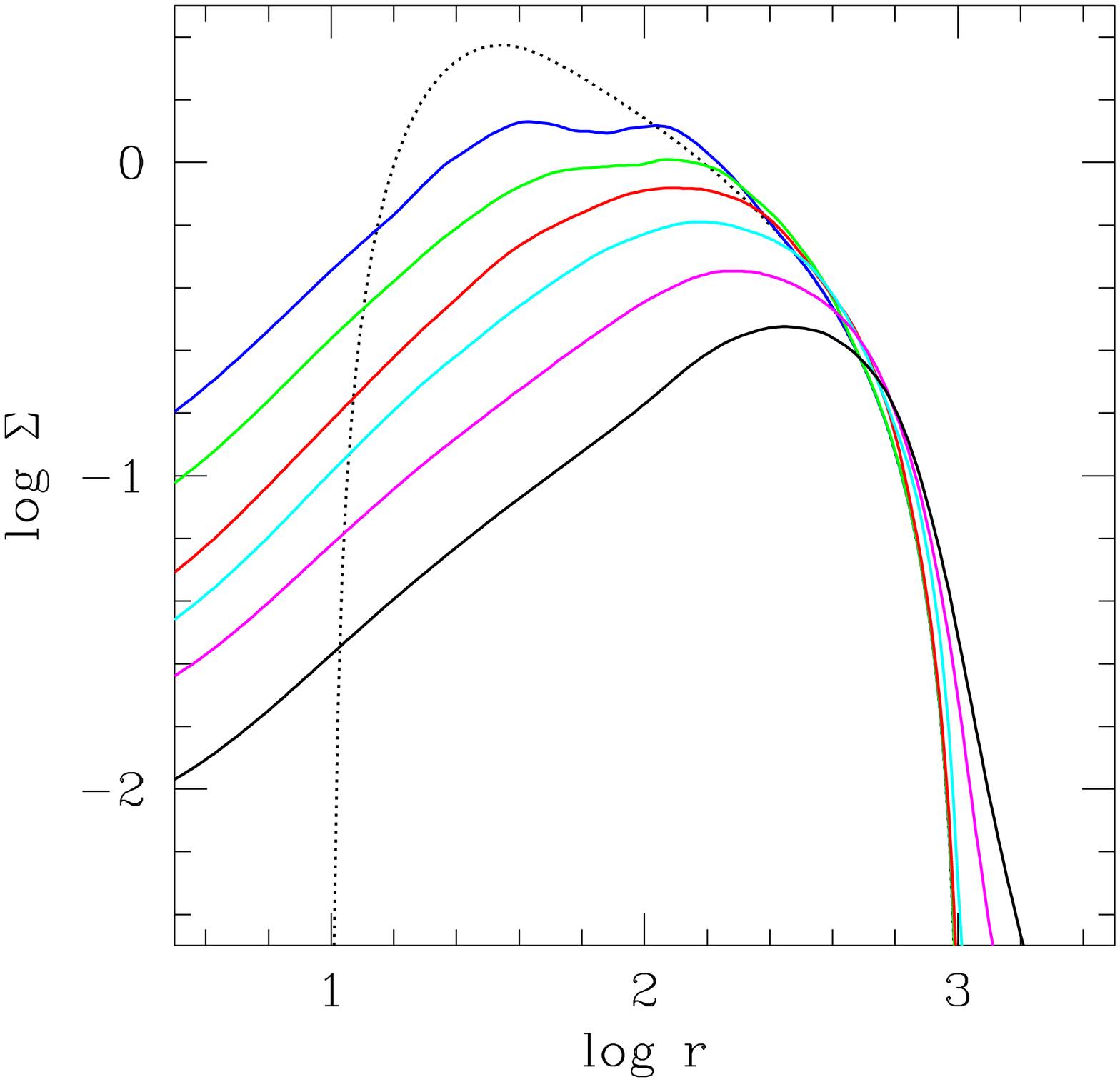}
\includegraphics[width=0.49\columnwidth]{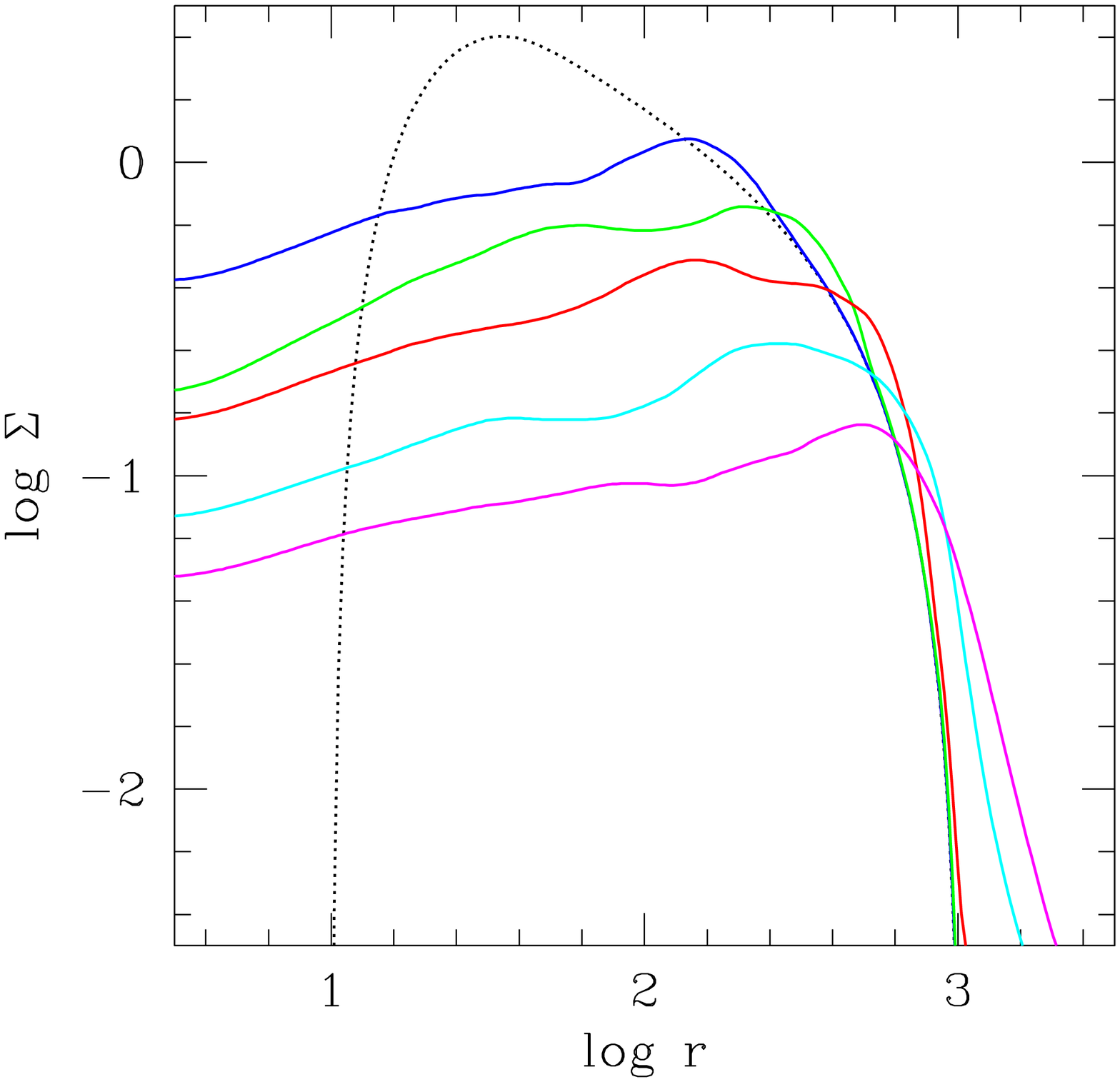}
\end{center}
\caption{Top Left: Shows the variation of the mean mass accretion rate
  $\dot{M}(r)$ vs $r$ in the ADAF/SANE simulation for the six
  independent time chunks S1--S6. The colour code is as follows: S1
  (blue), S2 (green), S3 (red), S4 (cyan), S5 (magenta), S6
  (black). The flat region of each curve identifies the range of $r$
  over which the accreting gas is in inflow equilibrium. This range
  increases monotonically with time, as one expects. Top Right:
  Similar plot for the ADAF/MAD simulation for the five time chunks
  M1--M5.  Colour code: M1 (blue), M2 (green), M3 (red), M4 (cyan), M5
  (magenta).  Bottom Left: An explanation for why the mass accretion
  rate shown in Fig.~\ref{fig:time_evolve} declines secularly with
  time in the ADAF/SANE simulation. In each time chunk, the surface
  density $\Sigma$ has to match smoothly to the $\Sigma$ profile of
  the initial torus (dotted curve). Therefore, the decrease in
  $\dot{M}$ is purely a consequence of the initial conditions.  Bottom
  Right: Similar plot for the ADAF/MAD simulation.}
\label{fig:Mdot_Sigma}
\end{figure}

\begin{table*}
\begin{minipage}{\columnwidth}
\caption{Time Chunks in the ADAF/SANE Simulation}
\begin{tabular}{|c|c|c|c|c|}

\hline
Chunk & Time Range ($M$) & $t_{\rm chunk}/M$ & $r_{\rm strict}/M$ & 
$r_{\rm loose}/M$ \\ 
\hline
S1 & 3000--6000 & 3000 & 19 & 23 \\
S2 & 6000--12000 & 6000 & 25 & 43 \\
S3 & 12000--25000 & 13000 & 29 & 45 \\
S4 & 25000--50000 & 25000 & 43 & 62 \\
S5 & 50000--100000 & 50000 & 66 & 92 \\
S6 & 100000--200000 & 100000 & 86 & 113 \\
\hline

\end{tabular}
\label{tab:ADAF/SANE}
\end{minipage}
\end{table*}

\begin{table*}
\begin{minipage}{\columnwidth}
\caption{Time Chunks in the ADAF/MAD Simulation}
\begin{tabular}{|c|c|c|c|c|}

\hline
Chunk & Time Range ($M$) & $t_{\rm chunk}/M$ & $r_{\rm strict}/M$ & 
$r_{\rm loose}/M$ \\ 
\hline
M1 & 3000--6000 & 3000& 35 & 52 \\
M2 & 6000--12000 & 6000 & 37 & 65 \\
M3 & 12000--25000 & 13000& 69 & 90 \\
M4 & 25000--50000 & 25000 & 109 & 128 \\
M5 & 50000--100000 & 50000 & 170 & 207 \\
\hline

\end{tabular}
\label{tab:ADAF/MAD}
\end{minipage}
\end{table*}

Before discussing the behavior of $\dot{M}$ in
Fig.~\ref{fig:time_evolve}, we first describe the method we use in the
rest of the paper to analyze the time evolution of quantities.  We
divide the data from each simulation into a number of ``time chunks''
which are logarithmically spaced in time. In the case of the ADAF/SANE
simulation we have six time chunks, S1--S6, with each successive chunk
being twice as long as the previous one
(Table~\ref{tab:ADAF/SANE}). This logarithmic spacing is well-suited
for the issues discussed in this paper since most of the quantities we
are interested in show power-law behavior as a function of both time
and radius. In the case of the shorter ADAF/MAD simulation we divide
the data into five time chunks, M1--M5 (Table
\ref{tab:ADAF/MAD}). Note that there is no overlap between chunks, and
hence each chunk provides independent information.

Returning to Fig.~\ref{fig:time_evolve}, we see that $\dot{M}$ shows a
large decrease with time in both simulations.
Fig.~\ref{fig:Mdot_Sigma} explains the reason for this.  Since the
accreting gas originates in the initial gas torus shown in
Figs.~\ref{fig:SANE_init} and \ref{fig:MAD_init}, the mass
distribution in the flow has to match smoothly to this mass
reservoir. With increasing time, the radius range over which the flow
achieves steady state increases (as discussed in greater detail in the
following sections). At the boundary of the steady state region,
quantities like the surface density, $\Sigma = (1/2\pi)\int\int\rho\,
dA_{\theta\phi}$ (shown in Fig.~\ref{fig:Mdot_Sigma}), have to match
the corresponding values in the torus, and this fixes $\dot{M}$ for
that epoch. Since the torus has a prescribed variation of $\Sigma$
with $r$, we thus have a pre-determined variation of $\dot{M}$ with
time.  In hindsight, it might have been better to design the initial
torus so as to obtain a roughly constant $\dot{M}$ with time.  An
alternate approach, pioneered by \citet{INA03}, is to inject mass
steadily at some outer radius rather than to start with a fixed total
mass in a torus.

\subsection{Resolving the MRI}

Following \citet*{HGK11}, we determine how well the MRI is resolved in
our simulations by computing the parameters
\begin{equation}
Q_{\hat{\theta}} = \frac{2 \pi}{\Omega dx^{\hat{\theta}}}
          \frac{|b^{\hat{\theta}}|}{\sqrt{4\pi \rho}}, \quad 
Q_{\hat{\phi}} = \frac{2 \pi}{\Omega dx^{\hat{\phi}}}
          \frac{|b^{\hat{\phi}}|}{\sqrt{4\pi \rho}}.          
\end{equation}
Here, the grid cell sizes, $dx^{\hat{\theta}}$, $dx^{\hat{\phi}}$, and
the magnetic field components, $b^{\hat{\theta}}$, $b^{\hat{\phi}}$,
are evaluated in the orthonormal fluid frame.  The fluid's angular
velocity is $\Omega$.  The parameter $Q_{\hat{\theta}}$ is defined
such that it becomes $\lambda_{\rm MRI}/d\hat{z}$ in the limit of a
vertical field, where $\lambda_{\rm MRI}$ is the wavelength of the
fastest growing mode of the linear MRI.

\citet{HGK11} considered a number of diagnostics, principally
$B_r^2/B_\phi^2$ and dimensionless viscosity parameter $\alpha$, but
also $B_z^2/B_\phi^2$ and plasma $\beta\equiv P_{\rm gas}/P_{\rm
  mag}$, as a function of numerical resolution. They studied both
local shearing boxes and global Newtonian discs and concluded that
simulations with $Q_{\hat{\theta}}\, \gsim\, 10$ and $Q_{\hat{\phi}}\,
\gsim\, 20$ are sufficiently well resolved to give quantitatively
converged results. They also state that simulations with smaller
values of $Q_{\hat{\phi}}$, but correspondingly larger values of
$Q_{\hat{\theta}}$, are equally good. Thus, we write their criterion
for convergence as $Q_{\hat{\theta}}Q_{\hat{\phi}}\,\gsim\, 200$. In
addition, they recommend that the ratio $dx^{\hat{\phi}}/dx^{\hat{r}}$
near the disc mid-plane should be no larger than 4.

In related work, \citet{Sorathia+12} simulated global (but
unstratified) Newtonian discs using a wide range of resolutions and
showed that the magnetic tilt angle, which is related to the ratio
$B_r^2/B_\phi^2$ mentioned above, is a good diagnostic for evaluating
convergence. On the basis of this diagnostic, they suggest that a
ratio $dx^{\hat{\phi}}/dx^{\hat{r}}\,\lsim\, 2$ is sufficient for
convergence, but a ratio of 4 tends to be somewhat under-resolved (see
their Fig.~11c). Thus, their criterion is stricter than the one
proposed by \citet{HGK11}.

Our simulations have $Q_{\hat{\theta}}\sim 10-20$ throughout the
initial magnetic loops.  The initial $Q_{\hat{\phi}}$ is zero because
the loops are poloidal.  For the ADAF/SANE run, the fluid inside
$r=100$ and within one density scale height of the midplane has
$Q_{\hat{\theta}}$ and $Q_{\hat{\phi}}$ between $10-20$, i.e.,
$Q_{\hat{\theta}}Q_{\hat{\phi}}\approx200$, which is sufficient
according to \citet{HGK11}. Our numerical grid has
$dx^{\hat{\phi}}/dx^{\hat{r}}\approx 3$ at the mid-plane, which is
safe according to \citet{HGK11} and borderline according to
\citet{Sorathia+12}.  Overall, we conclude that our ADAF/SANE
simulation is adequately resolved.  Our ADAF/MAD simulation has
$Q_{\hat{\theta}}>100$ and $Q_{\hat{\phi}}\sim 50$, so this simulation
is very well-resolved.

\begin{figure*}
\begin{center}
\includegraphics[width=0.449\textwidth]{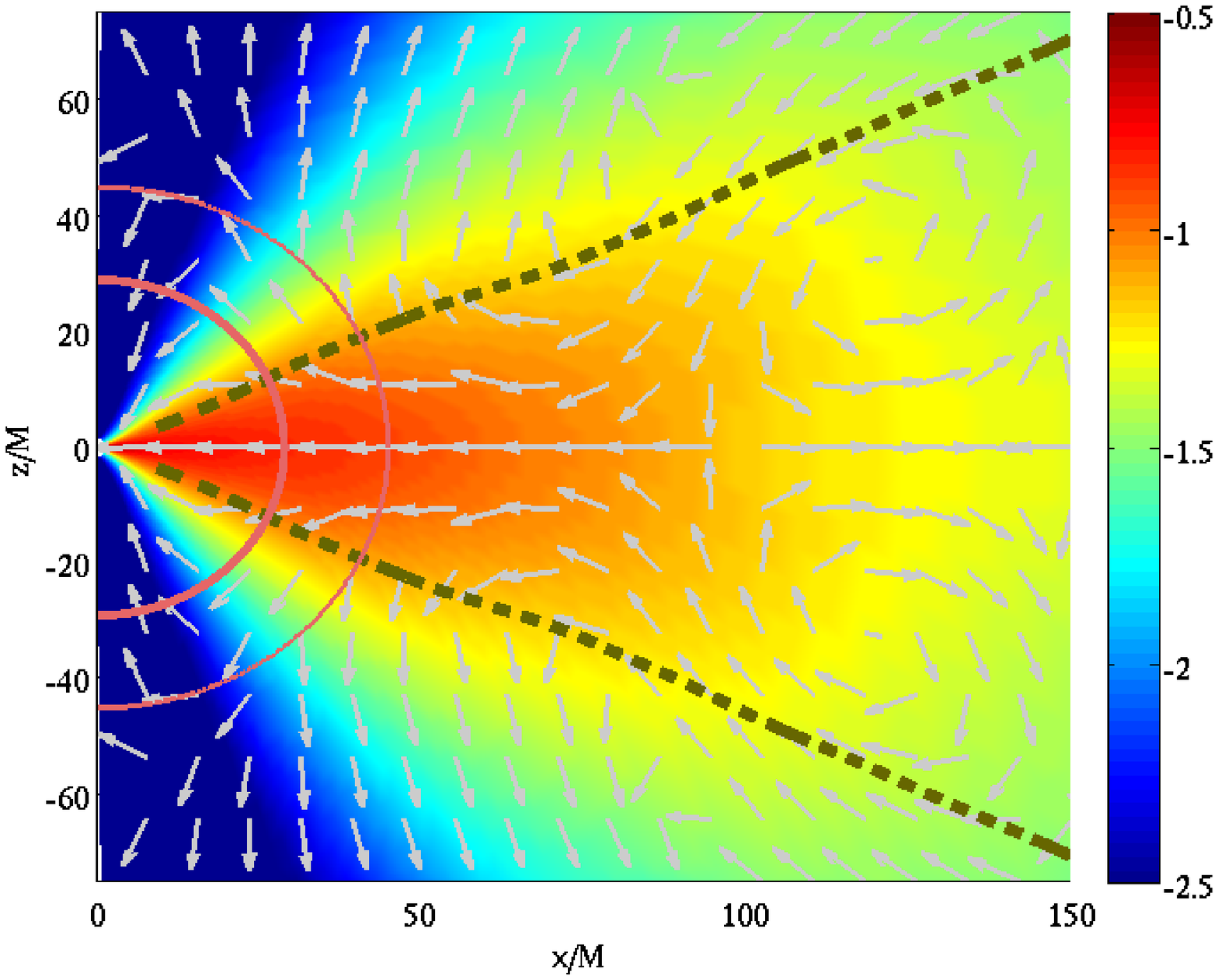}
\includegraphics[width=0.449\textwidth]{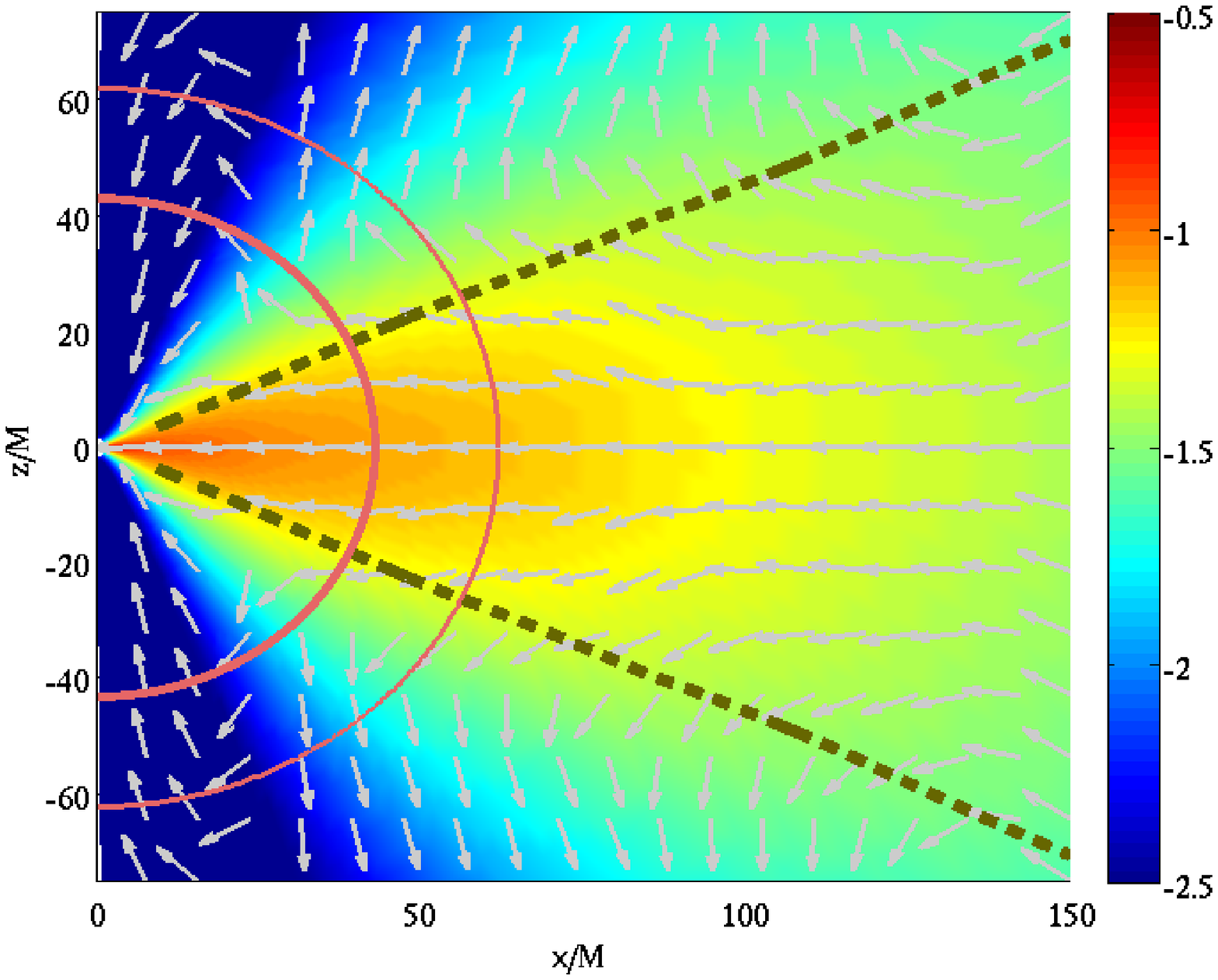}
\includegraphics[width=0.449\textwidth]{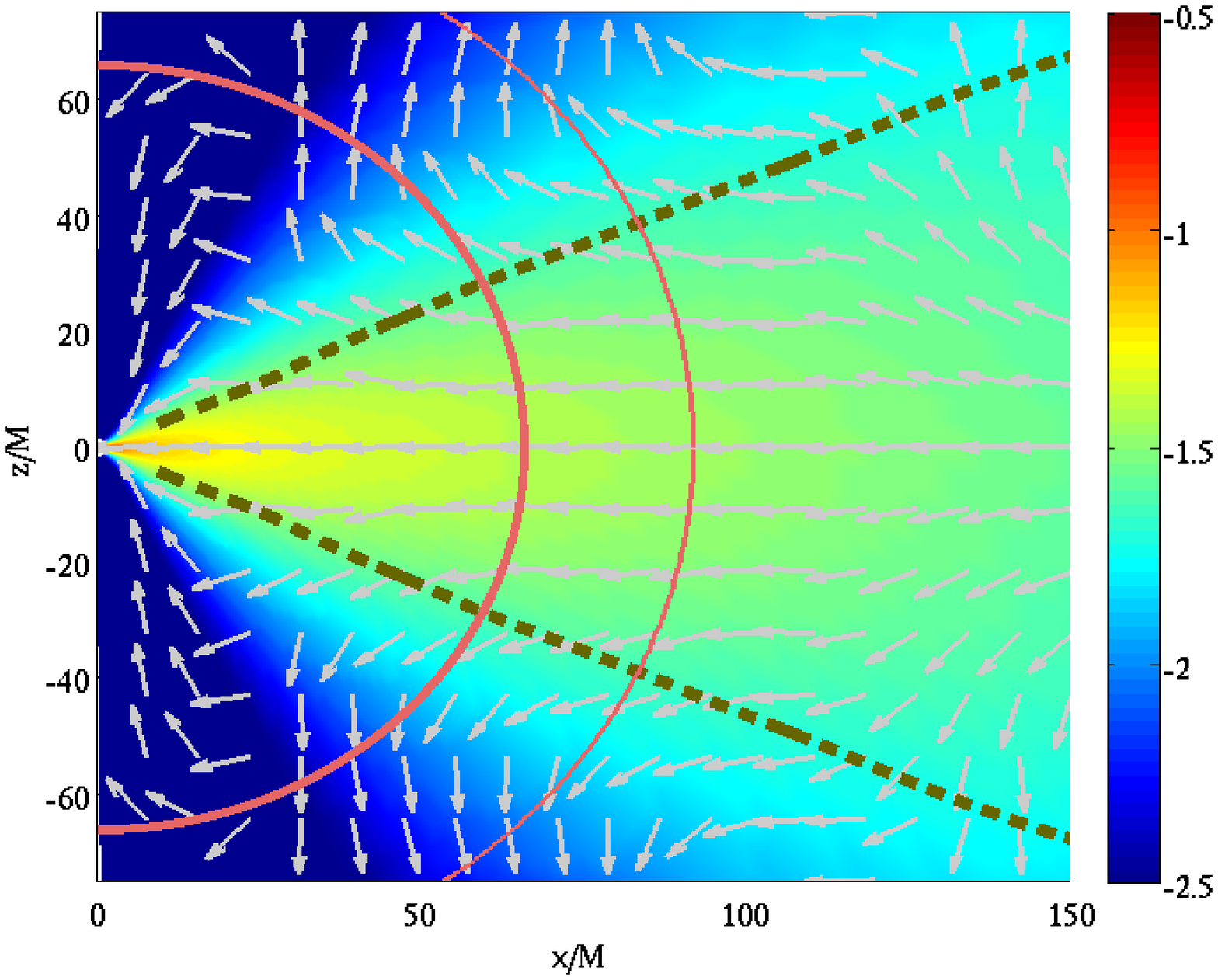}
\includegraphics[width=0.449\textwidth]{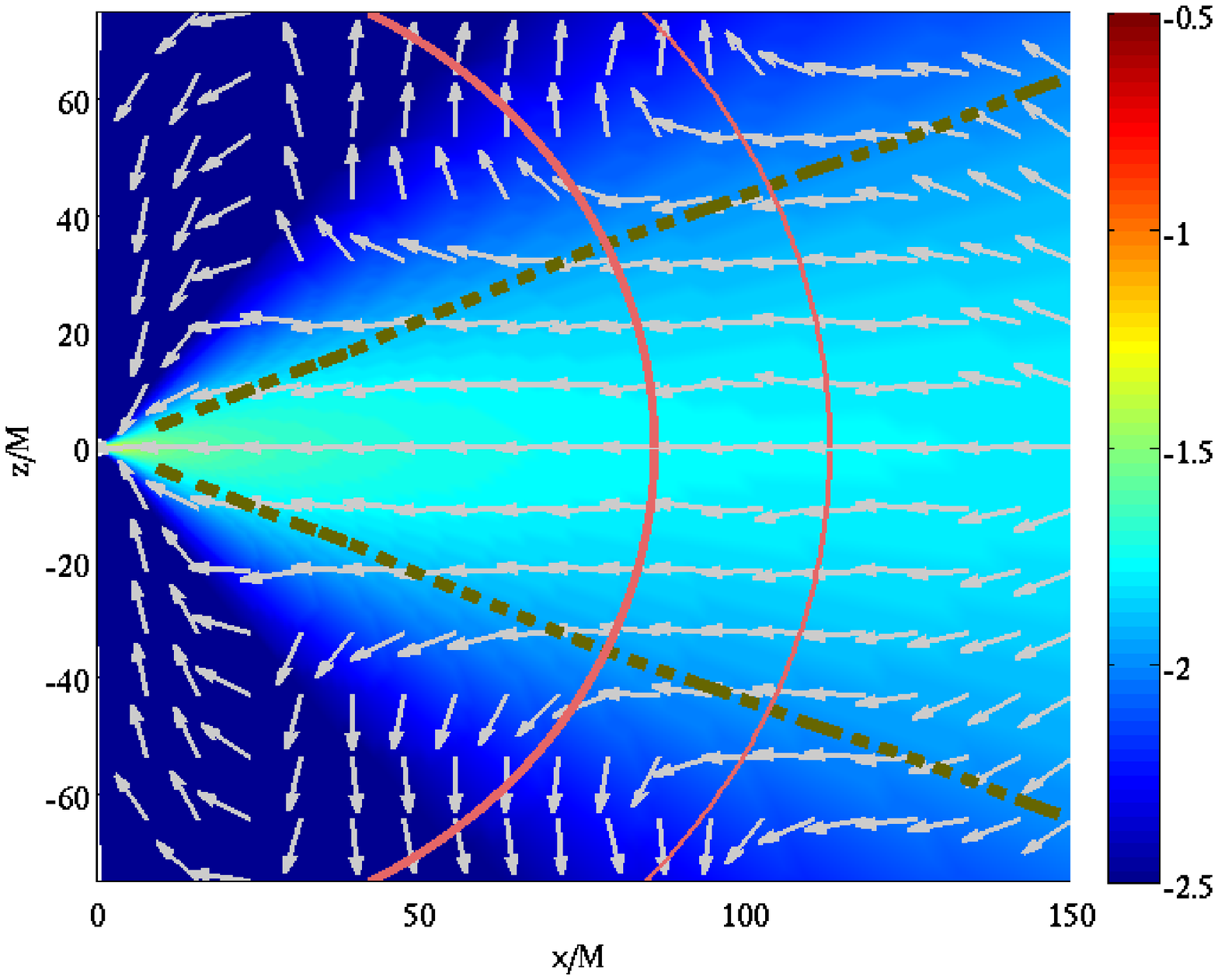}
\end{center}
\caption{Average flow properties of the ADAF/SANE simulation during
  chunks S3 (top left), S4 (top right), S5 (bottom left) and S6
  (bottom right). In each panel, the flow has been averaged over the
  duration of the chunk $t_{\rm chunk}$ (Table \ref{tab:ADAF/SANE}),
  over azimuthal angle $\phi$, and symmetrized around the
  mid-plane. Colour indicates $\log\rho$, arrows indicate direction
  (but not magnitude) of the mean velocity, and slanting dashed lines
  indicate the local density scale height. The two circular solid
  lines correspond to the steady state radius limits $r_{\rm strict}$
  (thick line) and $r_{\rm loose}$ (thin line), computed using the
  mean radial velocity within one scale height of the mid-plane (see
  text and Table \ref{tab:ADAF/SANE} for details).}
\label{fig:SANE_density}
\end{figure*}

\begin{figure*}
\begin{center}
\includegraphics[width=0.449\textwidth]{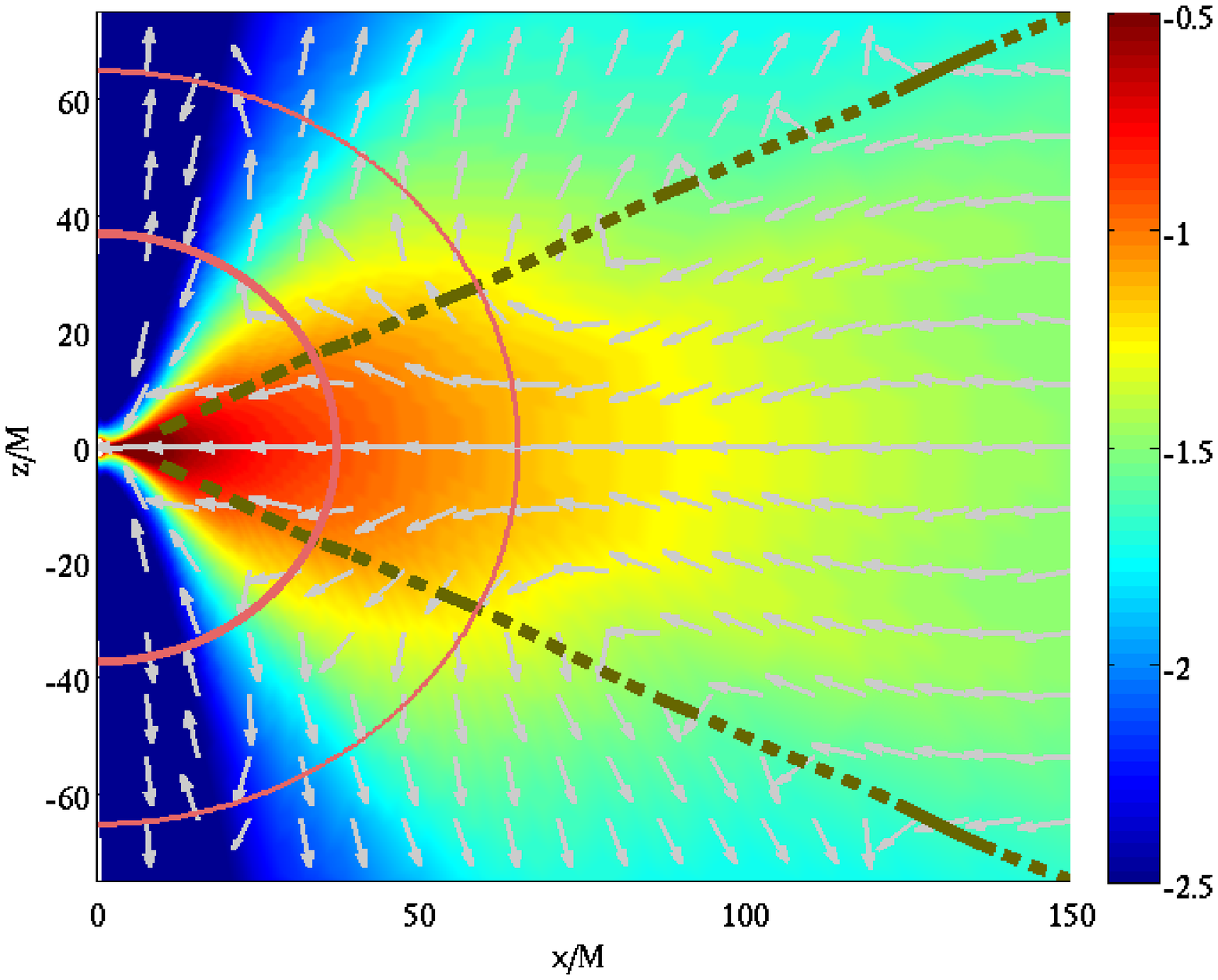}
\includegraphics[width=0.449\textwidth]{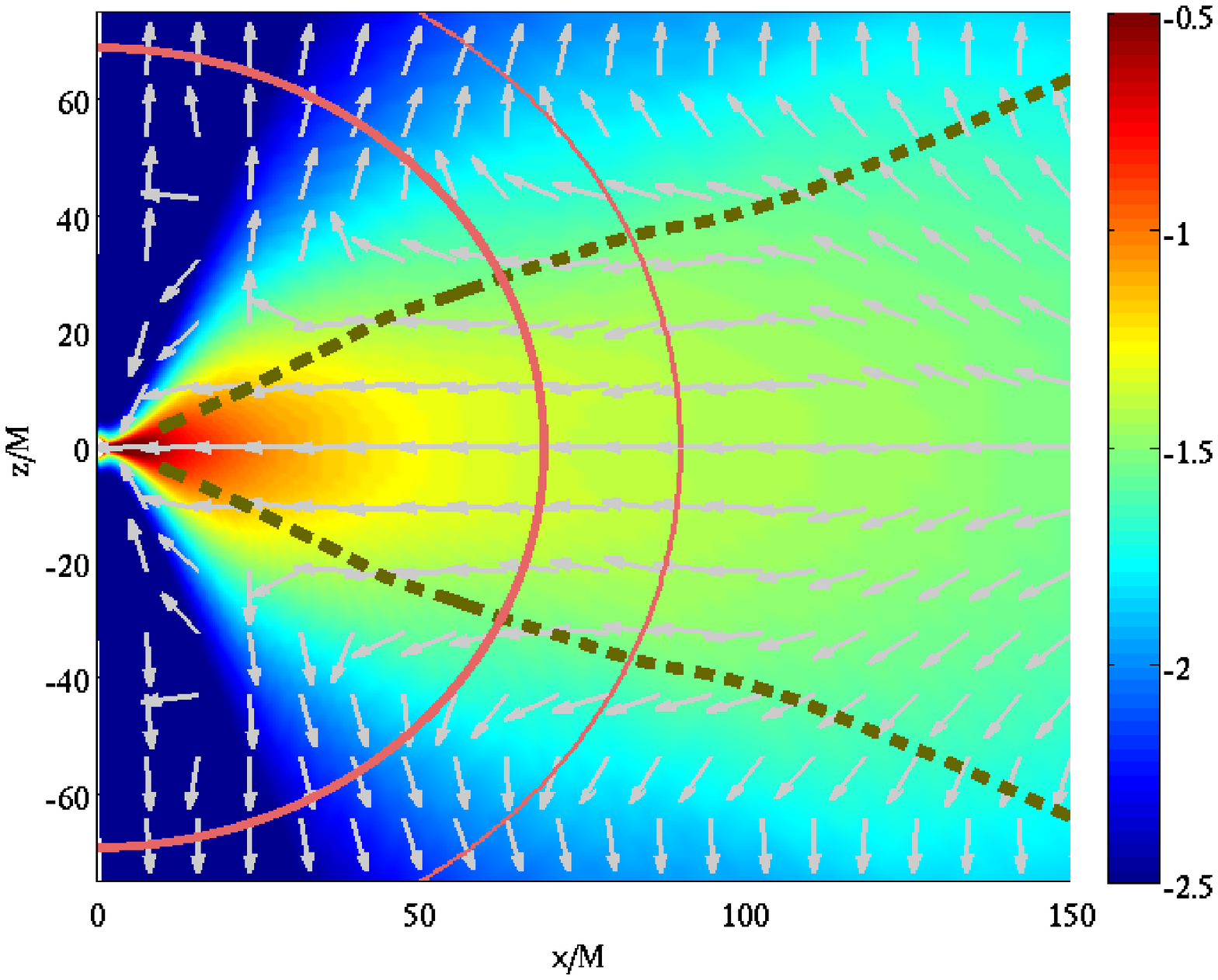}
\includegraphics[width=0.449\textwidth]{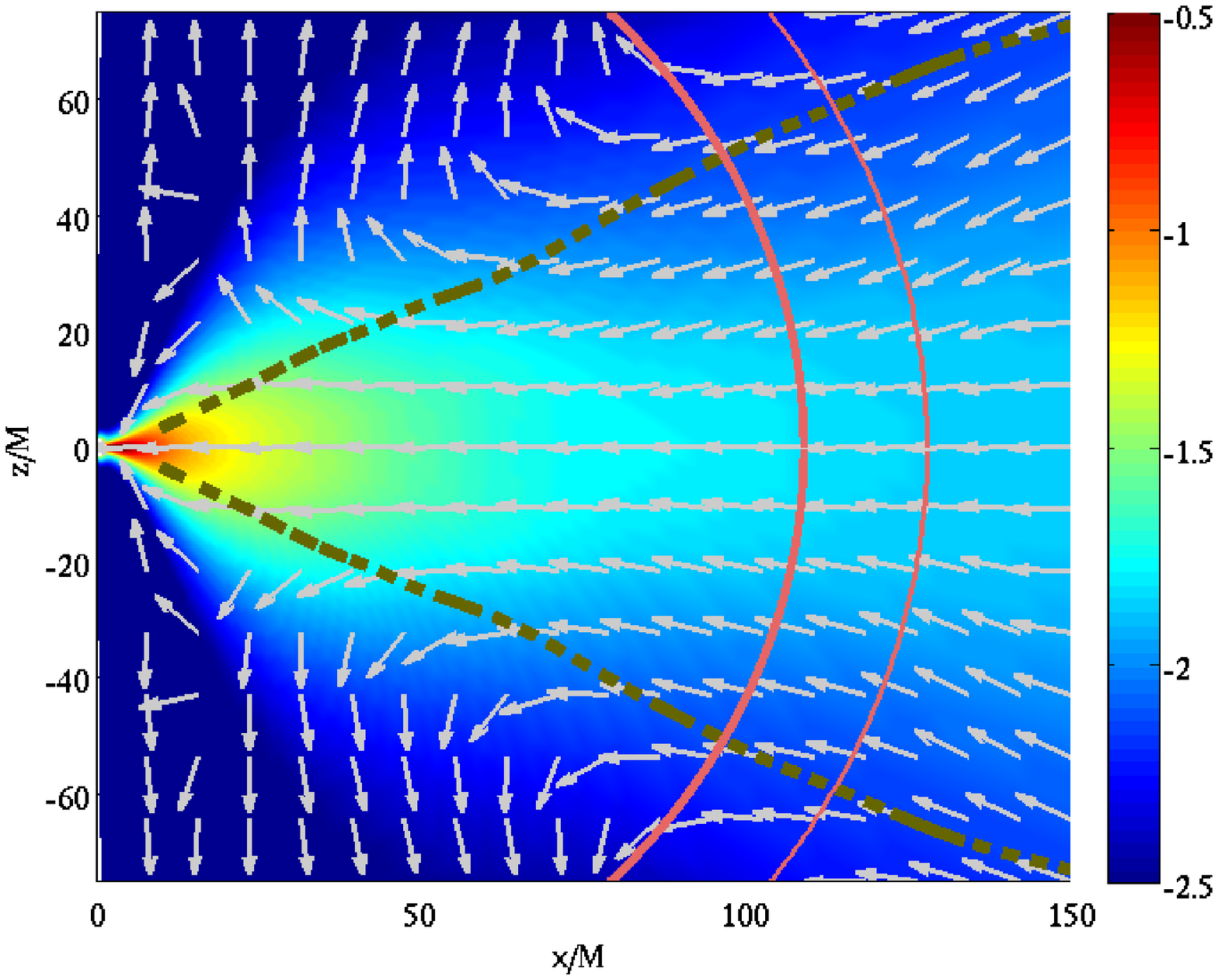}
\includegraphics[width=0.449\textwidth]{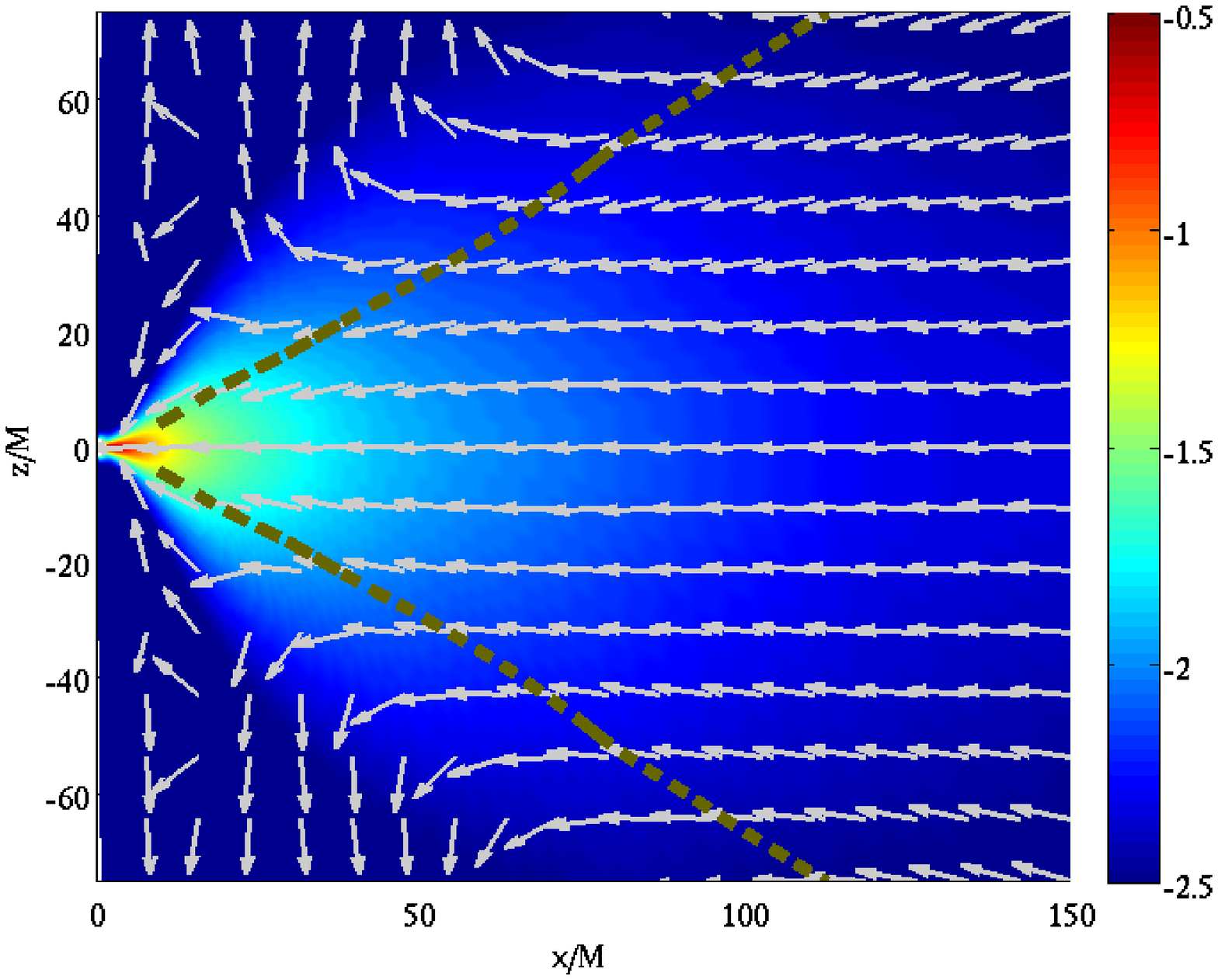}
\end{center}
\caption{Similar to Fig.~\ref{fig:SANE_density}, but for time chunks
  M2 (top left), M3 (top right), M4 (bottom left), M5 (bottom right)
  of the ADAF/MAD simulation. Note that in chunk M5 (lower right)
  $r_{\rm strict}$ and $r_{\rm loose}$ both lie outside the plotted
  area (see the numerical values given in Table \ref{tab:ADAF/MAD}).}
\label{fig:MAD_density}
\end{figure*}

Exploring the issue of convergence further, we note that the grid used
in the present study is very similar to the one employed by
\citet{Tchekhovskoy+11} for simulating their MAD models. These authors
tested convergence by increasing the number of cells in the $\phi$
direction by a factor of 2, i.e., using 128 cells over the range $\phi
= 0-2\pi$ instead of the fiducial 64 cells.  The results they obtained
with this increased resolution agreed with those from their fiducial
runs, indicating that 64 cells over $2\pi$ in $\phi$ (or 32 cells over
a wedge of angle $\pi$) are sufficient for convergence. Thus we are
confident that our ADAF/MAD run has sufficient resolution.

\citet{MTB12} describe a large number of simulations, of which one
sequence of models, A*BtN10, was initialized with a purely toroidal
field. These models, which evolve into configurations similar to our
ADAF/SANE simulation, used a resolution of $N_r=128$, $N_\theta=64$,
$N_\phi=128$, which is slightly different from, but generally similar
to, our resolution, $N_r=256$, $N_\theta=128$, $N_\phi=64$. In
addition, \citet{MTB12} considered one high-resolution toroidal-field
model, A0.94BtN10HR, with $N_r=256$, $N_\theta=128$,
$N_\phi=256$. Looking at the detailed results, it is not obvious that
their high-resolution model is distinctly superior to their standard
lower-resolution models.

Based on all of the above, we believe the two simulations described in
this paper are adequately resolved.

\section{Analysis and Results}
\label{sec:SANE/MAD}

\subsection{Criteria for Convergence and Steady State}
\label{sec:criteria}

Figure \ref{fig:SANE_density} shows time-averaged, $\phi$-averaged,
$z$-symmetrized results for the final four time chunks, S3, S4, S5,
S6, of the ADAF/SANE simulation.  The strong averaging of the
simulation data eliminates most of the turbulent fluctuations that
were evident in Fig.~\ref{fig:snapshot}, and enables us to focus on
mean properties of the flow. The accretion flow is geometrically
thick, as expected, and the gas velocity is predominantly inward
within one scale-height of the mid-plane.  At higher latitudes, many
velocity arrows point away from the BH, indicating that there is mass
outflow. At yet higher latitudes, as we approach the poles, the gas
appears again to flow in towards the BH. It is therefore not obvious
how much gas actually flows out to infinity. We discuss this question
in detail in the next subsection.

Figure \ref{fig:MAD_density} shows an equivalent plot for the ADAF/MAD
simulation, corresponding to the final four time chunks, M2, M3, M4,
M5. Comparing Figs. \ref{fig:SANE_density} and \ref{fig:MAD_density},
the flow streamlines in the MAD run show more well-organized outflow
behavior.  There are also outflowing streamlines along the axis,
suggesting some kind of polar jet. However, very little energy, and
practically no mass, flows along this jet. Therefore, for all
practical purposes, the simulation does not have a jet.

A critical issue for analyzing simulation data is knowing which
regions of the solution have had sufficient time to settle down to a
state of ``inflow equilibrium'', and which regions are still in the
process of getting there.  One way to do this is by looking at plots
such as Fig.~\ref{fig:Mdot_Sigma} and estimating ``by eye'' the region
of steady state. However, a more objective criterion is preferable, so
we follow the prescription for inflow equilibrium described in
\citet{Penna+10}. For each time chunk, we compute the time-averaged
radial velocity profile $v_r(r)$ of the gas within one scale-height of
the mid-plane (the restriction to one scale-height is to enable us to
focus on the accretion flow rather than any mass outflow or jet). From
this, we estimate the viscous time as a function of radius $r$ in the
standard way:
\begin{equation}
t_{\rm visc}(r) \equiv \frac{r}{|v_r(r)|}.
\end{equation}
We then define two criteria, one ``strict'' and one ``loose'', to
estimate the radius range over which the flow has achieved inflow
equilibrium:
\begin{eqnarray}
t_{\rm visc}(r_{\rm strict}) &=& t_{\rm chunk}/2 ~=~ t_{\rm tot}/4, \label{eq:strict} \\ 
t_{\rm visc}(r_{\rm loose}) &=& t_{\rm chunk} ~=~ t_{\rm tot}/2. \label{eq:loose}
\end{eqnarray}
Here, $t_{\rm chunk}$ is the time duration of the chunk under
consideration, and $t_{\rm tot}$ is the total run time from the
beginning of the simulation up to the end of the current
chunk\footnote{Note that the chunks are so defined that the duration
  of each chunk is half the total run time of the simulation up to
  that point (Tables \ref{tab:ADAF/SANE}, \ref{tab:ADAF/MAD})}.  

The philosophy behind the above criteria is that we expect the flow to
reach inflow equilibrium on a time scale of order the viscous time.
Further, it takes a few viscous times to average out fluctuations.
The strict criterion has $t_{\rm tot}=2t_{\rm chunk}=4t_{\rm visc}$,
which is a fairly safe and conservative choice, while the loose
criterion takes a more optimistic view of how soon inflow equilibrium
is achieved. Note that \cite{Penna+10} defined inflow equilibrium by
the condition $t_{\rm tot}=2t_{\rm visc}$, which is the same as our
present loose criterion. The values of $t_{\rm chunk}$, $r_{\rm
  strict}$ and $r_{\rm loose}$ for the various time chunks are listed
in Tables \ref{tab:ADAF/SANE} and \ref{tab:ADAF/MAD}, and $r_{\rm
  strict}$ and $r_{\rm loose}$ are shown as circular solid lines in
Figs.~\ref{fig:SANE_density} and \ref{fig:MAD_density}.  It will be
noticed that the objectively determined $r_{\rm strict}$ and $r_{\rm
  loose}$ are compatible with values one might deduce by visual
inspection of Fig.~\ref{fig:Mdot_Sigma}.

In Figs.~\ref{fig:SANE_density} and \ref{fig:MAD_density}, the
time-averaged velocity streamlines are well-behaved within the
respective inflow equilibrium regions of the four panels.  Note also
that the steady state zone is much more extended in the MAD simulation
compared to the SANE simulation. For instance, MAD chunk M5, which has
run only half as long as SANE chunk S6 is converged out to a
substantially larger radius (compare the values of $r_{\rm strict}$,
$r_{\rm loose}$ in Tables \ref{tab:ADAF/SANE} and \ref{tab:ADAF/MAD}).
The reason is the larger radial velocity of the gas in the MAD
simulation (compare Figs. \ref{fig:SANE_velangmmtm} and
\ref{fig:MAD_velangmmtm}).

When the accretion flow has reached inflow equilibrium, we expect
$\theta$- and $\phi$-integrated fluxes of conserved quantities, as
defined in equations (\ref{eq:mdot})--(\ref{eq:j}), to be independent
of radius. Recall that there is no radiative cooling, hence there
ought to be strict conservation of not only mass, but also energy and
angular momentum.  As time proceeds, the range of $r$ over which these
fluxes are constant will increase, and should track $r_{\rm strict}$
or $r_{\rm loose}$ (depending on the degree of constancy one
requires). 

\begin{figure}
\begin{center}
\includegraphics[width=0.7\columnwidth]{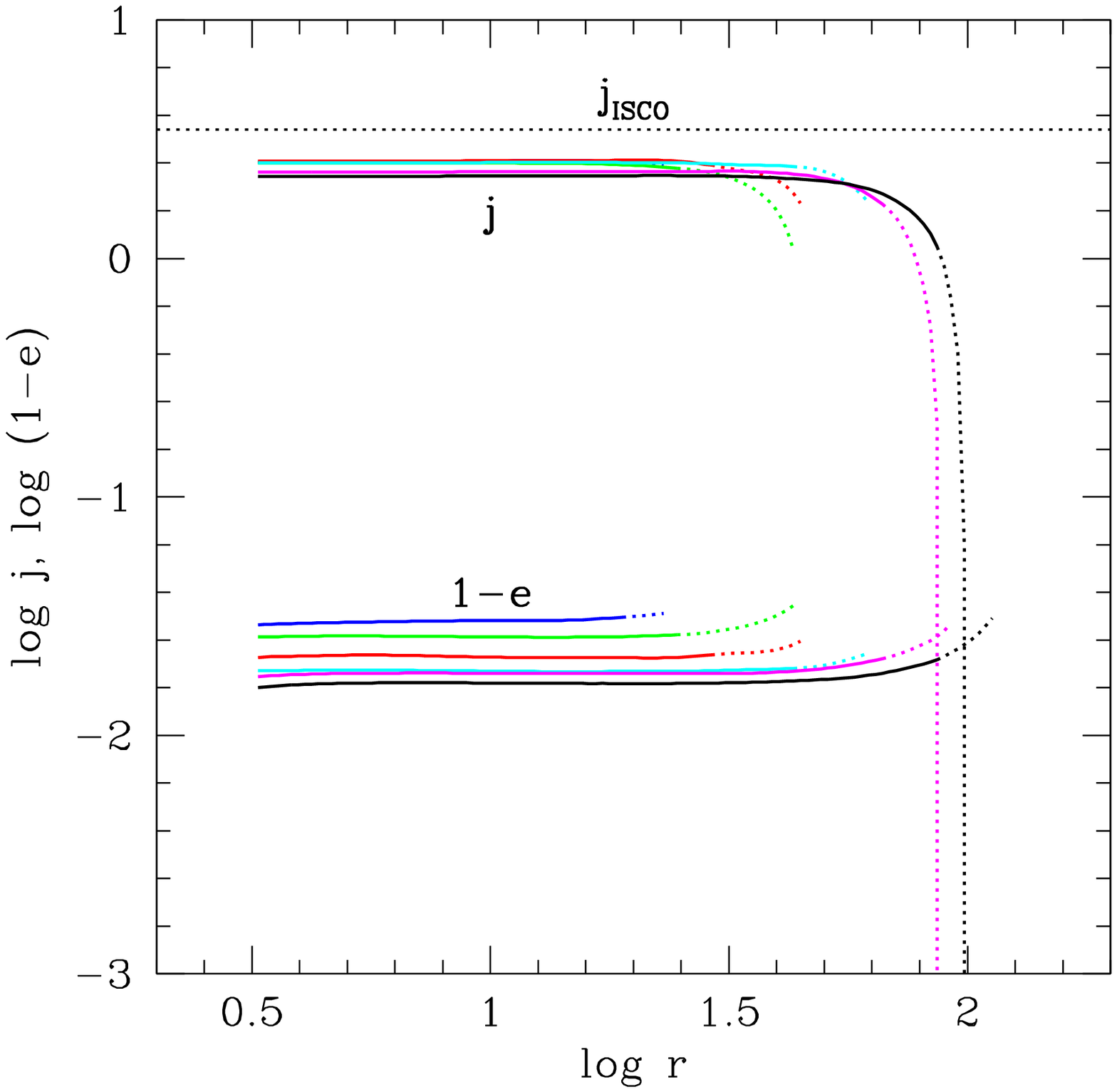}
\end{center}
\caption{The black dotted line at the top labeled $j_{\rm ISCO}$
  corresponds to the angular momentum of a Keplerian orbit at the
  radius of the ISCO.  This represents the specific angular momentum
  flowing into the BH in the case of a standard thin disc
  \citep{NT73}. The cluster of lines just below the dotted line shows
  the run of specific angular momentum flux with radius $j(r)$
  corresponding to chunks S1 (blue), S2 (green), S3 (red), S4 (cyan),
  S5 (magenta) and S6 (black) for the ADAF/SANE simulation.  All of
  these curves lie below the NT curve, indicating that the ADAF flow
  is sub-Keplerian, as predicted by theory. Each of the curves has a
  flat segment where the time-averaged flow shows excellent steady
  state convergence and a region at larger radii where $j$ deviates
  from steady state. The bottom set of lines (same colour coding)
  shows the specific binding energy flux $(1-e)$ for the same time
  chunks. For both sets of lines, the solid and dotted line segments
  correspond to $r\le r_{\rm strict}$ and $r\le r_{\rm loose}$,
  respectively (see text and Tables \ref{tab:ADAF/SANE},
  \ref{tab:ADAF/MAD}).}
\label{fig:SANE_jdotedot}
\end{figure}

\begin{figure}
\begin{center}
\includegraphics[width=0.7\columnwidth]{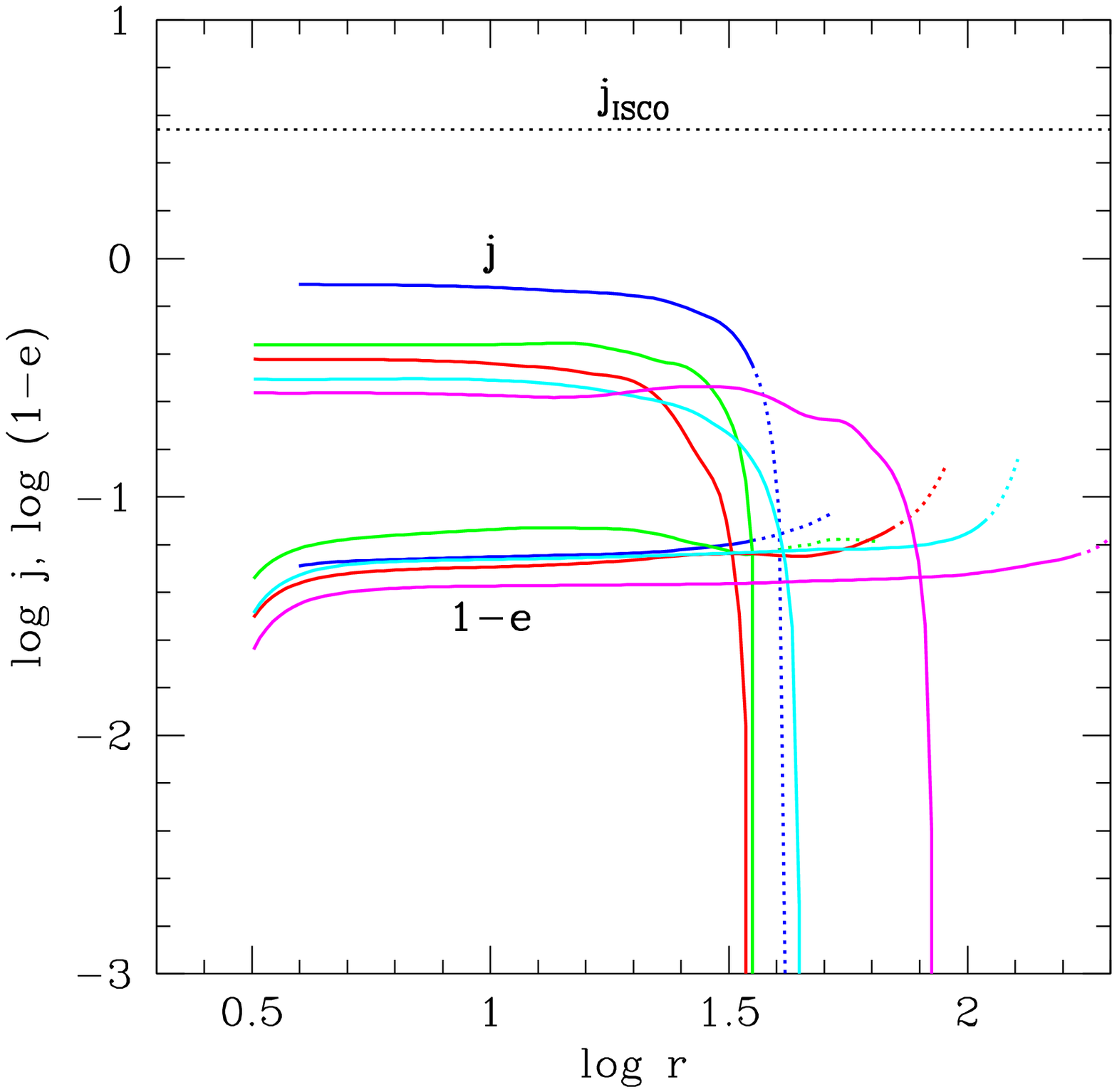}
\end{center}
\caption{Similar to Fig.~\ref{fig:SANE_jdotedot}, but for the ADAF/MAD
  simulation. The colour coding is: chunk M1 (blue), M2 (green), M3
  (red), M4 (cyan), M5 (magenta). }
\label{fig:MAD_jdotedot}
\end{figure}

Figure \ref{fig:SANE_jdotedot} shows the fluxes of specific angular
momentum $j$ and specific binding energy $(1-e)$ for the six time
chunks in the ADAF/SANE simulation. The range of radius over which
these fluxes are in inflow equilibrium increases from time chunk S1 to
S6, i.e., with increasing time, as expected.  The solid line segments
in the plot correspond to the strict criterion $r \leq r_{\rm
  strict}$, and the dotted lines correspond to the loose criterion $r
\leq r_{\rm loose}$. This convention is adopted in all later plots.

Figure \ref{fig:SANE_jdotedot} highlights the difference in
convergence properties between the two criteria. Although the strict
criterion is not perfect, the fluxes do remain nearly constant over
the radius ranges defined by this criterion.  The loose criterion,
however, shows unacceptably large deviations from flux
constancy. Hereafter, we quote quantitative results only for regions
satisfying the strict criterion (the inner solid circles in
Fig.~\ref{fig:SANE_density}), though we plot results for
both\footnote{Obviously, more accurate results could be obtained by
  using an even stricter criterion, e.g., $t_{\rm visc} \leq t_{\rm
    chunk}/4$. However, this would reduce the range of $r$ so much
  that we would not have sufficient dynamic range to obtain any useful
  results.}.  Interestingly, the angular momentum flux shows larger
deviations from constancy than either the binding energy flux $(1-e)$
or the mass accretion rate (shown in Figs. \ref{fig:Mdot_Sigma} and
\ref{fig:SANE_outflow}). We are not sure why this is the case.

Figure \ref{fig:SANE_jdotedot} indicates that there is a slow secular
decrease in the converged values of both $j$ and $(1-e)$ with time;
the values for chunk S6 are smaller than those for S5, and so on.
This is similar to, though not as extreme as, the declining trend in
$\dot{M}$ already seen in Fig.~\ref{fig:Mdot_Sigma}. We suspect that,
in the case of $j$ and $(1-e)$, the reason for the decline is that the
SANE simulation is slowly approaching the MAD limit (despite our best
efforts to avoid it).

Figure \ref{fig:MAD_jdotedot} shows equivalent results for the
ADAF/MAD simulation. Here, $j$ and $(1-e)$ are less well-behaved than
in the SANE simulation.  In fact, it appears that even $r_{\rm
  strict}$ may overestimate the actual radius out to which inflow
equilibrium has been achieved.  The binding energy flux $(1-e)$ is a
few times larger for the MAD simulation compared to the SANE
simulation. This implies that the MAD accretion flow returns
mechanical and magnetic energy to infinity more efficiently compared
to the SANE simulation. In essence, the outflowing gas carries more
energy per unit mass.  The angular momentum flux $j$ is substantially
smaller in the MAD simulation compared to the SANE run. Indeed $j$
appears secularly to approach zero with increasing time, as seen also
in the highly sub-Keplerian values of $u_\phi$ (compare
Figs. \ref{fig:SANE_velangmmtm} and \ref{fig:MAD_velangmmtm}). In
fact, it seems that BH spinup via an ADAF/MAD accretion flow is highly
inefficient. This agrees with the results reported in
\citet{Tchekhovskoy+12} and \citet{MTB12}.

\begin{figure}
\begin{center}
\includegraphics[width=0.49\columnwidth]{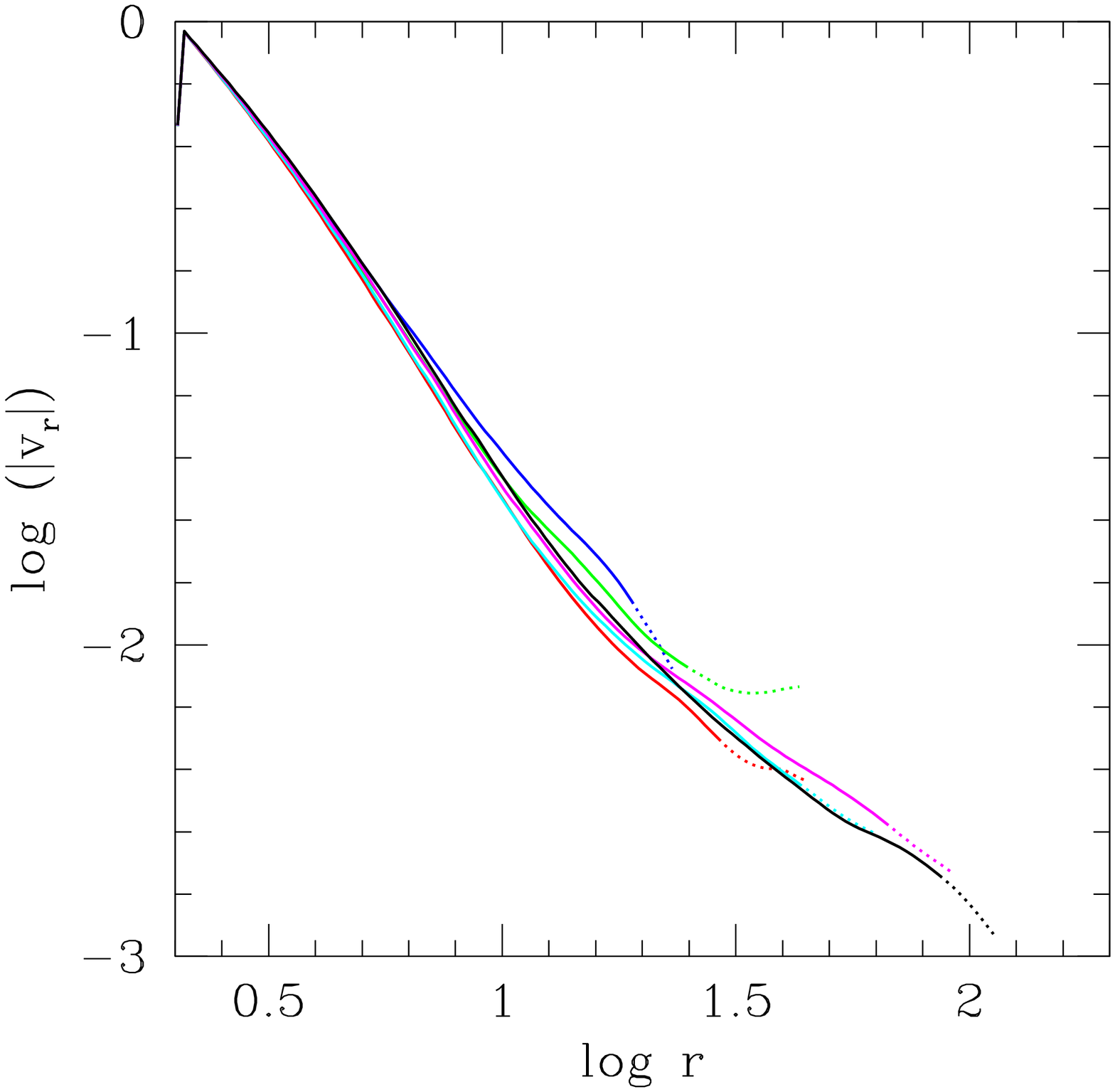}
\includegraphics[width=0.49\columnwidth]{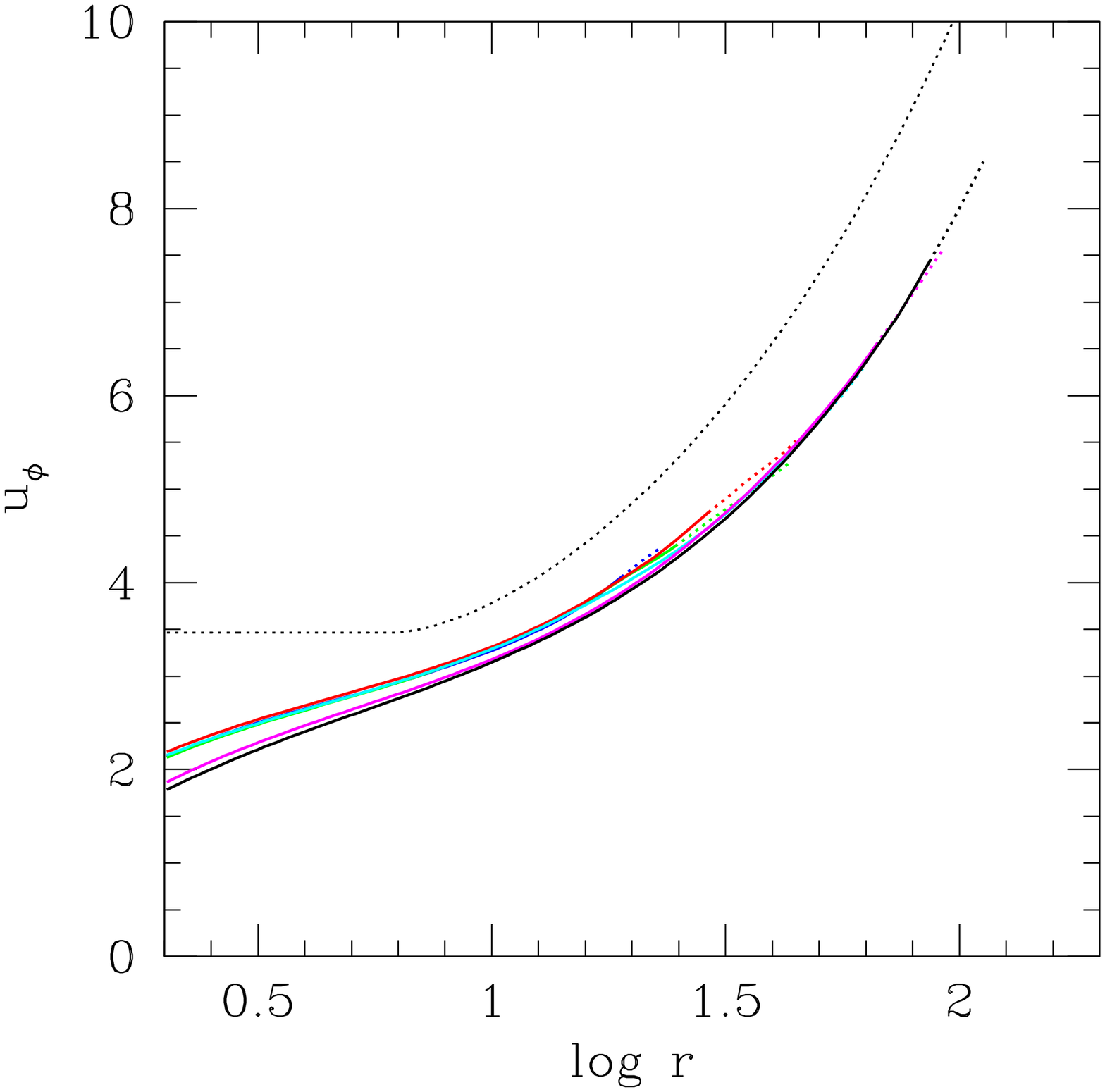}
\includegraphics[width=0.49\columnwidth]{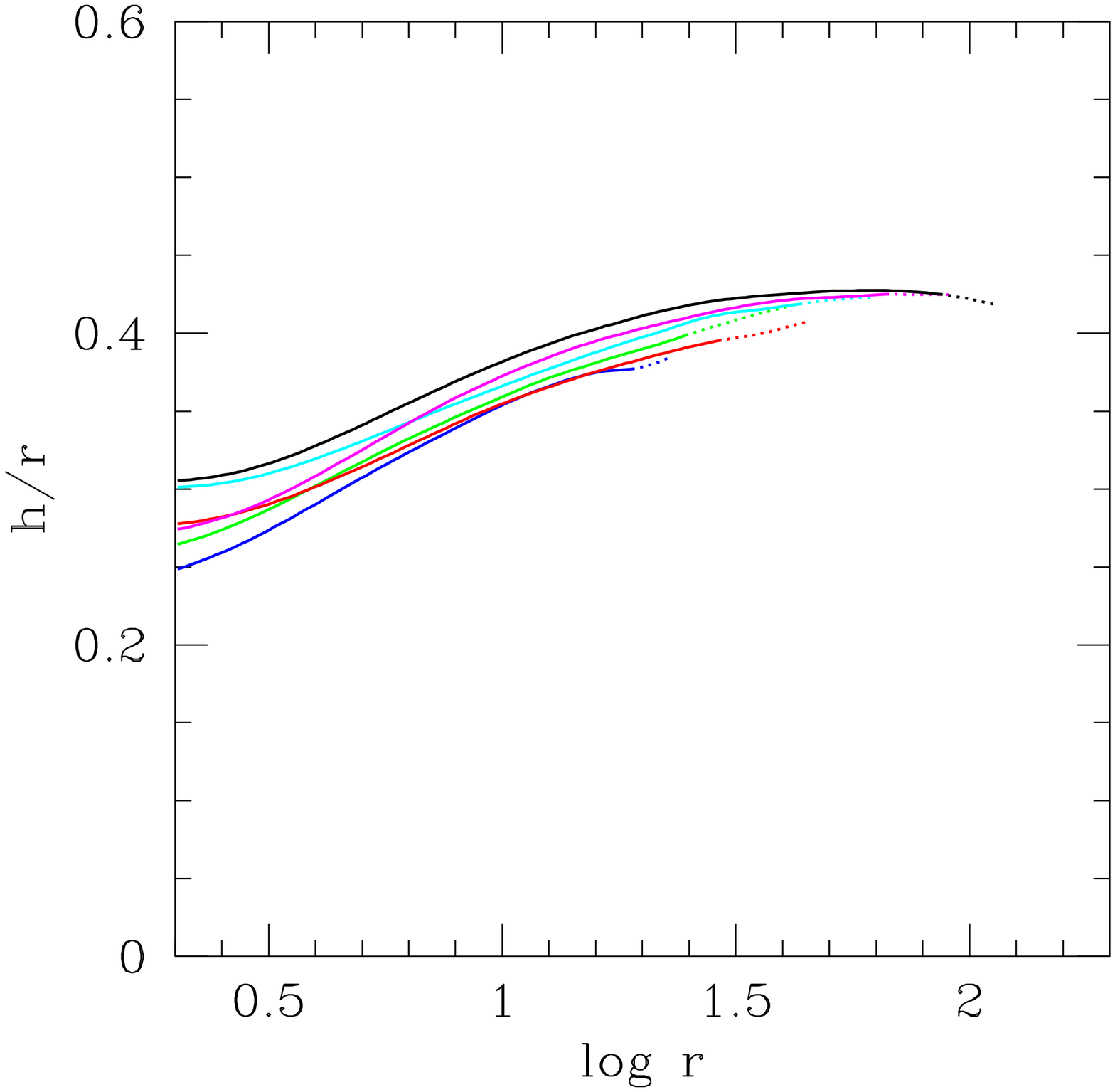}
\includegraphics[width=0.49\columnwidth]{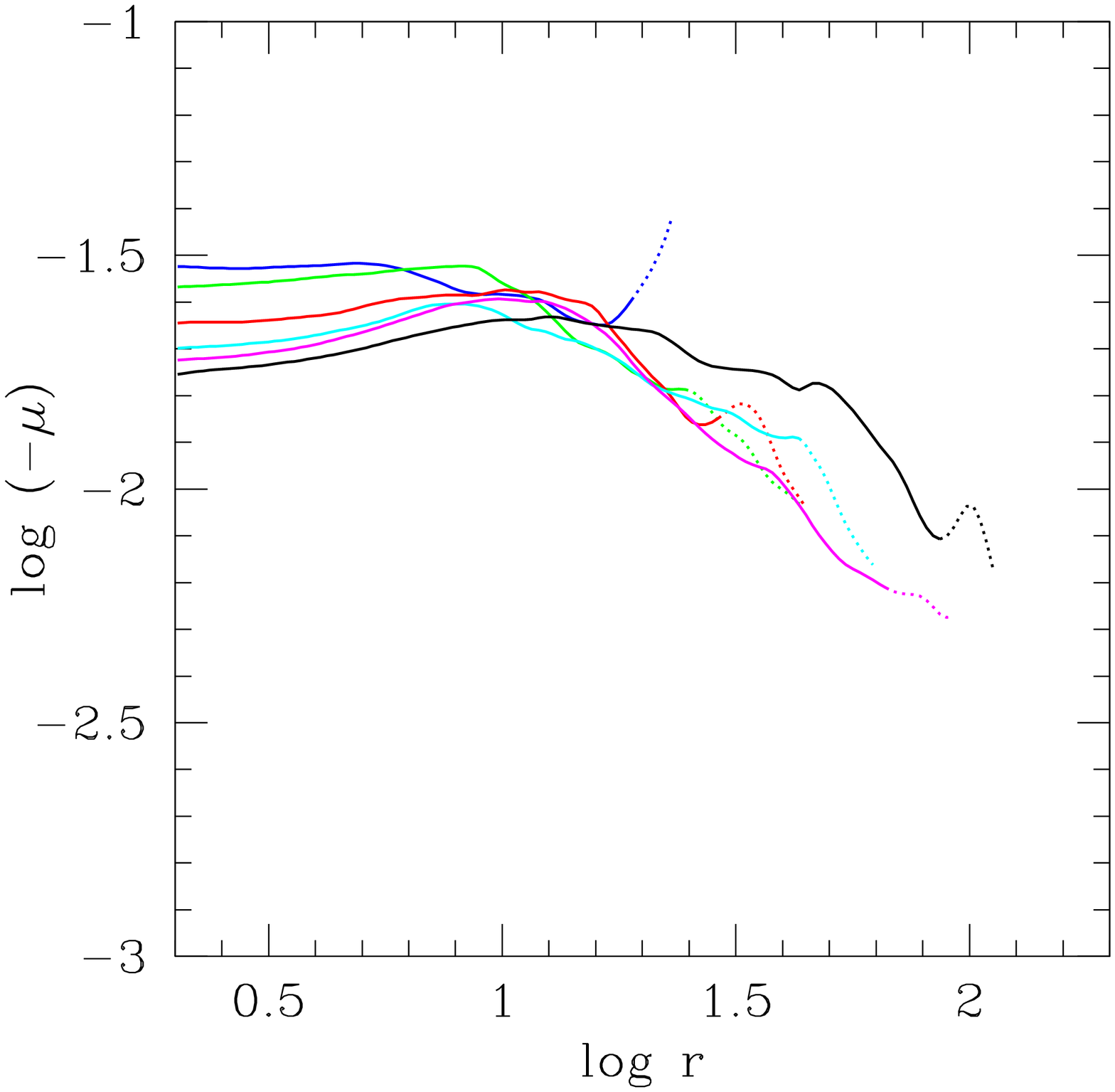}
\end{center}
\caption{Top Left: Shows the density-weighted mean radial velocity of
  the gas in the ADAF/SANE simulation within one scale height of the
  mid-plane during time chunks S1--S6.  The colour code and line types
  are the same as in Fig.~\ref{fig:SANE_jdotedot}. Top Right: A
  similar plot for the density-weighted specific angular momentum
  $u_\phi$ of the accreting gas. The black dotted line shows the
  Keplerian profile of angular momentum for a standard thin accretion
  disc \citep{NT73}. Bottom Left: Plot of the density scale height
  $h/r$ for the six time chunks. Bottom Right: Plot of the mid-plane
  values of $\mu$, which represents the normalized flux of the
  Bernoulli parameter (see eq.~\ref{eq:mu}). The fact that $\mu$ is negative indicates that
  the mid-plane gas is bound to the BH.}
\label{fig:SANE_velangmmtm}
\end{figure}

\begin{figure}
\begin{center}
\includegraphics[width=0.49\columnwidth]{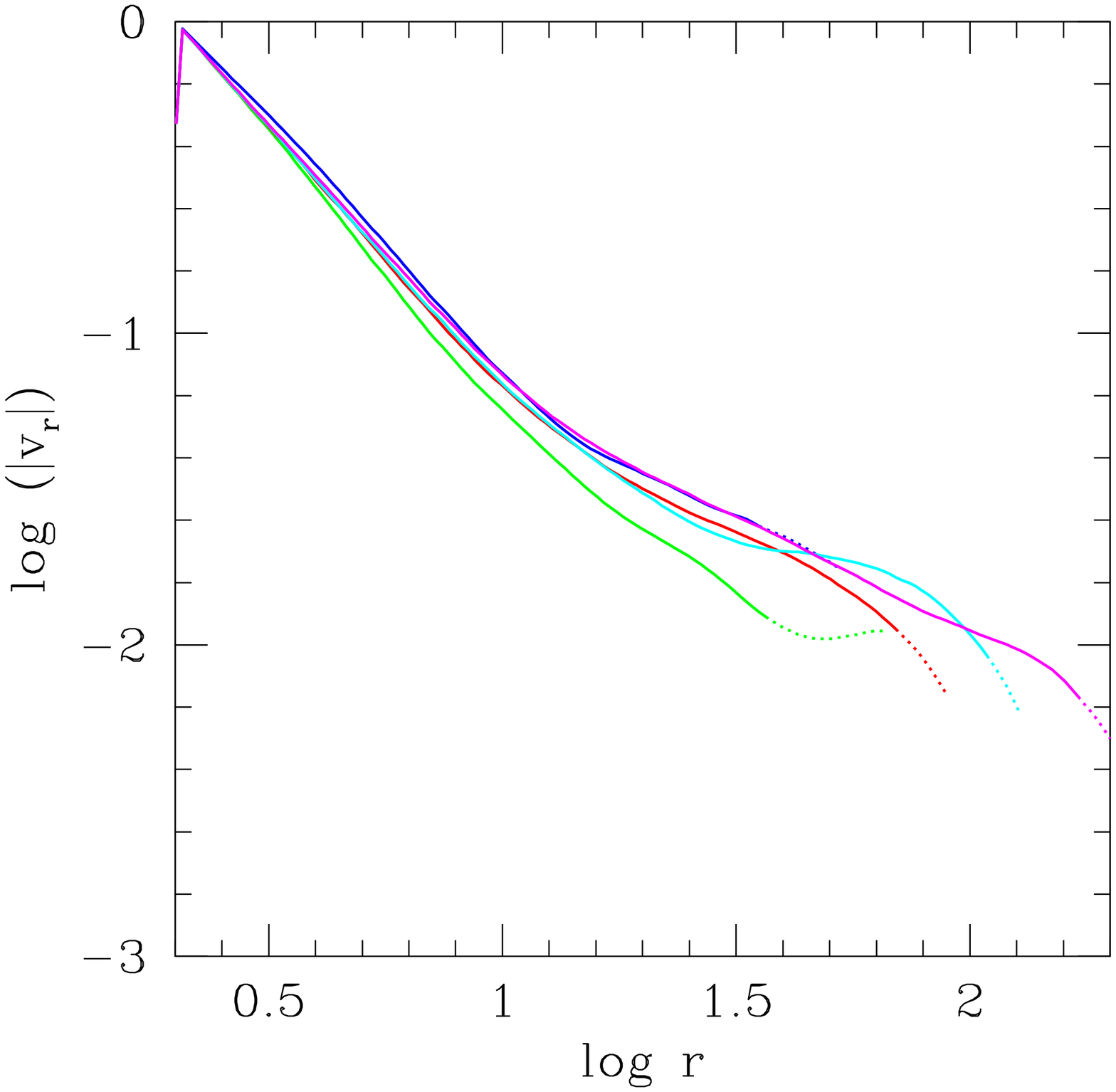}
\includegraphics[width=0.49\columnwidth]{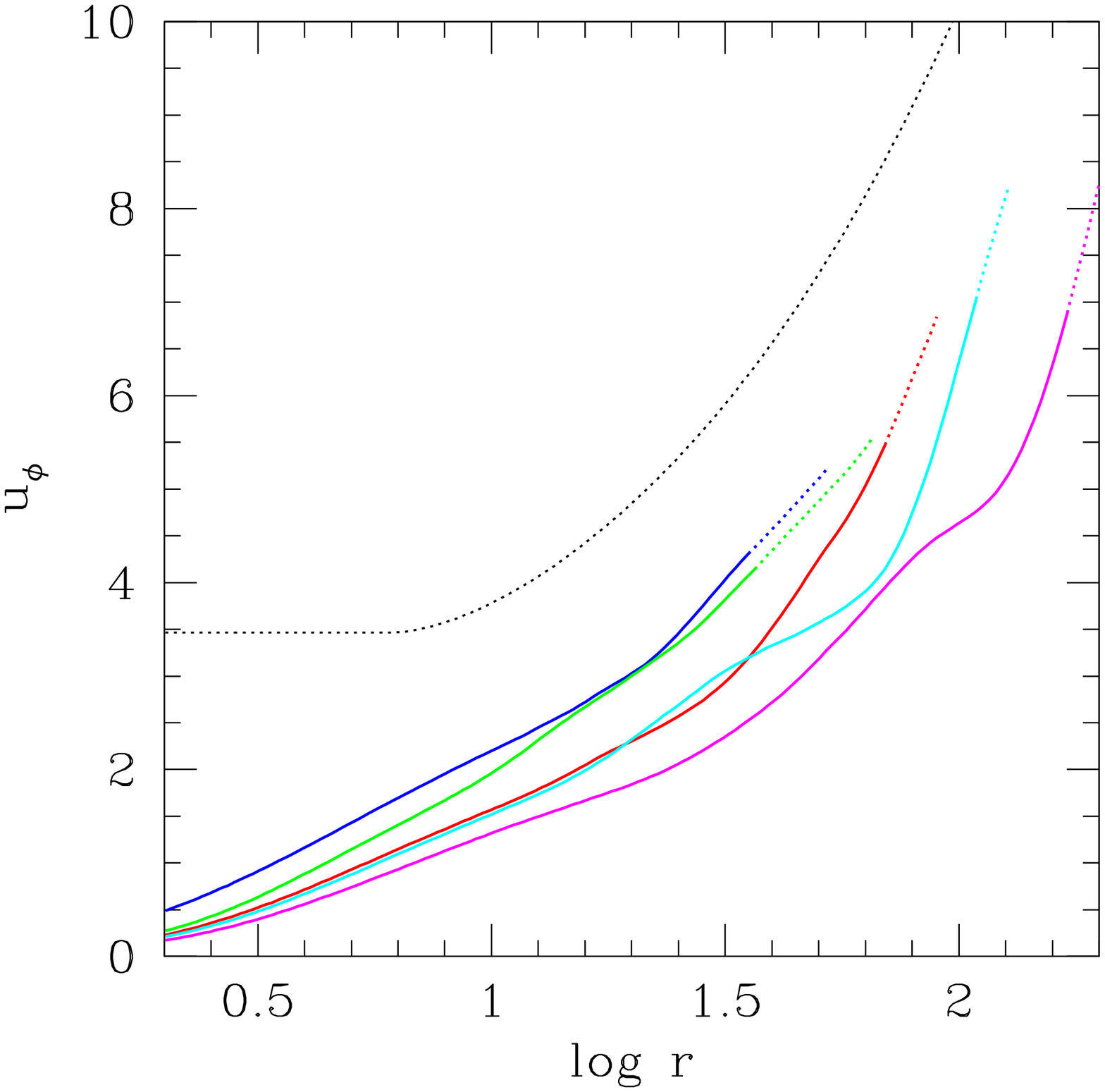}
\includegraphics[width=0.49\columnwidth]{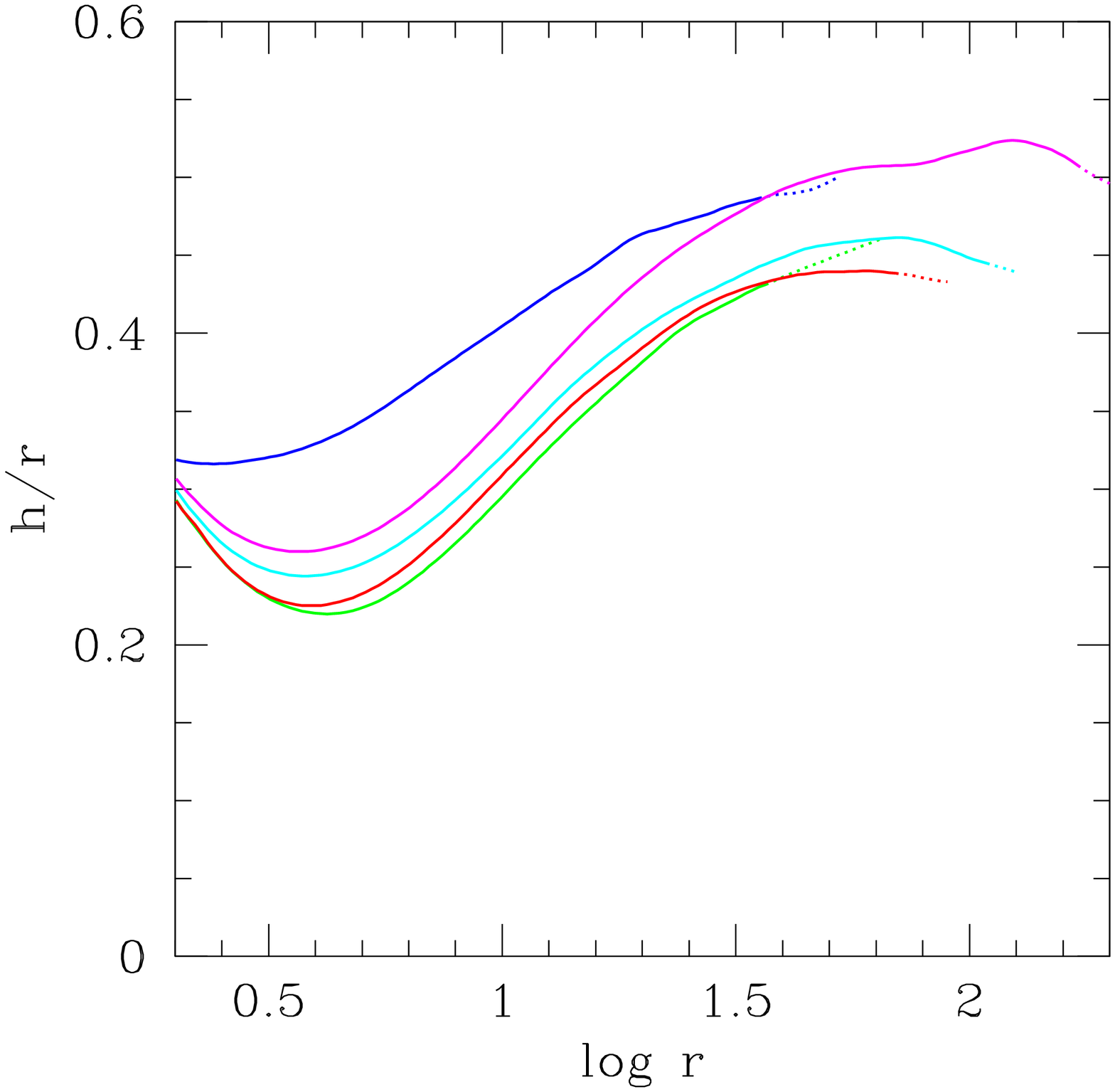}
\includegraphics[width=0.49\columnwidth]{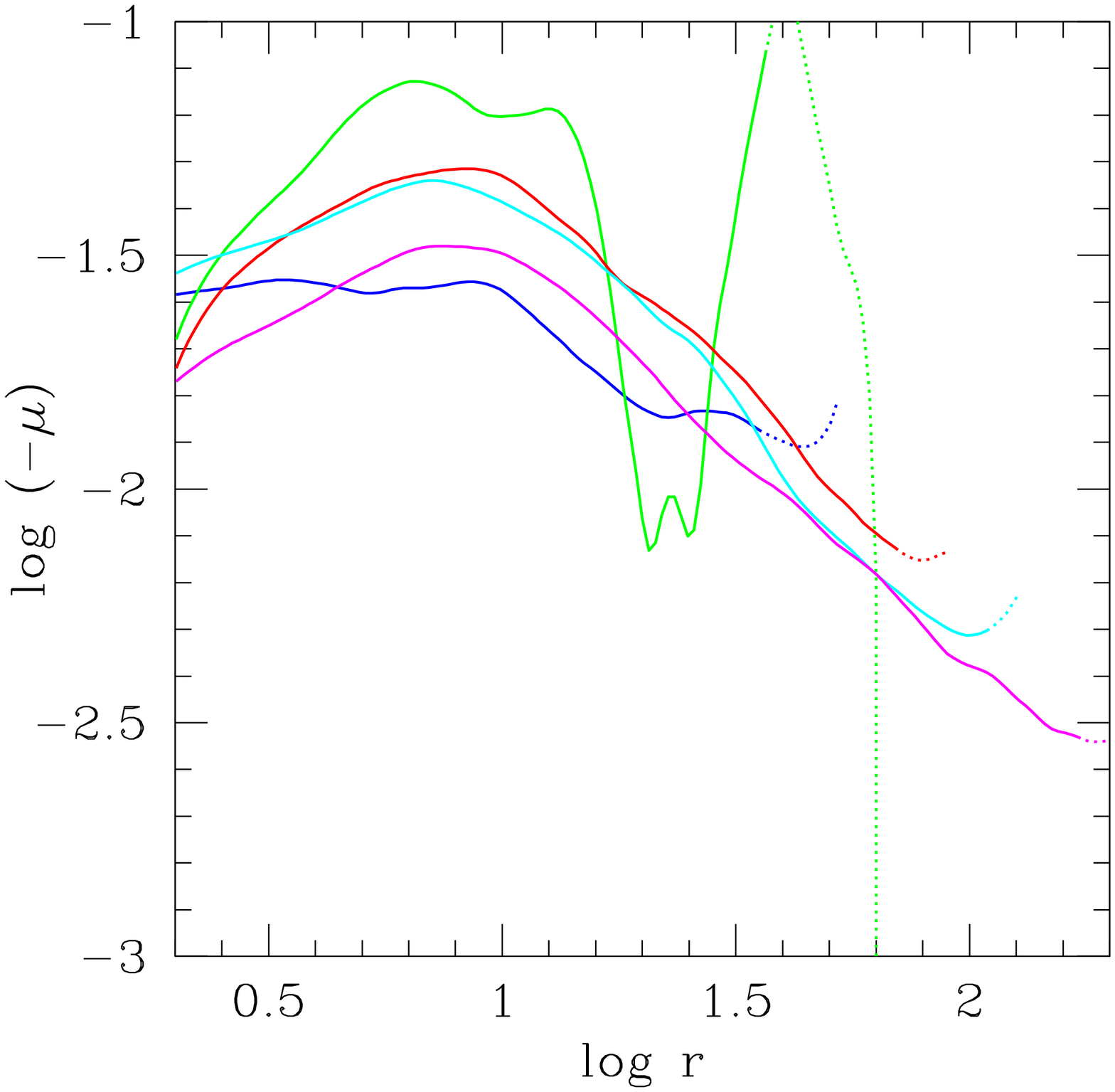}
\end{center}
\caption{Similar to Fig.~\ref{fig:SANE_velangmmtm}, but for the
  ADAF/MAD simulation. Colour coding is as in
  Fig.~\ref{fig:MAD_jdotedot}.}
\label{fig:MAD_velangmmtm}
\end{figure}

Figure \ref{fig:SANE_velangmmtm} shows the radial velocity $|v_r(r)|$,
the specific angular momentum $u_\phi(r)$ of the gas within one scale
height, and the normalized scale height $h/r$. There is good internal
consistency between the profiles from successive time chunks.  This is
especially true when we focus only on the regions that satisfy the
strict criterion for inflow equilibrium (the solid line segments).
Specifically, apart from a tendency for $h/r$ to increase slightly
with time, the profiles of various quantities in successive time
chunks line up well with one another, showing that we have a
well-behaved accretion flow.  We view the good agreement as a sign of
convergence in our results.

At $r=100$, we have $|v_r|\approx0.002$, which is far smaller than the
local free-fall velocity $v_{\rm ff} \approx 0.14$. This is to be
expected. The radial velocity in a viscous flow is $\sim\alpha(h/r)^2
v_{\rm ff}$, where $\alpha$ is the dimensionless viscosity parameter
and $(h/r)$ is the dimensionless geometrical thickness of the
disc. The simulated system has $h/r\sim0.4$ and $\alpha\sim 0.05$
(near $r\sim100$), and this explains the observed velocity.

The specific angular momentum $u_\phi$ of the accreting gas is
sub-Keplerian (as predicted by simple ADAF models).  Interestingly,
$u_\phi$ continues to decline with decreasing radius even in the
plunging region, i.e., inside the innermost stable circular orbit,
$r_{\rm ISCO}=6$. It appears that the dynamics of an ADAF are not
strongly modified when the gas crosses the ISCO. This is in contrast
to geometrically thin discs, where the angular momentum becomes nearly
constant once the gas flows inside the ISCO
\citep{Shafee+08,Penna+10}.

The fourth panel in Fig.~\ref{fig:SANE_velangmmtm} shows the
normalized Bernoulli-flux parameter $\mu$ (defined below in
eq.~\ref{eq:mu}) of the mid-plane gas. Recall that the initial gas in
the torus had Bernoulli in the range $10^{-2}-10^{-3}$. The mid-plane
gas in the accretion flow has a more negative value of $\mu$, which
means it is more tightly bound to the BH compared to the initial gas.
The profiles from the different time chunks agree reasonably well with
one another, but not perfectly. This is perhaps to be expected since
$\mu$ is computed as the difference of two quantities of order
unity. Note that the outflowing gas we consider in the next subsection
has a positive $\mu$. That gas has acquired extra energy in the
process of accretion, and it is the extra energy that drives the
outflow \citep{NY94}.

Figure \ref{fig:MAD_velangmmtm} shows the corresponding results for
the MAD simulation. The radial velocity is substantially larger
compared to the SANE simulation. Indeed, this is the reason for the
larger zone of inflow equilibrium in this simulation. Both disc
thickness $h/r$ and Bernoulli $Be$ show more fluctuations between
successive time chunks. This is part of a pattern --- fluctuations of
all quantities are generally larger in the MAD simulation.  The MAD
flow is slightly thicker than the SANE flow, $h/r \sim 0.5$ compared
to $\sim 0.4$, but it has roughly the same (negative) value of $Be$ at
the mid-plane.

\subsection{Mass Loss in an Outflow}
\label{sec:outflow}

The main motivation behind the present study is to evaluate the amount
of mass loss experienced by an ADAF through winds and
outflows. Figures~\ref{fig:SANE_density} and \ref{fig:MAD_density}
show that mass does flow out in both the SANE and MAD
simulations. However, just because a given parcel of gas moves away
from the BH does not necessarily mean that it escapes to infinity. The
gas might just move out for a certain distance, turn round and merge
with the inflowing gas.  We need a physical criterion other than mere
outward motion to determine whether or not mass is lost. Before
proceeding further we note that there is no sign of a relativistic
polar jet in our simulations, in agreement with the results of
\citet{MTB12} for their runs with non-spinning BHs. This is perhaps
not surprising since there is growing evidence that relativistic jets
are powered by BH spin \citep{Tchekhovskoy+11,NM12}.  In any case, the
discussion below is concerned with non-relativistic mass outflows, not
jets.

We work with gas properties averaged over the duration of a time chunk
$t_{\rm chunk}$ and azimuthal angle $\phi$, and symmetrized around the
mid-plane. We do this not only for quantities like density and
velocity, but for all other quantities mentioned below, e.g., $\rho
u_t$, $u u_t$, $b^2 u_t$, etc. As Figs.~\ref{fig:SANE_density} and
\ref{fig:MAD_density} show, such averaging eliminates all turbulent
fluctuations inside the region of inflow equilibrium, allowing us to
focus on the mean properties of the flow. This is important when
trying to evaluate the magnitude of outflows.

We have considered three criteria for deciding whether a gas
streamline escapes to infinity.  The first two criteria involve
variants of the Bernoulli parameter of the gas. This was the parameter
considered by \citet{NY94} in their original work in which they
identified mass loss as being potentially important in ADAFs. In
Newtonian hydrodynamics, $Be$ is the sum of kinetic energy, potential
energy and enthalpy. At large distance from the BH, the potential
energy vanishes. Since the other two terms are positive, gas at
infinity must have $Be \ge 0$.  Furthermore, in steady state and in
the absence of viscosity, $Be$ is conserved along streamlines.  Hence
any parcel of gas that flows out with a positive value of $Be$ can
potentially reach infinity. This was the crux of the argument proposed
by \citet{NY94}.

In our case, we have an MHD flow in a general relativistic space-time.
Here, the Bernoulli parameter may be written as \citep{penna2012}
\begin{equation}
Be = -\frac{\langle\rho u_t\rangle + \Gamma\langle u u_t\rangle
+ \langle b^2 u_t\rangle}{\langle\rho\rangle} - 1, \label{eq:Be}
\end{equation}
where $\langle\cdots\rangle$ indicates an average over time and
azimuth. We subtract unity to eliminate the rest mass energy of the
gas. Far from the BH, the expression in (\ref{eq:Be}) reduces to the
Newtonian quantity --- kinetic energy plus gas enthalpy plus magnetic
enthalpy --- which has to be positive. Therefore, gas in a given
poloidal cell of the simulation is likely to escape to infinity if the
time-averaged properties in that cell satisfy the following two
conditions: (1) the mean velocity has an outward radial component,
i.e., $\langle v_r \rangle >0$, and (2) the gas has $Be \ge 0$. This
is the first of three criteria we have considered.

Because magnetic stress is anisotropic, the contribution of the
magnetic field to the Bernoulli is not well-defined. Therefore, some
authors (e.g., \citealt{Tchekhovskoy+11,MTB12}) ignore the magnetic
term and consider the following modified Bernoulli parameter,
\begin{equation}
Be' = -\frac{\langle\rho u_t\rangle + \Gamma\langle u u_t\rangle}
{\langle\rho\rangle} - 1, \label{eq:Bep}
\end{equation}
This is arguably a more robust quantity, though it underestimates the
Bernoulli. The second criterion we have considered for identifying
outflowing gas is that it should satisfy (1) $\langle v_r\rangle>0$
and (2) $Be' \geq 0$.

Our third criterion involves a normalized energy outflow rate, similar
to the ratio $\mu$ of energy flux to rest mass flux discussed in
theories of magnetized relativistic jets (e.g.,
\citealt*{Tchekhovskoy+10a}). For our general relativistic MHD flow, we
define $\mu$ to be
\begin{equation}
\mu = \frac{\langle T_t^p\rangle}{\langle\rho u^p\rangle}-1, \label{eq:mu}
\end{equation}
where the index $p$ refers to ``poloidal'', and we subtract unity to
eliminate the contribution due to rest mass. Note that $\langle
T_t^p\rangle/\langle\rho u^p\rangle$ is just a local version of
$\dot{E}/\dot{M}$ in equation (\ref{eq:e}). Thus, $\mu$ measures the
flux of the Bernoulli (normalized by mass flux) and is the most
natural quantity for our analysis. In particular, it includes the
contribution of the magnetic shear stress (terms like $b^rb_\phi$ in
eq.~\ref{eq:Trphi}), which is not included in the definitions of $Be$
and $Be'$ above.  As before, we consider a parcel of gas to escape to
infinity from a given radius $r$ if (1) its average velocity at $r$ is
pointed outward, and (2) $\mu\ge 0$.  For a steady axisymmetric ideal
MHD flow, $\mu$ is conserved along an outflowing streamline.  Hence
this $\mu$-based criterion is arguably the most physically
well-motivated of the three criteria, and the one closest in spirit to
the original work of \citet{NY94}.

\begin{figure}
\begin{center}
\includegraphics[width=0.7\columnwidth]{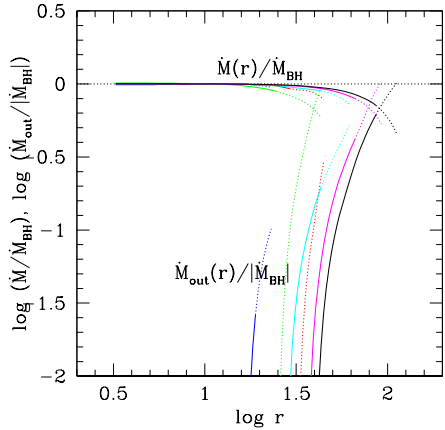}
\end{center}
\caption{The horizontal lines near the top of the plot show the net
  mass inflow rate $\dot{M}(r)$ for the six time chunks S1--S6 of the
  ADAF/SANE simulation, normalized by the net mass accretion rate on
  to the BH, $\dot{M}_{\rm BH}$.  The colours and line types are as in
  Fig.~\ref{fig:SANE_jdotedot}. The vertical lines near the bottom
  show the variation of the mass outflow rate $\dot{M}_{\rm out}(r)$
  according to the $\mu$ criterion (the results are similar to those
  obtained with the $Be$ or $Be'$ criteria), again normalized by
  $\dot{M}_{\rm BH}$.  There is poor convergence in the results for
  the outflow, since no two successive time chunks are consistent with
  one another. The deviations are systematic --- in the last three
  time chunks (S4:cyan, S5:magenta, S6:black), each successive time
  chunk gives a lower $\dot{M}_{\rm out}$ at a given $r$ compared to
  the previous chunk. Hence, the mass outflow rates shown here should
  be interpreted as upper limits.}
\label{fig:SANE_outflow}
\end{figure}

\begin{figure}
\begin{center}
\includegraphics[width=0.7\columnwidth]{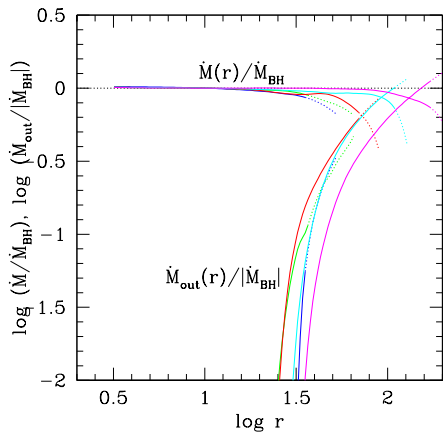}
\end{center}
\caption{Mass outflow rate in the ADAF/MAD simulation based on the
  $\mu$ criterion.  The colours and line types are as in
  Fig.~\ref{fig:MAD_jdotedot}. The last three chunks (M3:red, M4:cyan,
  M5:magenta) show large and systematic deviations, suggesting that
  (as in the case of the ADAF/SANE simulation) we do not have good
  convergence and the computed mass outflow estimates are upper
  limits.}
\label{fig:MAD_outflow}
\end{figure}

Using each of the three criteria described above, we have computed the
mass outflow rate $\dot{M}_{\rm out}(r)$ as a function of $r$ for each
of the time chunks in the ADAF/SANE and ADAF/MAD simulations.  The
results from the three criteria agree well with one another. We show
plots corresponding to only the $\mu$ criterion.

Figure~\ref{fig:SANE_outflow} shows for the ADAF/SANE simulation the
mass outflow rate $\dot{M}_{\rm out}(r)$ and the net mass inflow rate
$\dot{M}(r)$, both normalized by the net mass accretion rate on to the
BH, $\dot{M}_{\rm BH}$. The results for the mass inflow rate
$\dot{M}(r)$ are identical to those shown in the top left panel of
Fig.~\ref{fig:Mdot_Sigma}, except that the normalization by
$\dot{M}_{\rm BH}$ shifts the curves vertically and causes them to lie
on top of one another.

Surprisingly, the results for $\dot{M}_{\rm out}$ show very poor
convergence. Specifically, the $\dot{M}_{\rm out}$ profiles
corresponding to different time chunks deviate substantially from one
another.  Moreover, the deviations are systematic. In each time chunk,
the outflow appears to pick up just around the limiting radius for
inflow equilibrium. Since the latter moves out for later chunks, the
entire $\dot{M}_{\rm out}$ profile also moves out.  Apparently, at
each time, the current estimate of the mass outflow rate at a given
radius is an overestimate compared to the rate we would estimate at a
later time (compare in particular the last three time chunks shown in
cyan, magenta and black). Because of this, the outflow rate estimate
even from the last time chunk S6 (black curve) must be viewed only as
an upper limit.  Moreover, even this estimate corresponds to a mass
loss rate at $r\sim100$ no more than the net inflow rate $\dot{M}_{\rm
  BH}$ into the BH.  Given that it is an upper limit, we can state
with some confidence that mass outflow is unimportant for $\rH < r <
100$.

It is useful to compare our results with those obtained by
\citet{MTB12} for their model A0.0BtN10.  This model was initialized
with a toroidal field and is an excellent example of an ADAF/SANE
system. In Table 4 of their paper, the authors provide various
estimates of the mass outflow rate measured at a characteristic radius
$r_{\rm o}=50$.  Their quantity $\dot{M}_{\rm mw,o}$ is most relevant
since it focuses on unbound gas, defined as $Be'>0$.\footnote{The
  authors define a second quantity, $\dot{M}_{\rm w,o}$, which
  represents all outflowing gas, regardless of whether the Bernoulli
  is positive or negative. It is less relevant for us since most of
  this gas is bound to the BH and cannot escape to infinity. We thank
  J.~McKinney (private communication) for clarifying the definitions
  of $\dot{M}_{\rm mw,o}$ and $\dot{M}_{\rm w,o}$.}  The normalized
mass outflow rate, $\dot{M}_{\rm mw,o}/\dot{M}_{\rm H}$, that
\citet{MTB12} find at $r=50$ in model A0.0BtN10 is essentially zero,
in good agreement with our result, $\dot{M}_{\rm out}/\dot{M}_{\rm
  BH}=0.07$ at $r=50$ in chunk S6; at $r=r_{\rm strict}=86$, our
outflow rate is $\dot{M}_{\rm out}/\dot{M}_{\rm BH}=0.6$.  It should
be noted that $\dot{M}_{\rm mw,o}$ includes additional constraints,
viz., that the escaping gas should have $b^2/\rho<1$ and gas to
magnetic pressure ratio $\beta<2$. Our mass outflow criteria do not
include these constraints. When we include them, we find that our mass
outflow rate is zero at $r=50$ and $86$. Apart from these details,
both the present work and model A0.0BtN10 in \citet{MTB12} agree on
the following key result: out to radii $\sim50-100$, ADAF/SANE systems
have negligible mass outflow.

Figure \ref{fig:MAD_outflow} shows mass outflow estimates obtained via
the $\mu$ criterion for the ADAF/MAD simulation. As in the case of the
ADAF/SANE simulation, the convergence behavior is poor.  In
particular, the results from chunks M3 (red), M4 (cyan) and M5
(magenta) do not agree well with one another. Thus, once again, we
believe the mass outflow rates we estimate from this simulation should
be viewed as upper limits.

Despite the unsatisfactory convergence, if we take the results at face
value, we find for time chunk M5, $\dot{M}_{\rm out}/\dot{M}_{\rm BH}
\approx 0.2$, $0.6$, $1.1$, at radii $r=50$, $100$, $170$ ($=r_{\rm
  strict}$), respectively.  Two of the simulations described in
\citet{MTB12}, A0.0BfN10 and A0.0N100, correspond to MAD flows around
non-spinning BHs and are good comparisons (though our simulation has
run significantly longer). At radius $r_{\rm o}=50$, A0.0BfN10 has
essentially zero outflow, i.e., $\dot{M}_{\rm mw,o}/\dot{M}_{\rm
  H}\approx 0$, while A0.0N100 has $\dot{M}_{\rm mw,o}/\dot{M}_{\rm
  H}\approx 0.4$. Our estimate, $\dot{M}_{\rm out}/\dot{M}_{\rm BH}
\approx 0.2$, agrees well\footnote{As mentioned earlier, \citet{MTB12}
  require several conditions to be satisfied, viz., $v_r>0$, $Be'>0$,
  $b^2/\rho<1$, $\beta<2$, before they include a particular gas
  streamline in their estimate of $\dot{M}_{\rm mw,o}$. When we apply
  the same conditions on our ADAF/MAD simulation, we estimate the mass
  outflow rate at $r=50$ to be $0.06$, still in good agreement with
  their outflow rates.}.

\begin{figure}
\begin{center}
\includegraphics[width=0.49\columnwidth]{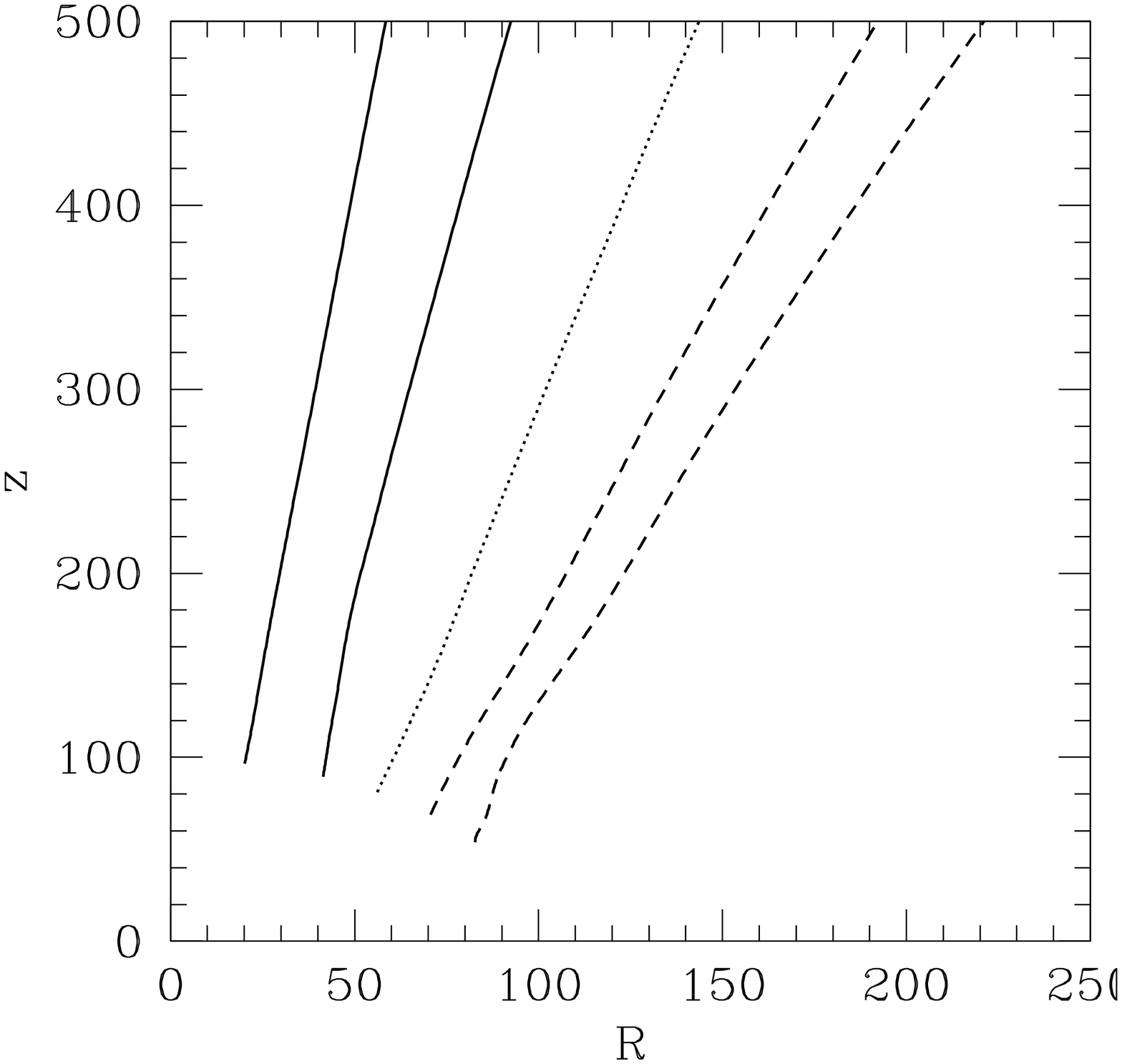}
\includegraphics[width=0.49\columnwidth]{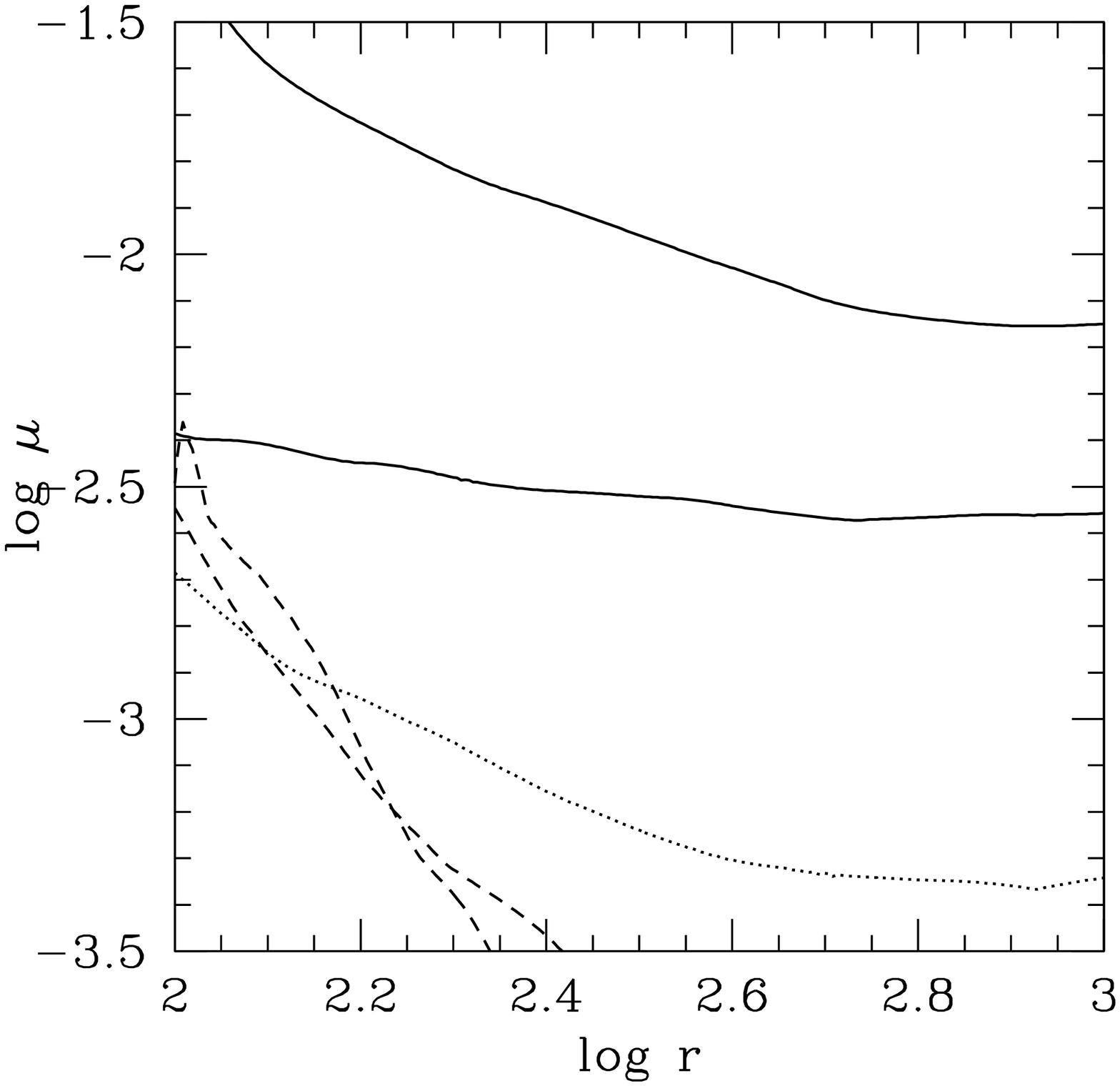}
\end{center}
\caption{Left: Shows five outflowing streamlines in time chunk S6 of
  the ADAF/SANE simulation. The streamlines have their footpoints at
  $(r,\theta) = (86,0.2)$, $(86,0.4)$, $(86,0.6)$, $(86,0.8)$,
  $(86,1.0)$. All five streamlines have positive values of $\mu$ at
  their footpoints. Right: The variation of $\mu$ along each of the
  streamlines in the left panel, using the same line types. Note that
  $\mu$ shows large deviations from constancy for the last two
  streamlines.}
\label{fig:streamlines}
\end{figure}

\begin{figure}
\begin{center}
\includegraphics[width=0.49\columnwidth]{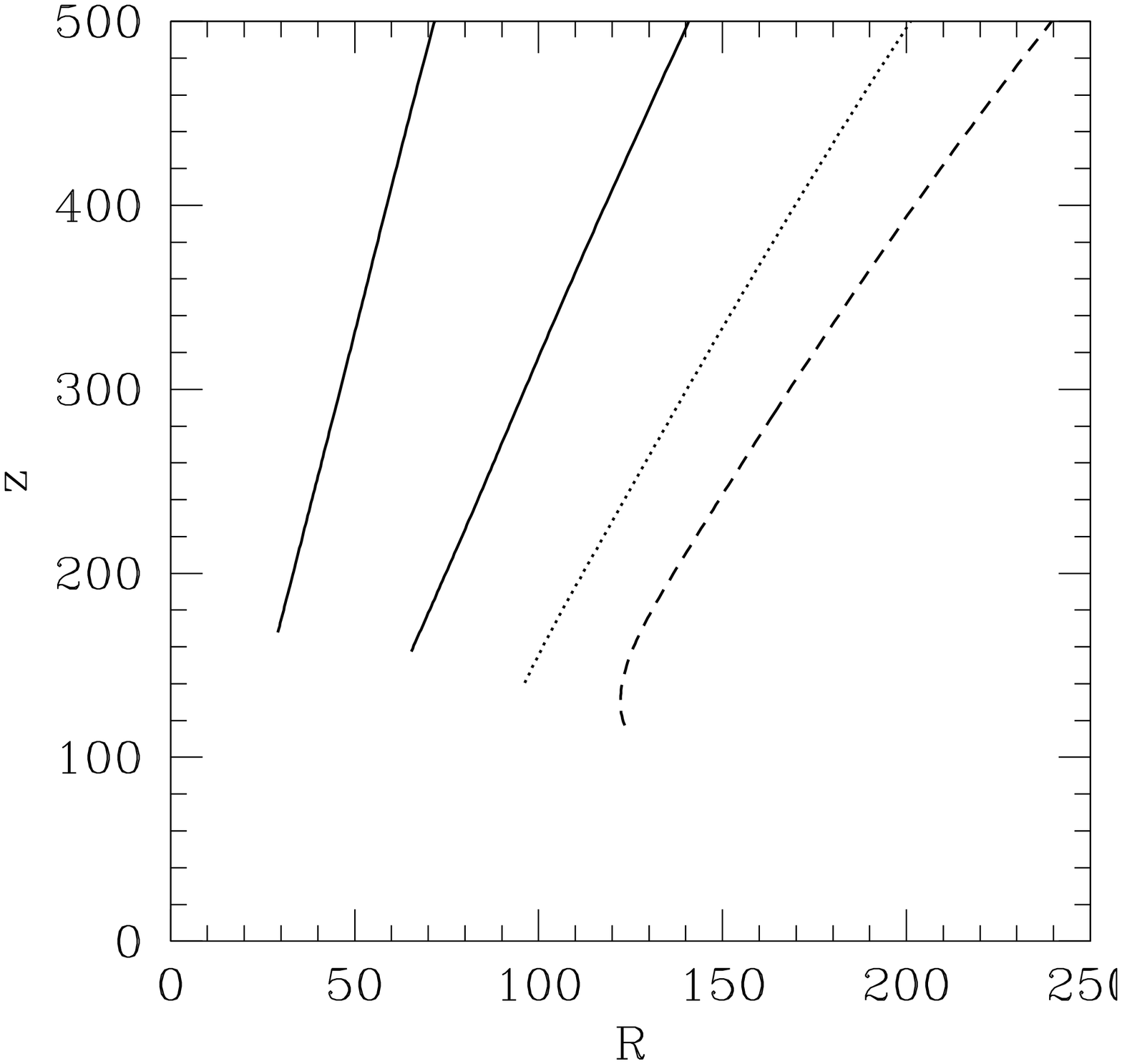}
\includegraphics[width=0.49\columnwidth]{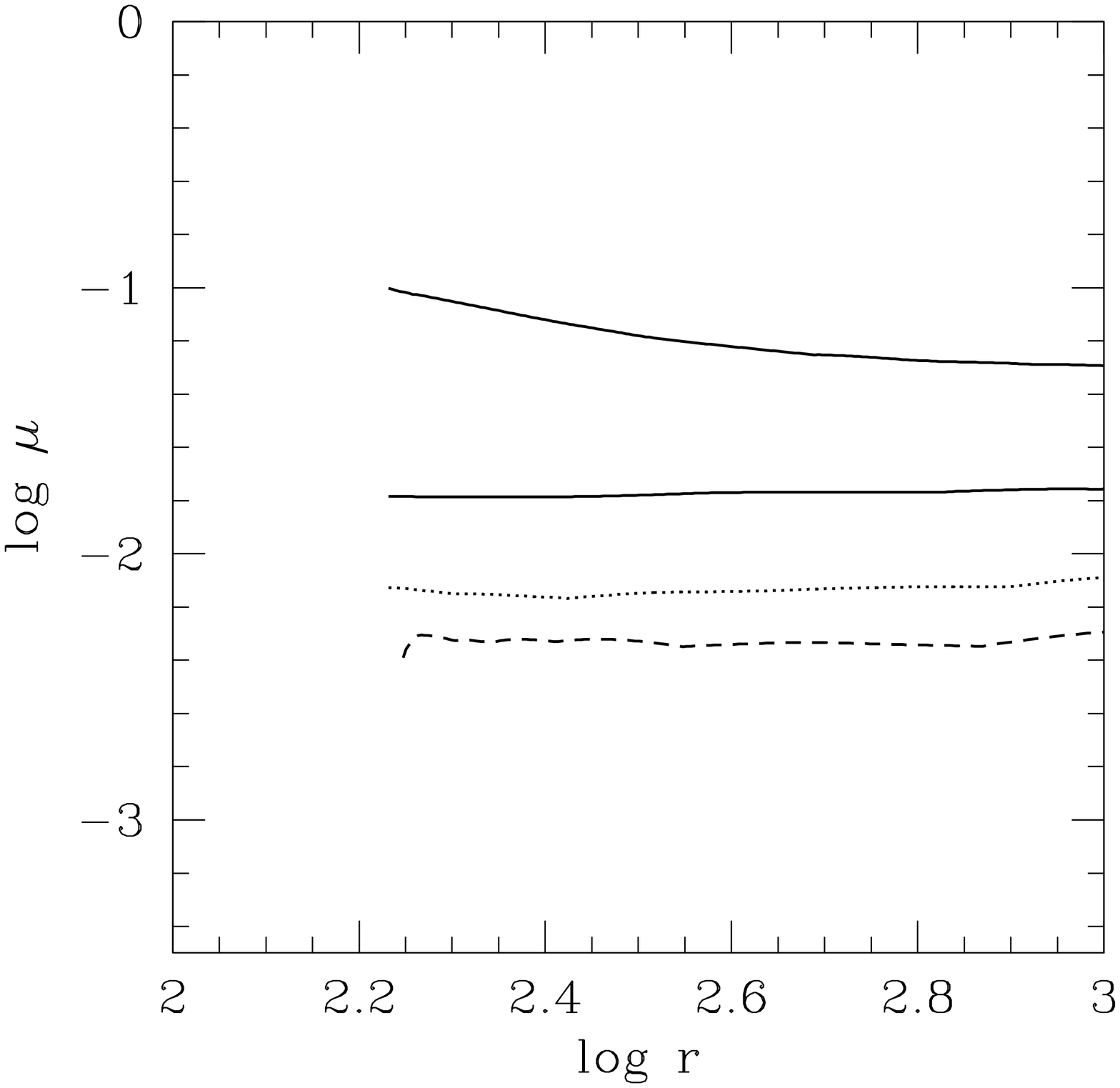}
\end{center}
\caption{Similar to Fig.~\ref{fig:streamlines}, but for the ADAF/MAD
  simulation. The streamline footpoints are at $(r,\theta)=
  (170,0.2)$, $(170,0.4)$, $(170,0.6)$, $(170,0.8)$. All four
  streamlines have positive $\mu$ at their footpoints, and all show
  good conservation of $\mu$.}
\label{fig:MAD_streamlines}
\end{figure}

We have looked a little deeper into why the $\dot{M}_{\rm out}(r)$
profiles we obtain from our simulations show poor convergence.  Figure
\ref{fig:streamlines} shows results corresponding to five streamlines
in time chunk S6 of the ADAF/SANE simulation. These streamlines have
footpoints at $r=r_{\rm strict}=86$ and $\theta=0.2$, 0.4, 0.6, 0.8,
1.0 rad, respectively. All these streamlines have a positive value of
$\mu$ at their footpoints. Since $\mu$ is supposed to be conserved
along each streamline, all of this gas ought to escape.  The right
panel of Fig.~\ref{fig:streamlines} shows the variation of $\mu$ along
each streamline as the gas moves away from the BH. We see that $\mu$
is approximately constant and positive for the the three streamlines
closest to the pole.  However, the two streamlines closer to the disc
show a sudden drop in the value of $\mu$ as one moves outward.
Clearly these streamlines have not reached steady state, since $\mu$
would then be constant. It seems likely that the positive value of
$\mu$ for these streamlines is a transient feature. Unfortunately,
these suspect streamlines carry the most mass.

Figure \ref{fig:MAD_streamlines} shows similar results for four
outflowing streamlines in the ADAF/MAD simulation. Here the
conservation of $\mu$ along outgoing streamlines is satisfied much
better. In addition, the value of $\mu$ is generally larger, which
indicates that the outflowing gas carries more energy per unit rest
mass.

\subsection{Convection}
\label{sec:convection}

A secondary goal of this study is to investigate the importance of
convection in magnetized ADAFs. It is well-known that the entropy
profile in an ADAF has a large negative gradient, making the flow
highly unstable by the Schwarzschild criterion. However, an ADAF also
has angular momentum increasing outward, which has a stabilizing
effect on convection.

For axisymmetric rotating flows, the two Hoiland criteria determine
whether or not gas is convectively unstable.  The same criteria are
likely to remain approximately valid also in magnetized flows, so long
as the field is reasonably weak, since the long-wavelength convective
modes are effectively hydrodynamical \citep{Narayan+02}.  In addition,
since convection is a local instability, the relativistic versions of
the Hoiland criteria \citep{seguin1975} carry over directly to general
relativity by the equivalence principle.

\begin{figure}
\begin{center}
\includegraphics[width=\columnwidth]{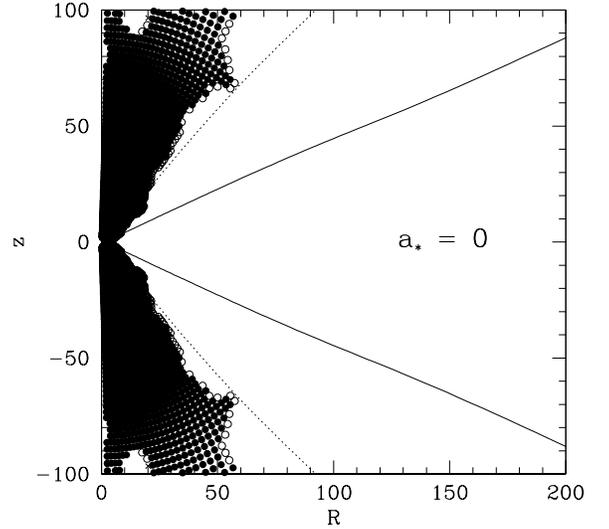}
\end{center}
\caption{Analysis of convective stability of the ADAF/SANE
  simulation. Results are shown for time chunk S6 using time- and
  azimuth-averaged, symmetrized, simulation data. At each point
  $(R,z)$ in the poloidal plane, the maximum growth rate $\gamma$
  according to the two Hoiland criteria are computed. Stable regions
  are shown by blank areas. Unstable regions where $\gamma <
  \Omega_K/30$ are indicated by crosses, regions with $\Omega_K/30 \le
  \gamma < \Omega_K/10$ are indicated by open circles, and regions
  with $\gamma\ge\Omega_K/10$ are indicated by filled circles. The
  solid and dotted lines correspond to one and two density scale
  heights, respectively. Note that the accretion flow is stable to
  convection over the entire inflow region. The instability near the
  poles is not significant since the analysis is not valid there.}
\label{fig:SANE_conv}
\end{figure}

\begin{figure}
\begin{center}
\includegraphics[width=\columnwidth]{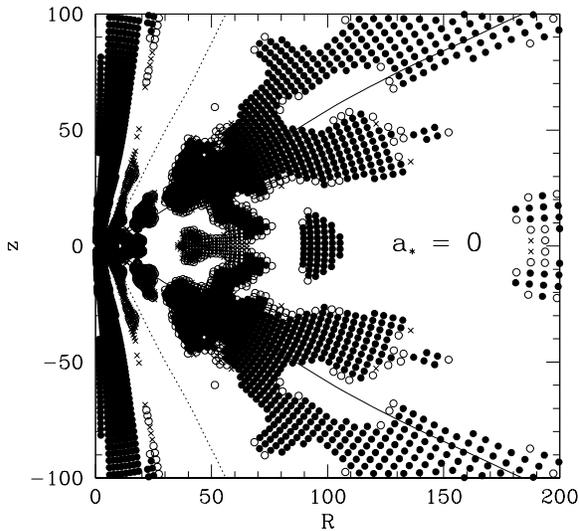}
\end{center}
\caption{Similar to Fig.~\ref{fig:SANE_conv} but for the ADAF/MAD
  simulation.}
\label{fig:MAD_conv}
\end{figure}

We have analyzed the final time chunk S6 in the ADAF/SANE simulation
to determine the level of convective instability in the accretion
flow.  Figure \ref{fig:SANE_conv} shows the result. In brief, all the
fluid within two scale heights of the mid-plane appears to be
convectively stable. The gas is certainly turbulent (see
Fig.~\ref{fig:snapshot}) -- this is what enables it to accrete -- but
it is apparently not convective, at least by the Hoiland criteria.
Rather, the turbulence seems to be entirely the result of the
MRI. Could magnetic fields be confusing the issue? We think this is
unlikely. Analytical studies of convection in the presence of magnetic
fields \citep{BH02,Narayan+02} show that magnetic fields generally act
in such a way as to stabilize convection.  That is, a fluid
configuration that is convectively unstable could be made stable by a
suitable field, but not the other way round. Of course, the magnetic
field might induce its own instability, e.g., MRI, but this can no
longer be considered convection.  We intend to explore this question
in greater depth in the future.

Figure \ref{fig:MAD_conv} shows the convection properties of the
ADAF/MAD simulation. Based on the Hoiland criteria, it appears that
the MAD simulation is more unstable to convection compared to the
ADAF/SANE simulation. This is not surprising. The gas rotates much
more slowly and hence the stabilizing effect of rotation, which we
think is the primary reason for the lack of convection in the
ADAF/SANE simulation, is no longer effective.  We caution, however,
that the magnetic stress is larger in the MAD simulation, and the
Hoiland criteria do not include the effect of this stress.  By the
argument in the previous paragraph, the magnetic field might well be
strong enough to switch off the convective instability even in those
regions where the Hoiland criteria indicate instability.  The
accreting gas in the MAD simulation has very little turbulence, so it
certainly does not manifest any of the usual features of turbulent
convection. We suspect that the flow is in a state of frustrated
convection as proposed by \citet{Pen+03}.

\section{ADAF or CDAF or ADIOS?}
\label{sec:ADAFor}

As originally defined, an ADAF is any accretion flow in which energy
advection is more important than energy loss through radiation. In
this sense, the term is all-inclusive. However, sometimes the name
ADAF is used in a more restrictive sense, where the flow is not only
advection-dominated but also has negligible mass loss through a wind
and is not strongly convective. If we further restrict ourselves to a
flow that shows self-similar behavior, we have the classic ADAF
scalings \citep{NY94,NMQ98},
\begin{equation}
v_r \sim -\alpha\, r^{-1/2}, ~~ \rho \sim \dot{M} \alpha^{-1}
r^{-3/2}, ~~ \Omega \sim (5/3-\Gamma)^{1/2}\,r^{-3/2},
\label{eq:ADAF}
\end{equation}
where $\dot{M}$ is the steady mass accretion rate, $\alpha$ is the
viscosity parameter, $\Omega$ is the angular velocity, and $\Gamma$ is
the adiabatic index.  These scalings follow from basic conservation
laws and the $\alpha$ prescription for viscosity. By assumption, there
is no mass outflow.

In the same spirit, the convection-dominated accretion flow (CDAF,
\citealt{NIA00,QG00b}) is an accretion flow in which the dynamics are
determined by conservation laws plus a steady outward flux of energy
carried by convection. This requirement gives the following CDAF
scalings,
\begin{equation}
v_r \sim - r^{-3/2}, ~~ \rho \sim \dot{M} r^{-1/2}, ~~ \Omega \sim
r^{-3/2}.
\label{eq:CDAF}
\end{equation}
Once again, there is no mass outflow.

Finally, the advection-dominated inflow-outflow solution (ADIOS,
\citealt{BB99}) describes a system in which a strong wind carries away
mass, angular momentum and energy.  Nothing is conserved in this
model, so there is considerable freedom in the form of the
solution. It is generally assumed that quantities behave as power-laws
of radius, which motivates the following ADIOS scalings,
\begin{equation}
v_r \sim - \alpha\, r^{-1/2}, ~~
\rho \sim r^{-3/2+s}, ~~
\Omega \sim r^{-3/2},
\label{eq:ADIOS}
\end{equation}
where $s$ is a free index which can have a value anywhere between 0
(self-similar ADAF) and 1 (maximal ADIOS). The mass outflow rate in
this model scales as $\dot{M}_{\rm out} \propto r^s$. Recently,
\citet{Begelman12} has presented arguments suggesting that
$s\approx1$.

All of the above models are based on a fluid description, without
allowing explicitly for magnetic fields. We believe this is
reasonable, at least for the ADAF/SANE simulation, where the magnetic
stress behaves to a good approximation like viscosity, and the
magnetic pressure is not very important relative to gas
pressure. \citet{AF06} have developed self-similar solutions for
magnetized ADAFs.  However, they assume a purely toroidal field (no
shear stress) and consequently have to invoke
$\alpha$-viscosity. Moreover, their solutions are similar to the
ADAF/ADIOS solutions mentioned above so long as the magnetic pressure
is modest, as in the ADAF/SANE simulation.  This last condition may
not be true for the ADAF/MAD simulation.  However, even for a MAD
flow, the model of \citet{AF06} is not appropriate since it assumes a
toroidal field, whereas the key feature of the MAD solution is a
strong poloidal field.

We have shown in \S\ref{sec:SANE/MAD} that the ADAF/SANE and ADAF/MAD
simulations appear not to be convective, to the extent we can tell
from the Hoiland criteria. We did not include the effect of the
magnetic field, so we cannot make any firm statements regarding
convection. Nevertheless, for the present, we will assume that neither
simulation is a full-fledged CDAF.  Also, neither flow has significant
mass outflow up to $r\sim100$.  We can thus say that the simulations
are best described as ``basic'' ADAFs\footnote{By ``basic ADAF'' we
  simply mean an ADAF that has no convection and no significant
  outflows. Systems with convection (CDAFs) and strong outflows
  (ADIOS) are still ADAFs in the general sense of the term, but they
  are not ``basic ADAFs''.}  over this radius range, though it is
possible that they are just beginning to make a transition to the
ADIOS state beyond $r=100$.  From equations (\ref{eq:ADAF}) and
(\ref{eq:ADIOS}), we see that both solutions predict $|v_r| \sim
\alpha r^{-1/2}$, which can be checked.

\begin{figure}
\begin{center}
\includegraphics[width=0.49\columnwidth]{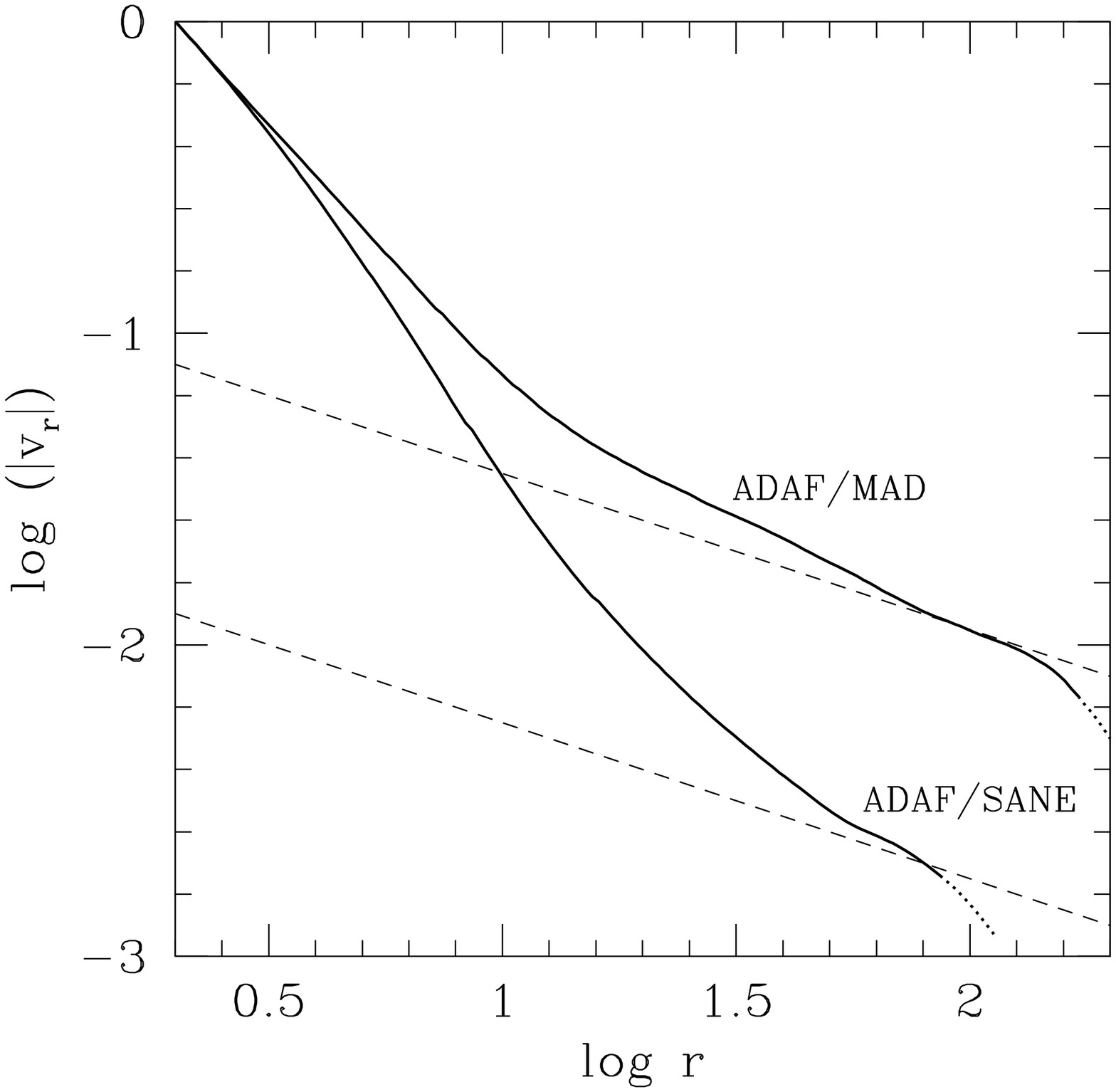}
\includegraphics[width=0.49\columnwidth]{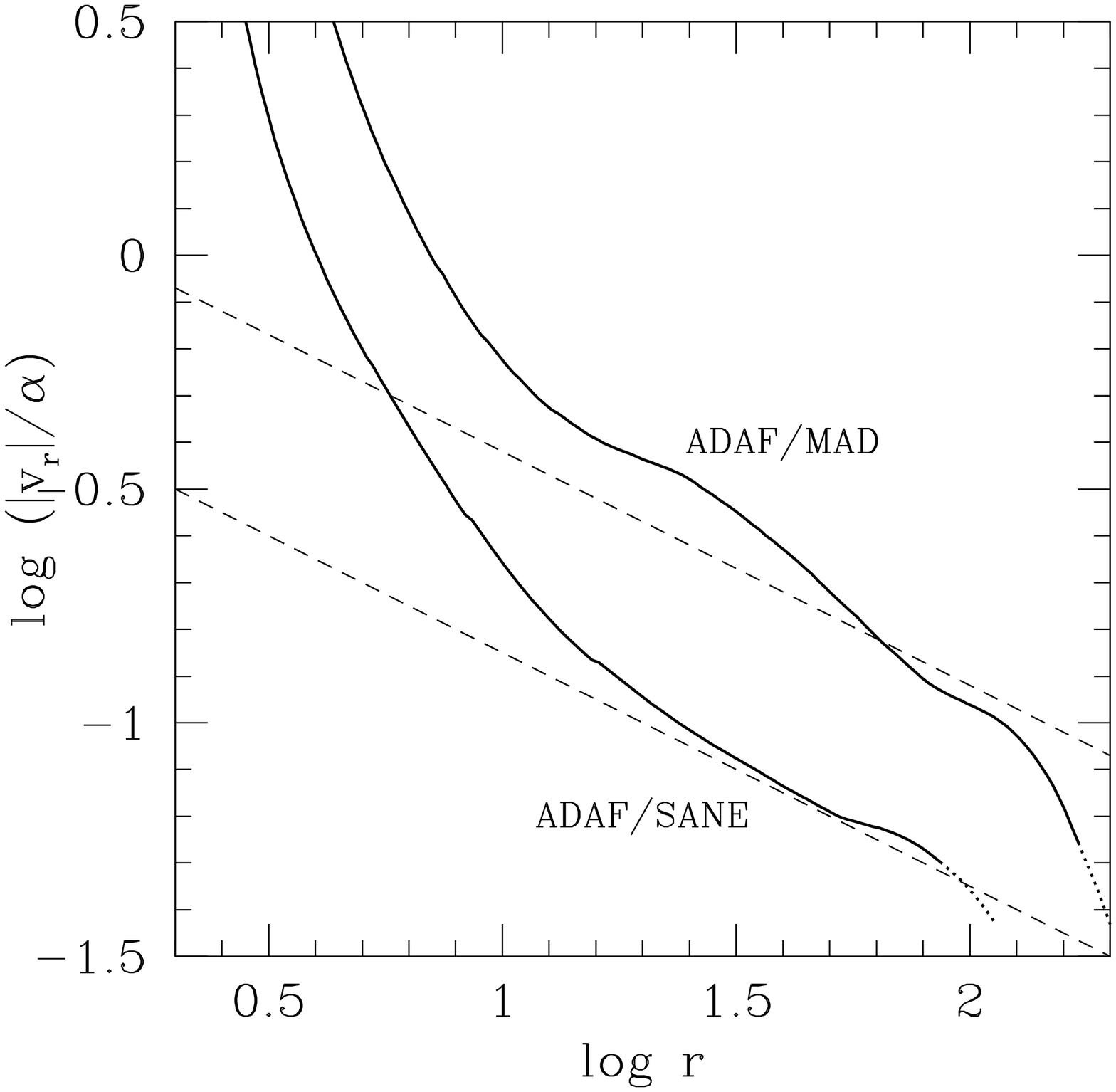}
\end{center}
\caption{Left: Radial velocity $|v_r(r)|$ of the gas in time chunk S6
  of the ADAF/SANE simulation (from Fig.~\ref{fig:SANE_velangmmtm})
  and time chunk M5 of the ADAF/MAD simulation (from
  Fig.~\ref{fig:MAD_velangmmtm}).  The two dashed lines have slope
  equal to $-1/2$, the value expected in the self-similar regime for
  both a basic ADAF and an ADIOS. Over most of the volume, the
  velocity varies more rapidly with radius than expected for a
  self-similar solution. Right: Similar to the previous panel, but
  showing the quantity $|v_r(r)|/\alpha(r)$. Note that the ADAF/SANE
  model agrees much better with the self-similar model, except as the
  gas approaches the ISCO ($r_{\rm ISCO}=6$).}
\label{fig:vel_SS}
\end{figure}

The left panel in Fig.~\ref{fig:vel_SS} shows the velocity profiles in
the final time chunks, S6 and M5, of the ADAF/SANE and ADAF/MAD
simulations. There is some indication that, at the outermost radii of
the respective converged regions, the velocity is settling to the
expected $r^{-1/2}$ dependence.  However, over most of the flow, the
velocity varies more steeply with radius. Part of the explanation is
that, in the self-similar regime, the radial velocity of an ADAF is
approximately given by $|v_r| \sim \alpha (h/r)^2 v_{\rm ff} \sim
10^{-2} v_{\rm ff}$. However, at the BH horizon the gas must have
$|v_r| = v_{\rm ff} = c$. The radial velocity thus has to transition
from its self-similar value to the free-fall value. It takes a
substantial range of $r$ to achieve this, especially in the ADAF/SANE
simulation.  The radial velocity in the MAD simulation is larger,
$|v_r| \sim 0.1 v_{\rm ff}$, so this flow is able to follow the
self-similar scaling closer to the BH.

A second effect is also in operation, viz., the effective $\alpha$ of
the accreting gas varies with $r$.  The right panel in
Fig.~\ref{fig:vel_SS} corrects for this by plotting $|v_r|/\alpha$,
where $\alpha(r)$ is estimated directly from simulation data for gas
within one density scale-height of the mid-plane.  The ADAF/SANE
simulation now shows satisfactory self-similar behavior over a wider
range of $r$.  Removing the $\alpha$ scaling does not improve things
much for the ADAF/MAD simulation.

All of this discussion is based on the radial velocity $v_r(r)$, which
we feel is the natural dynamical variable to consider. Most previous
authors have focused instead on the density profile $\rho(r)$. In
steady state the two quantities are simply related: $\dot{M} \sim \rho
v_r r^2 (h/r) \sim$\,constant. The mid-plane density profiles in our
two simulations are roughly compatible with the velocity results shown
in Fig.~\ref{fig:vel_SS}. Many authors, notably \citet{YWB12}, find
that the density follows a single power-law over a wide range of
radius. The velocity does not show this property
(Fig.~\ref{fig:vel_SS}).

\begin{figure}
\begin{center}
\includegraphics[width=0.49\columnwidth]{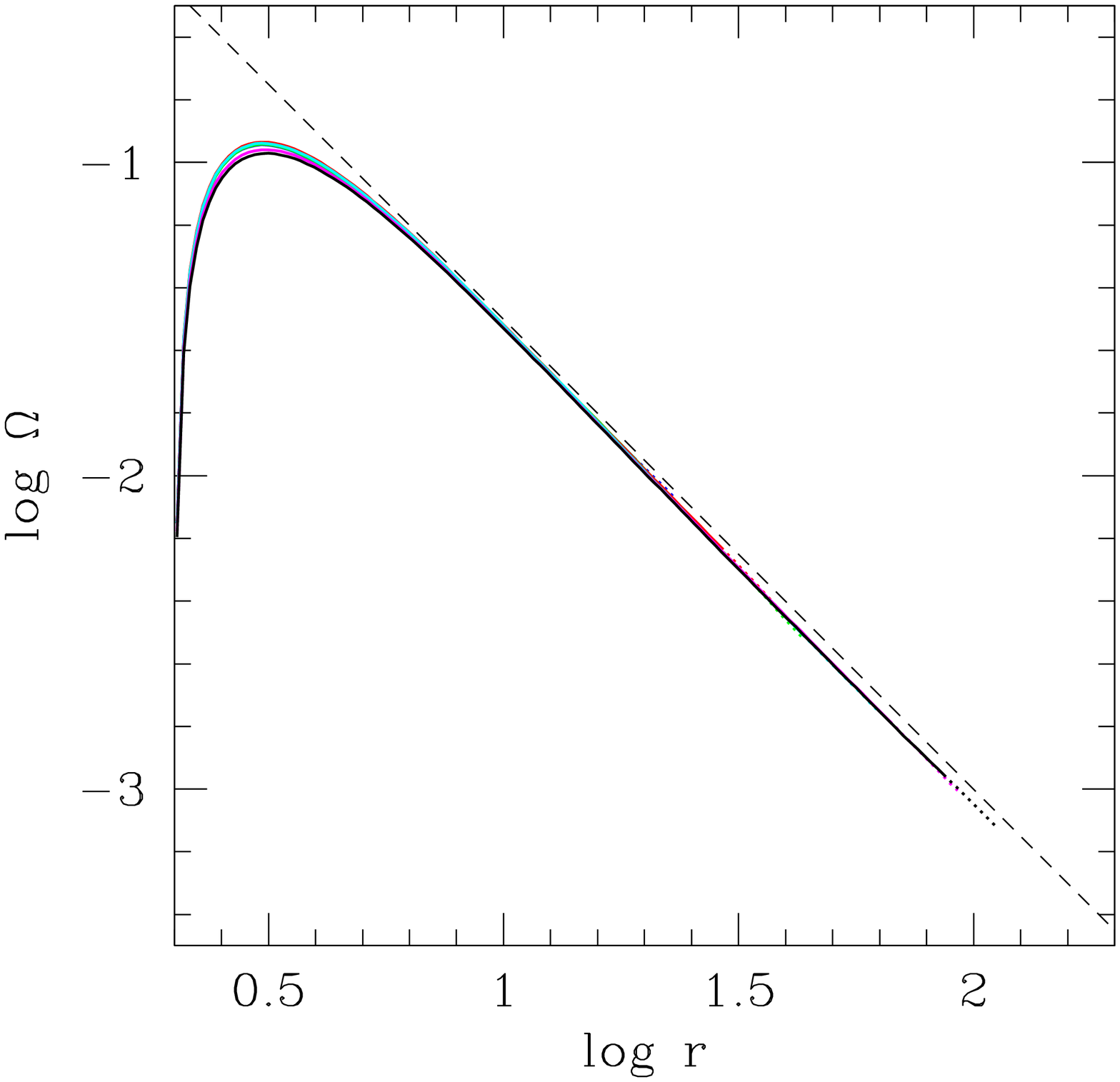}
\includegraphics[width=0.49\columnwidth]{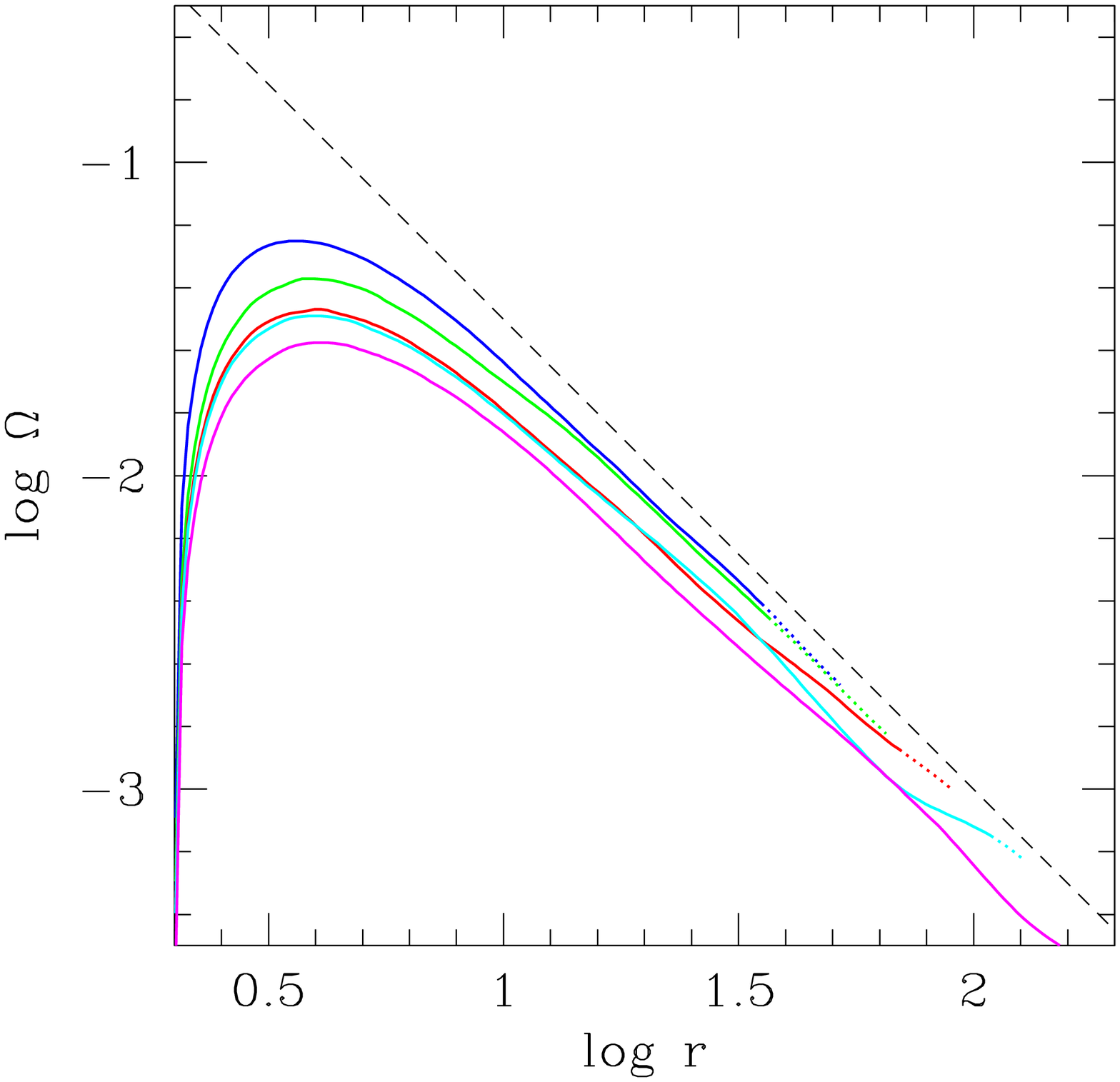}
\end{center}
\caption{Left: Angular velocity $\Omega(r)$ of the gas in time chunks
  S1--S6 of the ADAF/SANE simulation.  The dashed line has a slope
  equal to the self-similar value of $-3/2$. Right: Similar plot
  corresponding to the five time chunks M1--M5 of the ADAF/MAD
  simulation.}
\label{fig:omega}
\end{figure}

Figure \ref{fig:omega} shows the dependence of the gas angular
velocity $\Omega$ in our two simulations. The ADAF/SANE simulation
shows excellent convergence in the sense that the $\Omega(r)$ curves
from different time chunks agree very well with one another. Moreover,
the angular velocity follows the analytical $r^{-3/2}$ scaling quite
accurately.  However, the normalization is not correct. Since
$\Gamma=5/3$, the self-similar ADAF model predicts $\Omega\sim0$ (see
eq.~\ref{eq:ADAF}), whereas we find distinctly non-zero rotation in
our simulation.

The likely explanation is that the simulation behaves, not like the
steady state self-similar solution of \citet{NY94}, but rather like
the similarity solution derived by \citet{Ogilvie99}. The latter
solution describes the evolution of an advection-dominated flow as a
function of both $r$ and $t$, starting from an initial narrow ring of
material.  With increasing time, the flow evolves in a self-similar
fashion. Most interestingly, in Ogilvie's solution, the angular
velocity does not go to zero anywhere except in the region $r\to0$. In
fact, over most of the volume, the rotation rate remains a substantial
fraction of the Keplerian rate, exactly as in our simulations. Since
we started our simulations with an initial torus of material, the
similarity solution is a better point of reference than the
self-similar solution; the latter is valid only at asymptotically late
time when the flow has reached steady state at all $r$.

\begin{figure}
\begin{center}
\includegraphics[width=0.49\columnwidth]{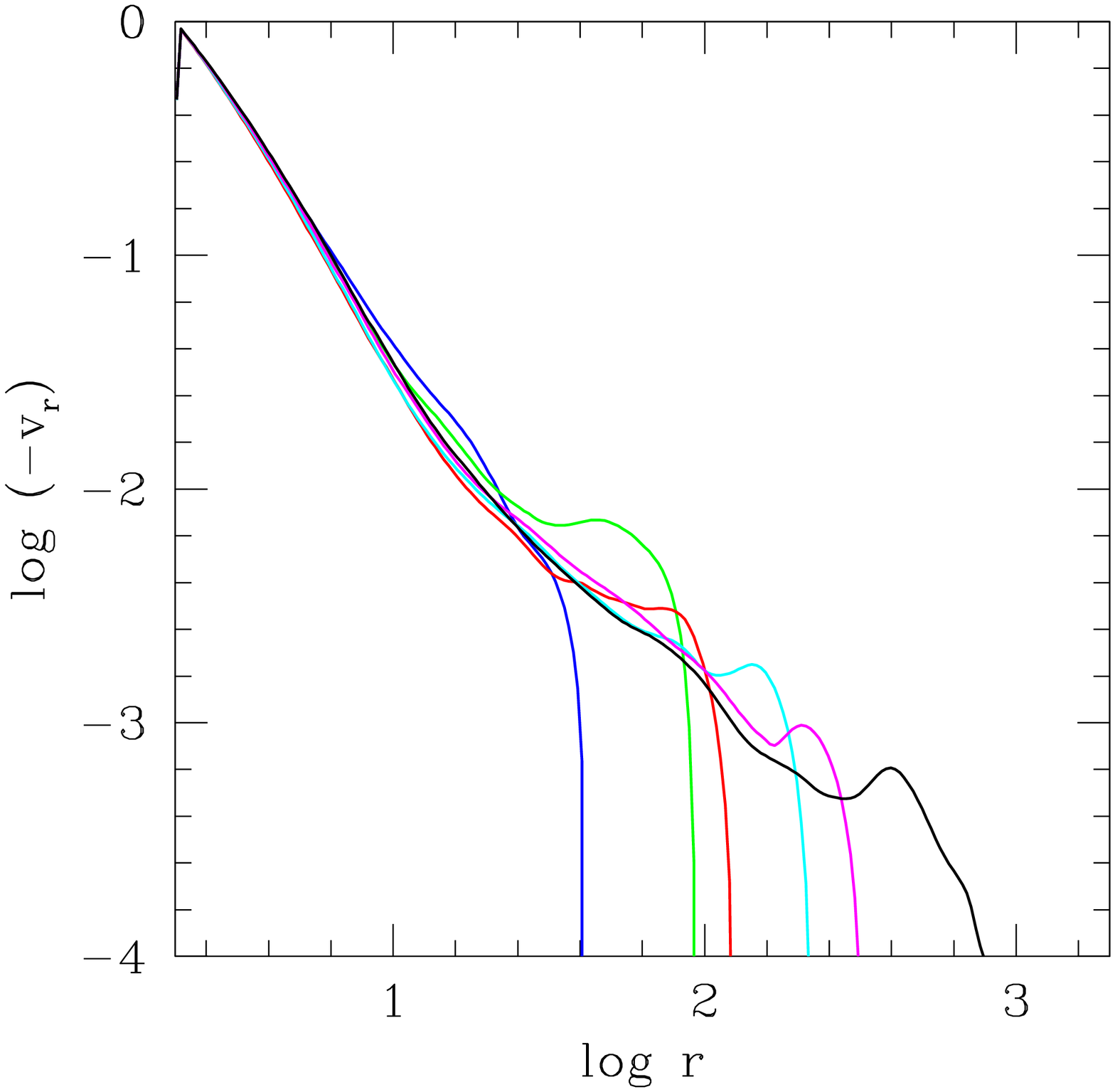}
\includegraphics[width=0.49\columnwidth]{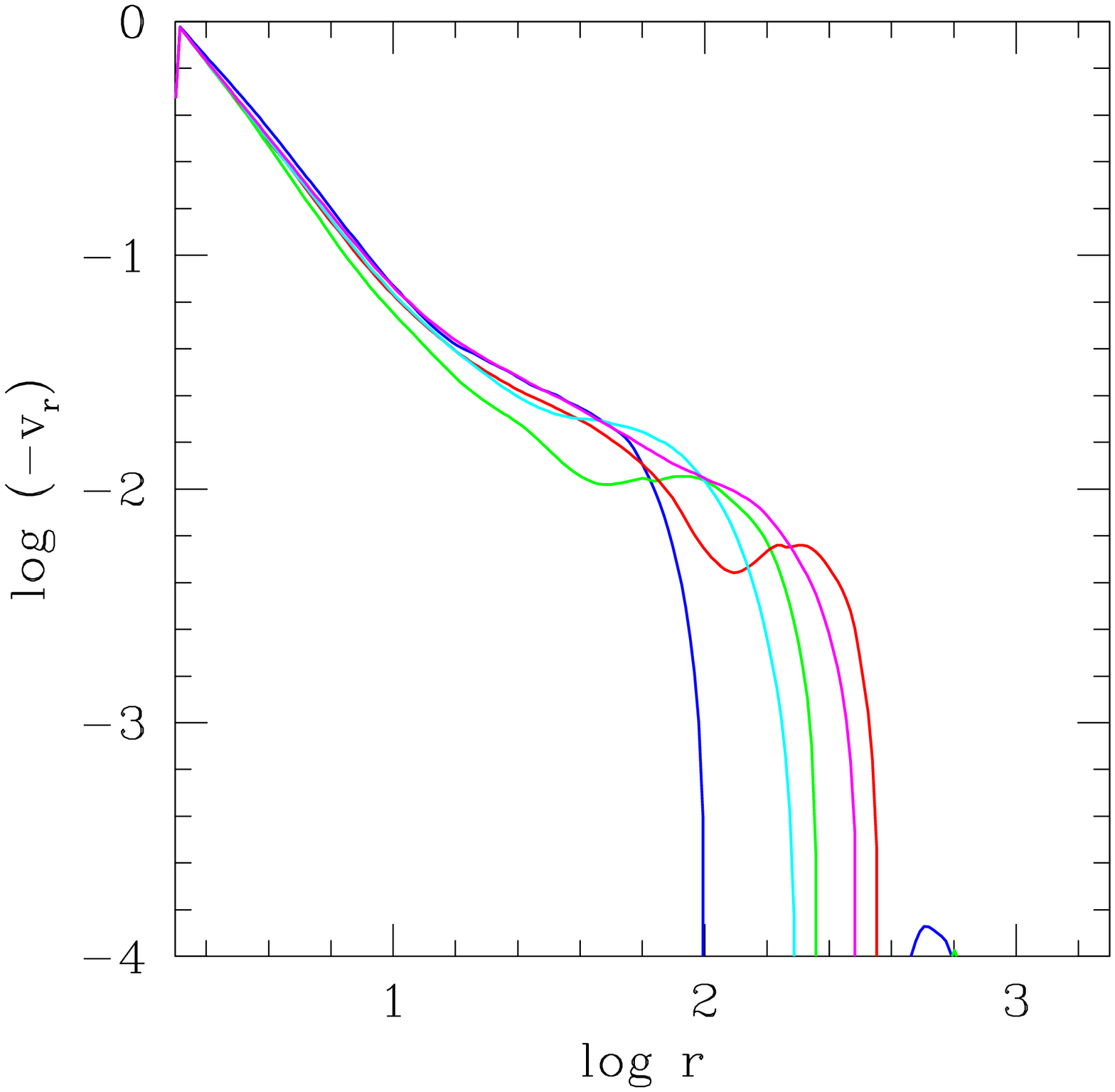}
\end{center}
\caption{Left: Radial velocities vs $r$ for time chunks S1--S6 of the
  ADAF/SANE simulation. The colour code is the same as in
  Fig.~\ref{fig:SANE_jdotedot}. Note that each curve dives down
  suddenly at a certain radius. This is the stagnation radius for that
  time chunk. Beyond this radius, the mean velocity is outward because
  of the viscous relaxation of the initial torus.  Right:
  Corresponding results for the ADAF/MAD simulation, with colour code
  as in Fig.~\ref{fig:MAD_jdotedot}.}
\label{fig:velstag}
\end{figure}

As a further comparison between the ADAF/SANE simulation and Ogilvie's
(1999) similarity solution, Fig.~\ref{fig:velstag} displays again the
radial velocity profiles for different time chunks, but now shown over
an extended range of radius. The velocity in each profile dives
suddenly to zero and becomes negative at a ``stagnation'' radius
$r_{\rm stag}$. We see that $r_{\rm stag}$ increases with increasing
time, as expected for the similarity solution.  The analytical
solution predicts $r_{\rm stag} \propto t^{2/3}$, which means that
$r_{\rm stag}$ should increase by a factor $\sim10$ between chunks S1
and S6. The actual increase is a factor of 20. We view this as good
agreement.

The ADAF/MAD simulation results shown in the right panels of
Figs. \ref{fig:omega} and \ref{fig:velstag} are less convincing.  This
simulation has a strong magnetic field and an arrested mode of
accretion which, based on the evidence of all the diagnostics plotted
in various figures, makes the flow behave more
erratically. Analytically, the MAD regime is sufficiently different
from the SANE regime that we cannot expect either the self-similar
ADAF solution or Ogilvie's similarity solution to be a good
description. 

As already stated, there is a hint near the outer edges of the
ADAF/SANE and ADAF/MAD simulations that ADIOS-like behavior is
beginning to take hold. If we had a larger range of radius in inflow
equilibrium, it might be possible to estimate how the outflow rate
varies with radius and thereby determine the index $s$ in the scaling
$\dot{M}_{\rm out} \propto r^s$. Unfortunately, this is out of reach
with our current simulations. \citet{YWB12} estimate from their large
dynamic range 2D hydrodynamic simulations that $s \sim0.4-0.5$.

\section{Summary and Discussion}
\label{sec:summary}

The main highlights of the present work are: (1) We have run our
simulations for an unusually long time in an effort to approach a
steady state ADAF as closely as possible over a wide range of radius.
(2) We have explored the role of the initial magnetic field topology.
With respect to the latter, we have considered two very different
limits: (1) an ADAF/SANE simulation (SANE = ``standard and normal
evolution''), which is a good proxy for an ADAF model in which the
magnetic field is merely an agent that causes angular momentum
transport (``viscosity'') but plays no important dynamical role, and
(2) an ADAF/MAD simulation (MAD = ``magnetically arrested disc''),
where the magnetic field is strong enough to alter substantially the
dynamics of the gas and to drive the system to a magnetically arrested
state \citep{INA03,NIA03,Tchekhovskoy+11,MTB12}.

Our key result is that, for radii out to $r\approx100$ (gravitational
units, $GM/c^2$), there is not much mass loss to an outflow.
Turbulence certainly leads to both inward and outward gas motions.
However, when we consider the time-averaged gas flow and how much gas
flows out with enough energy to escape from the gravitational
potential of the BH, it turns out to be only a fraction of the net
mass accretion rate $\dot{M}_{\rm BH}$ into the BH. Quantitatively, at
$r\approx100$, we find $\dot{M}_{\rm out} \approx 0.6\dot{M}_{\rm BH}$
for both simulations. Furthermore we view these estimates as upper
limits since the simulations reveal poor convergence in $\dot{M}_{\rm
  out}$ (see Figs.~\ref{fig:SANE_outflow}, \ref{fig:MAD_outflow}).

Because of the very long run times of our simulations, we are unable
to run multiple realizations of the SANE and MAD configurations to
explore variability from one realization to another.  On the other
hand, the long run time allows us to explore convergence as a function
of time within each simulation. We do this by dividing the simulation
data into a number of independent chunks in $\log t$
(\S\ref{sec:preliminary} and Tables \ref{tab:ADAF/SANE},
\ref{tab:ADAF/MAD}).  By comparing different time chunks and checking
how any quantity of interest varies from one chunk to the next, we are
able to decide how reliable the results are for that quantity.

A second important issue is the range of $r$ over which each time
chunk has reached inflow equilibrium. We use two different criteria, a
strict one (eq.~\ref{eq:strict}) and a loose one (eq.~\ref{eq:loose}),
and estimate for a given chunk the limiting radii, $r_{\rm strict}$
and $r_{\rm loose}$, corresponding to each of these criteria (Tables
\ref{tab:ADAF/SANE}, \ref{tab:ADAF/MAD}).  Many properties of the gas
show good convergence among different time chunks when we limit our
attention to radii $r\leq r_{\rm strict}$. The results are less
convincing with the loose criterion.  However, even with the strict
criterion, we find that some questions such as the amount of mass loss
in outflows cannot be answered with confidence.

We initialized the ADAF/SANE simulation with a number of poloidal
magnetic loops (Fig.~\ref{fig:SANE_init}) in an attempt to achieve an
accretion flow with very little net flux at each radius. By and large
this simulation behaved the way we hoped it would. In particular, the
magnetic flux at the BH horizon, measured by the parameter $\phi_{\rm
  BH}$, did not come close to the limiting MAD value of 50 (except for
one brief glitch at time $t \sim 140,000$, see
Fig.~\ref{fig:time_evolve}). Thus we believe the ADAF/SANE simulation
is a believable representation of an ADAF system. We could have
avoided the MAD regime more effectively by starting the simulation
with a purely toroidal field, as in Model A of \citet{INA03} or Model
A0.0BtN10 of \citet{MTB12}. This option is worth exploring in the
future.

The ADAF/SANE simulation shows good convergence and behaves as
expected. The radial velocity, angular velocity, angular momentum and
disc thickness profiles as a function of $r$ agree well between
different time chunks (Figs.~\ref{fig:SANE_velangmmtm},
\ref{fig:omega}). At large radii, the radial velocity falls well below
free-fall (Fig.~\ref{fig:vel_SS}). This is expected since accretion is
mediated by ``viscous'' angular momentum transport which causes the
velocity to be suppressed by a factor of $\alpha$ relative to
free-fall; there is also a factor of $(h/r)^2$ which causes a further
decrease in the velocity. Interestingly, as discussed in
\S\ref{sec:ADAFor}, the ADAF/SANE simulation is better described by
the similarity solution of \citet{Ogilvie99} than the original
self-similar solution of \citet{NY94}. Nevertheless, the radial
dependence of velocity follows the self-similar solution quite well
(Fig.~\ref{fig:vel_SS}, right panel).

The ADAF/MAD simulation shows quite different behavior compared to the
ADAF/SANE simulation. The inflow velocity is substantially larger and
the angular momentum and angular velocity are substantially smaller
(Figs.~\ref{fig:MAD_velangmmtm}, \ref{fig:omega}). The latter appears
to be an important characteristic of MAD flows. As discussed in
\citet{Tchekhovskoy+12}, the gas brings in very little angular
momentum to the BH and therefore induces little spin-up even for a
non-spinning BH. In the case of a spinning BH, a MAD flow actually
causes spin-down.  The reduced rotation rate of the gas means that
there is less centrifugal support. Consequently, the radial dynamics
are dominated by balance between gravity, gas pressure and magnetic
stress. We find that the gas accretes at about a tenth of the
free-fall speed, which is a factor of several larger than the velocity
in the ADAF/SANE simulation.

Because of the larger radial velocity, the ADAF/MAD simulation reaches
inflow equilibrium over a substantially larger range of radius at a
given time relative to the ADAF/SANE simulation (compare Tables
\ref{tab:ADAF/MAD} and \ref{tab:ADAF/SANE}). On the other hand,
convergence in the sense of agreement between different time chunks is
less convincing. We suspect that the reason is the large-scale ordered
magnetic field in the MAD simulation, which imposes coherent
long-lived structure in the flow.

In terms of the amount of mass outflow, the ADAF/SANE and ADAF/MAD
simulations behave rather similarly. We tried three different criteria
to determine how much gas escapes to infinity at a given radius: one
criterion was based on the Bernoulli parameter $Be$ (eq.~\ref{eq:Be}),
a second on a different Bernoulli $Be'$ that ignores the magnetic
contribution (eq.~\ref{eq:Bep}), and a third on the normalized energy
flux $\mu$ (eq.~\ref{eq:mu}).  The results are nearly identical with
all three criteria, which is reassuring.  Unfortunately, the results
show poor convergence with time.  In particular, the radial variation
of $\dot{M}_{\rm out}(r)$ for the last few time chunks (S4--S6 and
M3--M5) differ by much more than we would expect for a converged
simulation.  Nevertheless, taking the results at face value, we
conclude that the mass outflow rate $\dot{M}_{\rm out}$ becomes
comparable to the net inflow rate $\dot{M}_{\rm BH}$ into the BH at a
radius $r\sim 120 = 60\rH$ in the ADAF/SANE simulation and $r\sim 160
= 80\rH$ in the ADAF/MAD simulation. These radii are fairly far from
the BH. In fact, since our mass outflow rates are upper limits, the
critical radii where mass outflows begin to dominate could be
substantially larger.

Our result that outflows are weak out to $r\,\gsim\,100$ disagrees
strongly with previous work. Many simulations of ADAFs have been
described in the literature (see \S1 for a brief review), and most of
these studies have concluded that there are powerful mass outflows at
radii well below $r=100$. On investigation, it appears that there is a
significant methodological difference between our approach and that
used by previous authors. As explained in \S\ref{sec:outflow}, all of
our calculations are based on time- and azimuth-averaged quantities in
which fluctuations due to turbulence have been eliminated. Only if the
average velocity of gas in a grid cell has a positive radial
component, and furthermore if the gas has enough energy to escape from
the system ($\mu>0$), do we consider the particular gas packet to be
part of an outflow. Most other authors have focused on individual
snapshots of their simulations and counted any gas that happened to be
moving away from the BH as outflow. Since turbulence causes gas to
move to and fro, a good fraction of the gas in any snapshot would be
moving out simply as part of turbulent eddies. However, very little of
this gas would actually leave the system since the velocity vector is
likely to turn round on an eddy time. Moreover, much of the gas would
probably have insufficient energy ($\mu<0$) to climb out of the BH
potential. Indeed, several previous authors have noted, after
presenting very large estimates for the mass outflow rate, that most
of the gas in their ``outflows'' has a negative Bernoulli.

The distinction between the approach taken in previous papers and in
the present work can be appreciated by comparing
Fig.~\ref{fig:snapshot} and Fig.~\ref{fig:SANE_density}.  The snapshot
of the ADAF/SANE simulation in the left panel of
Fig.~\ref{fig:snapshot} shows turbulent eddies down to quite small
radii. A fraction of the gas in each of these eddies is temporarily
moving outward, but none of it is likely to escape to
infinity. However, in the standard approach used to estimate the mass
outflow rate, the outward-moving part of each eddy would be included
as part of $\dot{M}_{\rm out}$. This is likely to lead to a large
overestimate of the mass outflow rate. In contrast, our calculations
use the average flow streamlines shown in
Fig.~\ref{fig:SANE_density}. Consider the final time chunk S6 (lower
right panel). Inside $r\sim30-40$, there are no streamlines with
velocity vectors pointed away from the BH.  Therefore, when we compute
the mass outflow rate, we obtain vanishingly small values of
$\dot{M}_{\rm out}$ for radii $\lsim\,30$
(Fig.~\ref{fig:SANE_outflow}).

Because of the above major difference between our calculations and
those of previous authors, it is hard to compare our results. The one
exception is \citet{MTB12}, who, though basing their work on snapshot
data, explain their calculations in sufficient detail to enable a
comparison. Leaving aside jets, which are not relevant for the
non-spinning BHs considered here, \citet{MTB12} present two distinct
estimates of the mass outflow rate. One estimate is called
$\dot{M}_{\rm mw}$, and it focuses on outflowing gas with positive
$Be'$ (it also imposes a couple of other constraints, see
\S\ref{sec:outflow}). This quantity is closest to our
prescription for estimating the mass outflow. Their second outflow
estimate is called $\dot{M}_{\rm w}$, and it includes essentially all
outflowing gas in each snapshot, independent of $Be$.  This quantity
is close in spirit to mass outflow estimates in many other papers in
the literature, and is in our view an overestimate of the actual mass
loss rate because it includes gas churning in turbulent eddies.

For their Model A0.0BtN10, which is an excellent example of an
ADAF/SANE system around a non-spinning BH, \citet{MTB12} estimate
$\dot{M}_{\rm w}/\dot{M}_{\rm H} \sim 1.2$ at $r=50$ (here
$\dot{M}_{\rm H}$ is the net mass accretion rate into the BH, similar
to our $\dot{M}_{\rm BH}$), which suggests a strong outflow already at
this radius. However, they find $\dot{M}_{\rm mw}/\dot{M}_{\rm H}$ to
be essentially zero. In our ADAF/SANE simulation, at $r=50$ we find
$\dot{M}_{\rm out}/\dot{M}_{\rm BH}=0.07$, i.e., practically zero, in
good agreement with $\dot{M}_{\rm mw}$. In the case of their two
ADAF/MAD systems around non-spinning BHs, A0.0BfN10 and A0.0N100,
\citet{MTB12} find at $r=50$ that $\dot{M}_{\rm mw}/\dot{M}_{\rm H} =
0$, $0.4$, and $\dot{M}_{\rm w}/\dot{M}_{\rm H}=0.6$, $1.1$,
respectively. Our ADAF/MAD simulation gives $\dot{M}_{\rm
  out}/\dot{M}_{\rm BH} = 0.2$, in agreement with the $\dot{M}_{\rm
  mw}$ estimates.  It thus appears that our results are perfectly
compatible with the work of \citet{MTB12}. We are also in agreement
with \citet{Pang+11}, though the latter work is mostly concerned with
the accretion of slowly-rotating gas.

Some papers have argued for strong outflows based simply on the fact
that the radial profile of density and/or velocity do not follow the
standard ADAF scalings given in equation (\ref{eq:ADAF}). Focusing on
the radial velocity, the simulations generally show $|v_r|$ increasing
more rapidly with decreasing radius than expected in the self-similar
solution.  Our simulations too show this effect
(Fig.~\ref{fig:vel_SS}).  It turns out that two separate effects,
neither involving outflows, cause the velocity profile to be modified.

First, because the accreting gas makes a sonic transition as it
approaches the BH and switches to a free-fall mode inside this radius,
we have $|v_r| \sim v_{\rm ff}$ near the BH.  However, the velocity in
the self-similar regime is far below free-fall: $|v_r| \sim \alpha
(h/r)^2 v_{\rm ff}$. The flow needs a considerable range of $r$ to
adjust from one scaling to the other, and we believe this is a large
part of the reason why the velocity profiles seen in simulations look
so different from the simple power-law given in equation
(\ref{eq:ADAF}).  Clear examples of this effect may be seen in the
global 1D models of \citet*{NKH97}, where the non-self-similar zone
extends from the inner boundary to a few tens of gravitational radii.

Secondly, it is the quantity $v_r/\alpha$ that is expected to be
self-similar, not $v_r$ itself. Since $\alpha$ varies with radius in
our simulations (especially in the ADAF/SANE simulation), this causes
an additional deviation in $v_r(r)$. As Fig.~\ref{fig:vel_SS} shows,
removing the $\alpha$ dependence gives a better-behaved velocity
profile that agrees fairly well with the models shown in
\citet{NKH97}.

Another argument for strong outflows that is sometimes used in the
literature is to take the gas density at the outer radius of the
simulation, and to calculate from it the Bondi mass accretion rate
$\dot{M}_{\rm B}$. If the actual mass accretion rate $\dot{M}_{\rm
  BH}$ into the BH in the simulation is much smaller than
$\dot{M}_{\rm B}$, then it is claimed that the difference is because
most of the incoming gas was ejected in an outflow.  The problem with
this argument is that, for a given outer boundary condition on the
density, theory says that the accretion rate via an ADAF will be
smaller than $\dot{M}_{\rm B}$ by a factor $\sim\alpha(h/r)^2\sim$
few~\%.  Thus, having $\dot{M}_{\rm BH}\ll\dot{M}_{\rm B}$ is
perfectly natural for an ADAF; it does not imply strong outflows.
Note, however, that this explanation only goes so far. If it turns out
that $\dot{M}_{\rm BH}$ is much smaller than even
$\alpha(h/r)^2\dot{M}_{\rm B}$, then one has to look for other
explanations such as strong outflows or convection. To our knowledge,
no simulation to date has come close to violating this limit.

ADAFs in nature usually extend over many decades in radius. The ADAF
around Sgr A$^*$, for instance, extends from the BH out to the Bondi
radius at $r\,\gsim\,10^5$. Supermassive BHs in other low-luminosity
AGN similarly have ADAFs extending over 5 or more decades in radius.
In the case of stellar-mass BHs in X-ray binaries, the ADAF is usually
formed by evaporation from a thin disc on the outside \citep{NM08}.
For systems in quiesence, where the mass accretion rate is low, the
transition radius is typically $\sim10^3-10^4$.  In contrast,
simulations of ADAFs are generally restricted to a much smaller range
of radius (but see the recent work of \citealt{YWB12}). How relevant
are simulation results to real systems?

Our views on this question are driven by insights gained from global
1D models of ADAFs such as the ones shown in \citet{NKH97} and
\citet*{CAL97}. These solutions show three zones: an inner zone where
the flow adjusts to the free-fall boundary condition at the BH, an
outer zone where it adjusts to whatever outer boundary condition is
present in the system (Bondi or disc evaporation), and a middle zone
where the flow is more or less self-similar. If a simulation covers a
large enough radius range to capture some piece of the middle zone,
then it would be straightforward to stretch out the self-similar
regime to any radius range we require. We suspect that the two
simulations presented in this paper may have just managed to develop a
piece of the middle zone, but we do not have any proof of this. In any
case, we believe that only by obtaining inflow equilibrium over a
sufficiently large range of radius can we hope to use simulations to
make useful statements about real flows.

It should be noted that the properties of the self-similar middle zone
are fairly insensitive to parameters. There is an obvious dependence
on $\alpha$ (see eq.~\ref{eq:ADAF}) and a modest dependence on
$\Gamma$,\footnote{In the low-$\dot{M}$ RIAF branch of ADAFs, it is
  believed that the gas is two-temperature with non-relativistic ions
  and relativistic electrons \citep{NM08}. If we take $T_e/T_i =0.1$,
  a reasonable value for an ion-dominated ADAF, then we expect
  $\Gamma=1.61$. In the simulations presented here we have set
  $\Gamma=5/3$, which is close enough, although technically in the
  ``unphysical region'' discussed by \citet{MM07}. In the ADAF
  literature, $\Gamma=1.5$ is often used, but this is because those
  models wish to include the effect of a tangled magnetic field, which
  has an effective $\Gamma=4/3$. In numerical MHD simulations, the
  magnetic field is treated as an independent component, so we are
  only concerned with the gas. Any choice $\Gamma\,\gsim\, 1.6$ is
  probably reasonable.} but virtually nothing else matters.  In other
words, provided ADAF conditions are satisfied, the accretion flow will
head towards the particular disc thickness $h/r$ and Bernoulli $Be(r)$
it wants in the middle zone, regardless of the precise outer boundary
conditions.  This is demonstrated for instance in Fig.~5 of
\citet{NKH97}, where three very different outer boundary conditions on
the gas rotation and temperature all give pretty much identical
solutions in the middle zone. The same is true also for $Be$ (Fig.~7
of the same paper). \citet{YBW12} have carried out hydrodynamic
simulations of ADAFs where they find that $Be$ of the accreting gas is
mainly set by the outer boundary condition. It is possible that their
models do not extend over a large enough range of radius to sample the
self-similar zone.

All the results presented here refer to a non-spinning BH. This is the
simplest version of the ADAF problem, where there is no additional
complication from central energy injection by a spinning BH. It is
also the case that relates most directly to theoretical work as well
as to non-relativistic MHD simulations. In the case of ADAFs around
spinning BHs, although a large fraction of the energy from the BH
seems to go out in a relativistic jet \citep{Tchekhovskoy+11}, some of
it presumably propagates into the accreting flow. This energy very
likely will induce extra mass loss, as seen in the simulations
described by \citet{Tchekhovskoy+11} and \citet{MTB12}. Sorting out
the BH spin effect from the intrinsic effect due to ADAF physics is
left for future work.

In addition to outflows, we have also described in this paper a
preliminary analysis of convection. In brief, the ADAF/SANE simulation
shows no evidence of convective instability
(Fig.~\ref{fig:SANE_conv}), while the ADAF/MAD simulation is
apparently unstable by the Hoiland criteria over a part of its steady
state region (Fig.~\ref{fig:MAD_conv}). However, there is little
evidence in the MAD simulation for actual turbulent convection. Hence
we speculate that the ADAF/MAD simulation is probably in a state of
frustrated convection \citep{Pen+03}. Based on our current results, we
are inclined to think that convection is unimportant in ADAFs, whether
SANE or MAD, but this issue needs to be investigated in greater detail
before one can be certain. In particular, it is important to sort out
the effect of the magnetic stress, which is ignored in the Hoiland
criteria. Also, it is possible that the accretion flow is described by
something like the global 1D models in \citet{Abramowicz+02}, where
the flow behaves like a basic ADAF (no outflow, no convection) until a
radius $r\sim35\,\rH=70$, and then switches to a CDAF. We do not have
enough dynamic range in our ADAF/SANE simulation to rule this out.

We note that there are some observational indications against strong
mass loss in ADAFs. \citet{Allen+06} showed that a number of
low-luminosity AGN have radio jets with implied powers that are a
reasonable fraction of accretion energy at the Bondi rate from the
surrounding interstellar medium. In fact, \citet{McNamara+11}
identified systems with $P_{\rm jet} >\dot{M}_{\rm Bondi}c^2$, and
argued that these jets must be powered by BH spin. While it is true
that a rapidly spinning BH can produce a very strong jet, the jet
power is still linked to the accretion power; $P_{\rm jet}$ may be a
factor of a few larger than $\dot{M}_{\rm BH}c^2$, but not much more
\citep{Tchekhovskoy+11,MTB12}. Therefore, the observations mentioned
above mean that a good fraction of the available mass at the Bondi
radius must reach the BH \citep{NF11}.  If mass loss between the Bondi
radius and the BH is very large, as in some versions of the ADIOS
model \citep{BB99,Begelman12}, or if a CDAF is present over a wide
range of radius, there would not be sufficient mass near the BH to tap
the BH spin energy and power the observed jets. We believe that the
above observational evidence, assuming it holds up, drives us towards
one of the following descriptions of the accretion flow: (i) an ADAF
with a weak outflow, i.e., a value of the index $s$ close to 0, or
(ii) an ADAF with a strong outflow ($s\approx1$) but with the outflow
restricted to a small range of radius, say no more than one or two
decades, or (iii) a CDAF with properties and scalings rather different
from the analytical models in the literature \citep{NIA00,QG00b}, or
(iv) a perfectly spherically symmetric Bondi flow. We consider the
fourth possibility unlikely since it requires gas at the Bondi radius
to have an extremely low specific angular momentum.

The interesting differences we find between the ADAF/SANE and ADAF/MAD
simulations brings up the question of which is more relevant for real
systems. The defining feature of a MAD system is that accretion has
dragged in a considerable amount of magnetic flux and has caused the
field to accumulate around the BH. Whether or not accretion can drag
field so effectively has been much debated (e.g.,
\citealt*{Lovelace+11, GO12}, and references therein), but it is agreed that
field-dragging will be most efficient in thick accretion flows such as
ADAFs rather than in thin discs. Assuming that inward advection of
magnetic field does operate effectively in ADAFs, there is typically
more than enough magnetic field available in the external medium to
drive an accreting BH to the MAD state \citep{NIA03}

\section*{Acknowledgements}

The authors thank Jonathan McKinney, Re'em Sari, Alexander
Tchekhovskoy and Feng Yuan for comments on the paper. This work was
supported in part by NASA grant NNX11AE16G and NSF grant
AST-0805832. The simulations presented in this work were performed on
the Pleiades supercomputer, using resources provided by the NASA
High-End Computing (HEC) Program through the NASA Advanced
Supercomputing (NAS) Division at Ames Research Center.  We also
acknowledge NSF support via XSEDE resources at NICS Kraken and
LoneStar.

\bibliography{ms}

\end{document}